\documentclass[fleqn,usenatbib]{mnras}

\usepackage[T1]{fontenc}

\usepackage{graphicx}	% Including figure files
\usepackage{amsmath}	% Advanced maths commands
\usepackage{amssymb}	% Extra maths symbols
\usepackage{caption}
\usepackage{subcaption}
\usepackage{tabularx}
\usepackage[usenames,dvipsnames]{xcolor}
\usepackage{soul}
\usepackage{txfonts}

\title[Kinematics and stellar population of CRDs]{SDSS-IV MaNGA: Integral-field kinematics and stellar population of a sample of galaxies with counter-rotating stellar disks selected from about 4000 galaxies}

\author[D. Bevacqua et al.]{
Davide Bevacqua,$^{1}$\thanks{E-mail: davide.bevacqua@studio.unibo.it}
Michele Cappellari,$^{2}$
Silvia Pellegrini$^{1,3}$
\\
$^{1}$Department of Physics and Astronomy, University of Bologna, via P. Gobetti 93/2, 40129 Bologna, Italy \\
$^{2}$Sub-department of Astrophysics, Department of Physics, University of Oxford, Denys Wilkinson Building, Keble Road, Oxford, OX1 3RH, UK\\
$^{3}$INAF-OAS of Bologna, via P. Gobetti 93/3, 40129 Bologna, Italy
}

\date{Accepted XXX. Received YYY; in original form ZZZ}

\pubyear{2021}

\begin{document}
\label{firstpage}
\pagerange{\pageref{firstpage}--\pageref{lastpage}}
\maketitle

\begin{abstract}
We present the integral-field kinematics and stellar population properties of 64 galaxies (61 are Early-Type galaxies, ETGs) with Counter-Rotating stellar Disks (CRD) selected from about 4000 galaxies in the MaNGA survey, based on evidence of counter-rotation or two velocity dispersion peaks in the kinematic maps. For 17 CRDs, the counter-rotating components can also be separated spectroscopically. The frequency of CRDs in MaNGA is <5\% for ellipticals, <3\% for lenticulars and <1\% for spirals (at 95\% confidence level), consistent with previous estimates. We produced age and metallicity maps, and compared the stellar population properties to those of the general ETGs population. We found that CRDs have similar trends in age and metallicity to ETGs, but are less metallic at low masses, and show flatter age and steeper metallicity gradients, on average. A comparison of the velocity fields of the ionized gas and the stars reveals that in 33 cases the gas corotates with either the inner (15 cases) or outer (18 cases) stellar disk, in 9 cases it is misaligned. In most cases the gas corotates with the younger disk. Evidence of multimodality in the stellar population is found in 31 galaxies, while the 14 youngest and least massive galaxies show ongoing star formation; 14 galaxies, instead, exhibit unimodality, and are the oldest and most massive. As a general result, our work indicates that CRDs form primarily via gas accretion in retrograde rotation with respect to a pre-existing stellar disk.
\end{abstract}


%%%%%%%%%%%%%%%%%%%%%%%%%%%%%%%%%%%%%%%%%%%%%%%%%%

%%%%%%%%%%%%%%%%% BODY OF PAPER %%%%%%%%%%%%%%%%%%

\section{Introduction}

Counter-rotation in galaxies is a phenomenon that occurs between two stellar structures (e.g., two disks, the bulge and the disk, a bar and the disk, etc.), as well as between the gaseous and the stellar disks or two gaseous disks (see \citealt{Corsini_2014} for a review). In this paper, we focus on galaxies hosting two large-scale cospatial counter-rotating stellar disks. Studying this class of objects is interesting because they represent the tip of the iceberg of a (possibly large) class of galaxies in which some of the stars are formed by a separate mechanism from the rest. The detectability of the counter-rotating disk depends on the data quality as well as the characteristics of the two disks. This implies that a larger hidden population than observable must exist. Moreover, if counter-rotating disks are observed, a similar fraction of co-rotating ones must also exist which cannot be detected with the methods discussed in this paper.

The first galaxy with evidence of two counter-rotating stellar disks is the well-studied S0 galaxy NGC 4550 \citep{rubin4550, rix4550}. The presence of two counter-rotating stellar disks in this galaxy was later confirmed with integral-field spectroscopy \citep{Johnston_2012, Coccato_2012} and detailed dynamical modeling \citep{Cappellari_2007}.  Over the last three decades, many other counter-rotators have been found \citep{Krajnovic_2011}, but the overall census is still relatively small, and only few of them have been studied in detail (e.g., NGC 3593 \citealt{Bertola_1996, Coccato_2012} and NGC 5719 \citealt{Vergani_2007, Coccato_2011}). The first reason for the rarity of these objects is that it may be intrinsically difficult to build up such kinematic structures. A second reason is due to observational limits, both technical and of intrinsic nature \citep{Kujiken_1996,Pizzella_2004,katkov,Rubino_2021}; for example, the detection of the counter-rotation depends on the instrumental resolution, as well as on the inclination of the galaxy. Recent large galaxy surveys like Mapping Nearby Galaxies at APO (MaNGA; \citealt{Bundy_2015, Blanton_2017}) and Sydney-AAO Multi-object Integral field spectrograph (SAMI;  \citealt{sami}, \citealt{sami2}), are very promising to determine the incidence of the counter-rotating structures, as well as their origin and intrinsic nature. 

The most puzzling question regarding galaxies with counter-rotating stellar disks is their formation history. In fact, they present a variety of observed properties for the two disks, like their stellar population properties, their masses and extensions, and their thickness. Many different formation scenarios have been proposed so far.
The peculiar kinematics could be a consequence of internal processes, e.g. the dissolution of a bar \citep{Evans_Collett} or resonance capturing \citep{Tremaine_Yu}; however, the significant age difference measured in some cases for the two disks is a strong observational evidence against such scenarios. Recent simulations \citep{Fiteni_2021} also show that internally driven mechanisms are unable to produce counter-rotating disks.

Instead, the origin of counter-rotation may more likely reside in a retrograde accretion of gas from an external source. Simulations have shown that a disk galaxy can form a counter-rotating stellar disk by acquiring a sufficient amount of gas in a retrograde orbit from either a close companion \citep{Thakar_1996,Thakar_1998}, cosmological filaments \citep{Algorry_2014, Taylor_2018, Khoperskov_2020} or a minor merger with a gas-rich dwarf galaxy \citep{Thakar_1997,DiMatteo_2008}; in all these cases, the secondary disk is younger and corotating with the gaseous disk. Although they end up with the same kinematical structure, the formation of the counter-rotating component from these three different gas sources corresponds to different predicted properties for the secondary disk, as their extensions and chemical mixtures. Currently, observations \citep{Vergani_2007, Pizzella_2014, Coccato_2012, Coccato_2015, katkov, katkov_2016, Pizzella_2018}  do not favour any of these scenarios in particular.

Another scenario proposed the formation of the counter-rotating disk in a major merger. Even though most major mergers are highly disruptive events, since they enormously heat the progenitor systems and result in a final morphology far from disk-like, \cite{Puerari_Pfenniger} showed that, with a fine tuning of the initial conditions, a merger between two equally massive disk galaxies with opposite rotation can reproduce the kinematics of the observed counter-rotators. Simulations by \cite{Crocker_2009}, for example, reproduced the observed properties of NGC 4550 as a result of a major merger (also supported by observations of \citealt{Johnston_2012}). More recently, \cite{Martel_Richard} showed that a major merger between two gas-rich disk galaxies can produce counter-rotating stellar disks even though the progenitors rotate in the same direction.

To properly understand the nature of these objects, and validate or reject formation scenarios, a large sample of galaxies with counter-rotating stellar disks is needed. In this paper, we make use of the statistical power of the MaNGA survey, which has observed $\sim$ 10,000 galaxies, to build a large sample of counter-rotators. In section \ref{sect:data} we describe our method to classify galaxies with counter-rotating stellar disks by inspecting their kinematic maps obtained with MaNGA. After building the sample, in section \ref{sect:kinem} we present the procedures to extract the stellar kinematic maps, and look for spectroscopical evidences of the two counter-rotating stellar components; furthermore, we study the alignment or misalignment of the gaseous disk with respect to the stellar ones, by comparing the kinematic position angles.  In section \ref{sect:fitpop}, we study the stellar population properties of the sample. In particular, we present the procedure to extract the age and metallicity maps, and how we estimated the `global' stellar population properties, and gradients; then, we compare the age maps with the gaseous and stellar velocity maps to qualitatively determine whether the gas corotates with the younger or the older stellar disk; finally, we perform regularised fits, looking for the presence of single or multiple stellar populations. In section \ref{sect:results} we present the results, and discuss them in section \ref{sect:discussion}; in section \ref{sect:conclusions} we summarize our work.

\begin{figure}
\includegraphics[width=\columnwidth]{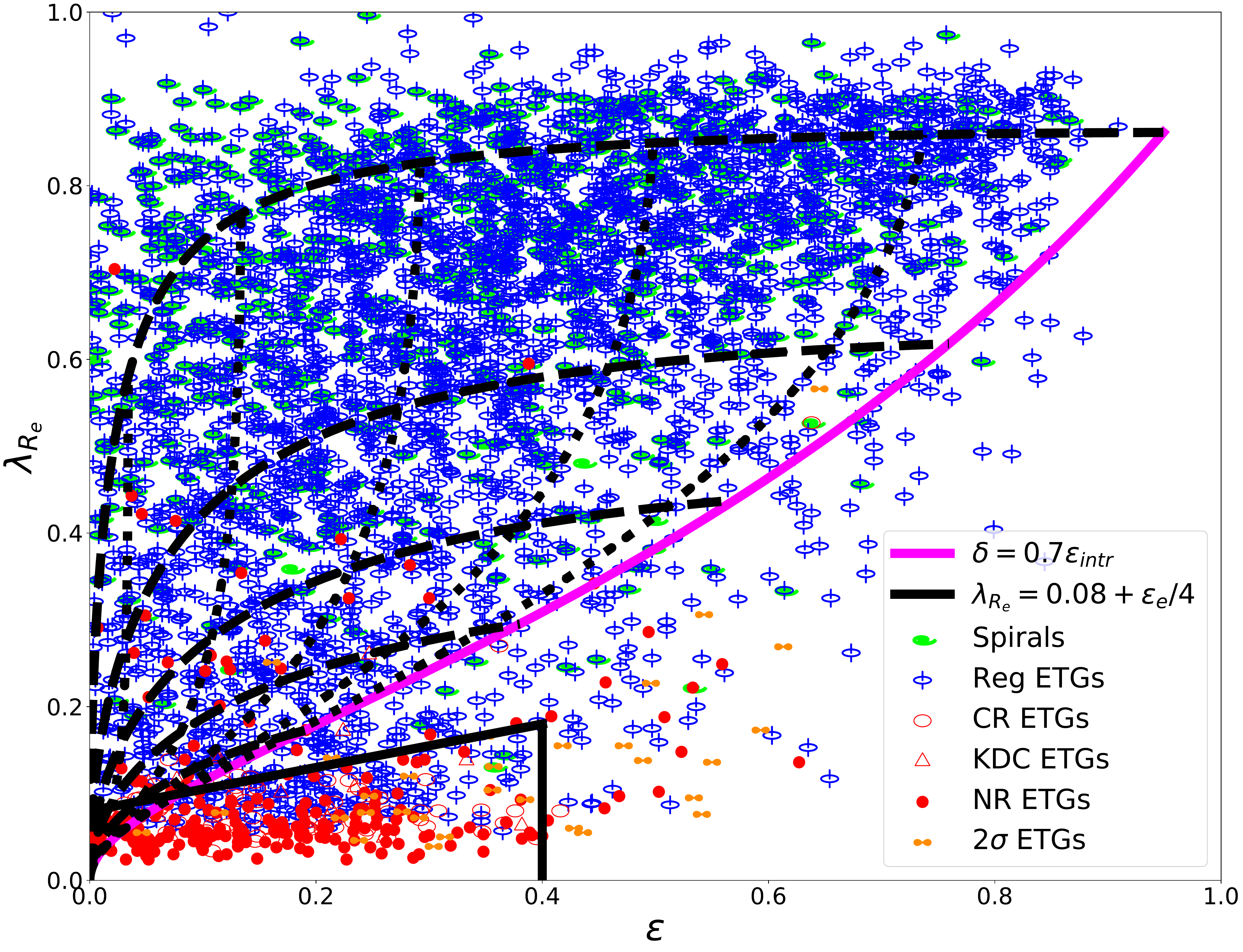}
\caption{Sample of about 4000 galaxies from the DR16 on the ($\lambda_{R_e}, \varepsilon$) diagram. The magenta line, introduced in \protect\cite{Cappellari_2007}, describes the approximate empirical boundary $\delta=0.7 \epsilon_{\rm intr}$ for the maximum anisotropy, at given $\epsilon$, of edge-on fast rotator galaxies. The black dashed/dotted lines show how the magenta line transforms with inclination. In general, fast rotator galaxies lie within the envelope produced by the inclination tracks, while `$2\sigma$-galaxies' (i.e. galaxies with counter-rotating stellar disks) are typically below the magenta line, because counter-rotation lowers the overall measured velocity (and thus $\lambda_{R_e}$). The black straight line is the approximate separation between slow rotators (inside) and fast rotators or $2\sigma$-galaxies (outside) proposed by C16. The plotted values were determined by \citet{G19benchmark}, who also provided the kinematic classification indicated by the symbols.}
\label{fig:lamelgraham}
\end{figure} 

\section{Data and sample selection}\label{sect:data}
\subsection{Starting sample from the MaNGA survey}\label{sect:manga}
The Integral-Field Spectroscopy (IFS) data used in this study are taken from the Data Release 16 (DR16) \citep{DR16} of the MaNGA survey. DR16 includes IFS data for 4597 unique galaxies observed in the redshift range 0.01 < z < 0.15, selected to have a flat number density distribution with respect to the i-band absolute magnitude, taken as a proxy for the stellar mass \citep{Wake_2017}. Light is fed to the two BOSS spectrographs \citep{Smee_2013}, in a spectral range of 360 to 1030 nm, with a median instrumental resolution of $\sim$ 72 km s$^{-1}$, corresponding to a spectral resolution $\sim$ 2000 \citep{Law_2016}. The physical properties of these 4597 galaxies (e.g. morphological classification, mass, ellipticity etc.) used in this paper were kindly provided by Mark Graham; an electronic table with these properties is included as supplementary material in the published version of this paper, and we refer the reader to \cite{G19benchmark} for a detailed description of how the properties have been  measured.

From the DR16 sample, we exclude IFS data that are either flagged bad or show signs of being problematic, merging galaxies, and galaxies too small for the MaNGA beam size, thus reducing to $\sim 4000$ galaxies, that we plot on the ($\lambda_{R_e}$,$\varepsilon$) diagram \citep{Emsellem_2007}, shown in Figure \ref{fig:lamelgraham}; here, $\varepsilon$ is the observed ellipticity, and $\lambda_{R_e}$ is the spin parameter, measured within the effective radius (R$_e$), defined as~\cite{Emsellem_2007}
\begin{equation}
\lambda_{R} = \frac{\sum_{n=1}^N F_n R_n \mid V_n \mid}{\sum_{n=1}^N F_n R_n \sqrt{V_n^2 + \sigma_n^2}}
\end{equation} 
where $N$ is the number of spatial bins, and $F_n, R_n, V_n, \sigma_n$ are the flux, distance to the center, velocity and velocity dispersion of the $n$-th bin, respectively. Galaxies are plotted according to the visual kinematic classification performed by \cite{G19benchmark}; however, we ignored this classification and searched for galaxies with counter-rotating stellar disks following the selection criteria described in the next section.

\begin{figure}
\centering
\includegraphics[width=\columnwidth]{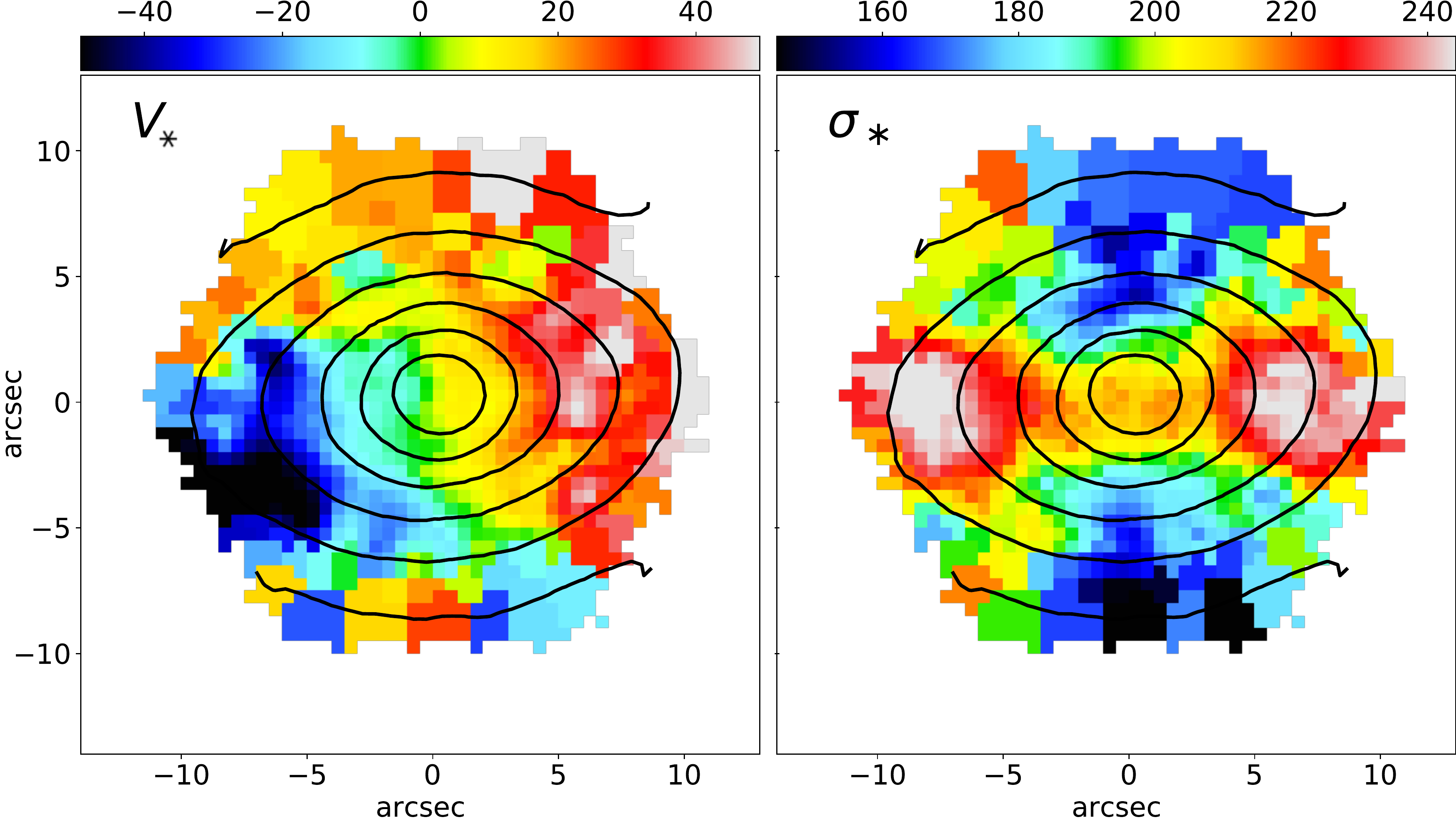}\\
\includegraphics[width=\columnwidth]{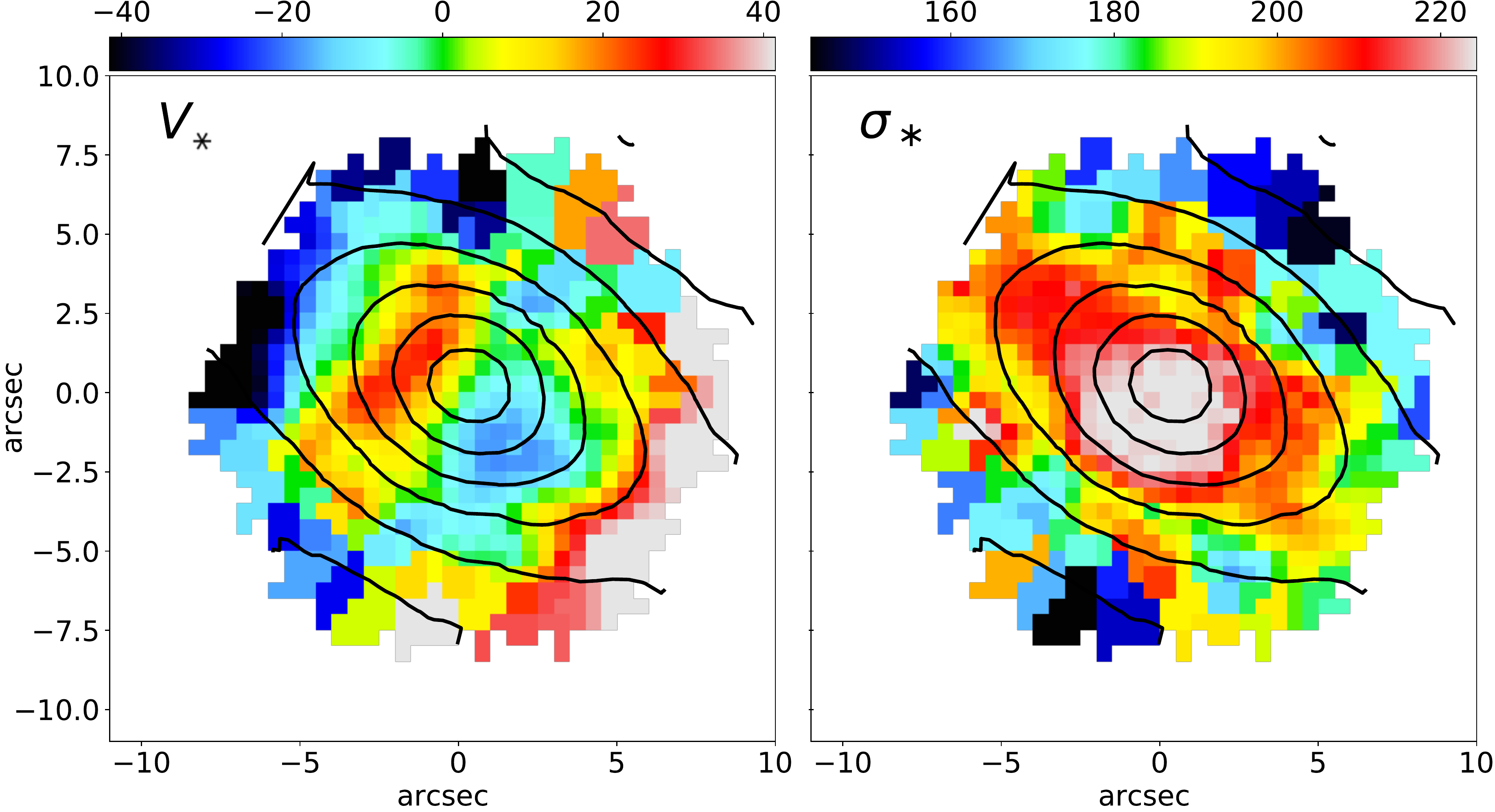}
\caption{Voronoi-binned velocity, $V_\ast$, and velocity dispersion, $\sigma_\ast$, maps, as produced by the MaNGA DAP, of a galaxy (MaNGA ID 1-248869) exhibiting the characteristic two peaks in $\sigma_\ast$ but no counter-rotation (upper panels), and of a galaxy (MaNGA ID 1-166613) exhibiting counter-rotation, but a single central peak in $\sigma_\ast$ (lower panels).}
\label{fig:basicselection}
\end{figure}

\subsection{Selection criteria}\label{sect:selection}

To build a sample of galaxies with counter-rotating stellar disks, we first performed a visual inspection of the kinematic maps, provided by the MaNGA Data Analysis Pipeline (DAP) \citep{Westfall_2019}, looking for evidences of counter-rotation in the mean stellar velocity ($V_\ast$) maps, and the presence of the two characteristic peaks in the stellar velocity dispersion ($\sigma_\ast$) maps. The ATLAS$^{\mbox{3D}}$ survey \citep{atlas_i} introduced the class of `2$\sigma$-galaxies' \citep{Krajnovic_2011} to describe the evidence of counter-rotation that produces two symmetric maxima of velocity dispersion along the galaxy major axis \citep{rubin4550, Bertola_1996}. However, given the modest spatial resolution of MaNGA, we do not expect the two kinematic features to be always detectable in the DAP maps; further, the appearance of one feature does not necessarily imply the other one. In fact, on one hand, if the two counter-rotating disks are indistinguishable (e.g. because of projection effect, for inclined galaxies close to face-on), the counter-rotation pattern does not appear in the velocity field, even when the two peaks in the dispersion map are present. On the other hand, sometimes, while the counter-rotation in the velocity map is clear, the dispersion map may show, for instance, a single central peak due to a prominent bulge. For these reasons, we include in our sample galaxies that show at least one of the two features; further, we do not demand two well-separated $\sigma_\ast$ peaks, and also include galaxies with a single peak elongated over the major axis. We then introduce the name `Counter-Rotating stellar Disks' (CRD) to designate galaxies that show evidences of counter-rotation \textit{or/and} the two $\sigma_\ast$ peaks; with this definition, 2$\sigma$-galaxies constitute a subset of CRDs. Two examples of a galaxy showing counter-rotation and lacking the two $\sigma_\ast$ peaks, and a galaxy with no counter-rotation but with the two $\sigma_\ast$ peaks, are shown in Figure \ref{fig:basicselection}.

\begin{figure}
\includegraphics[width=\columnwidth]{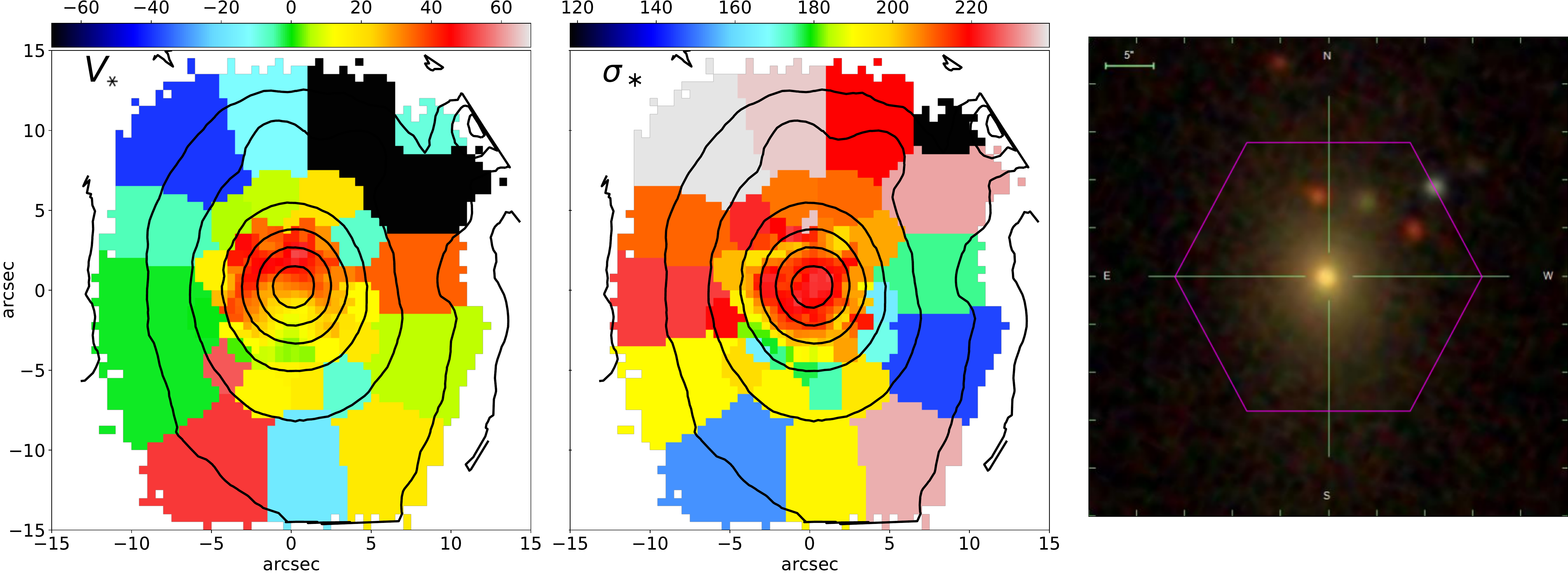}\\
\includegraphics[width=\columnwidth]{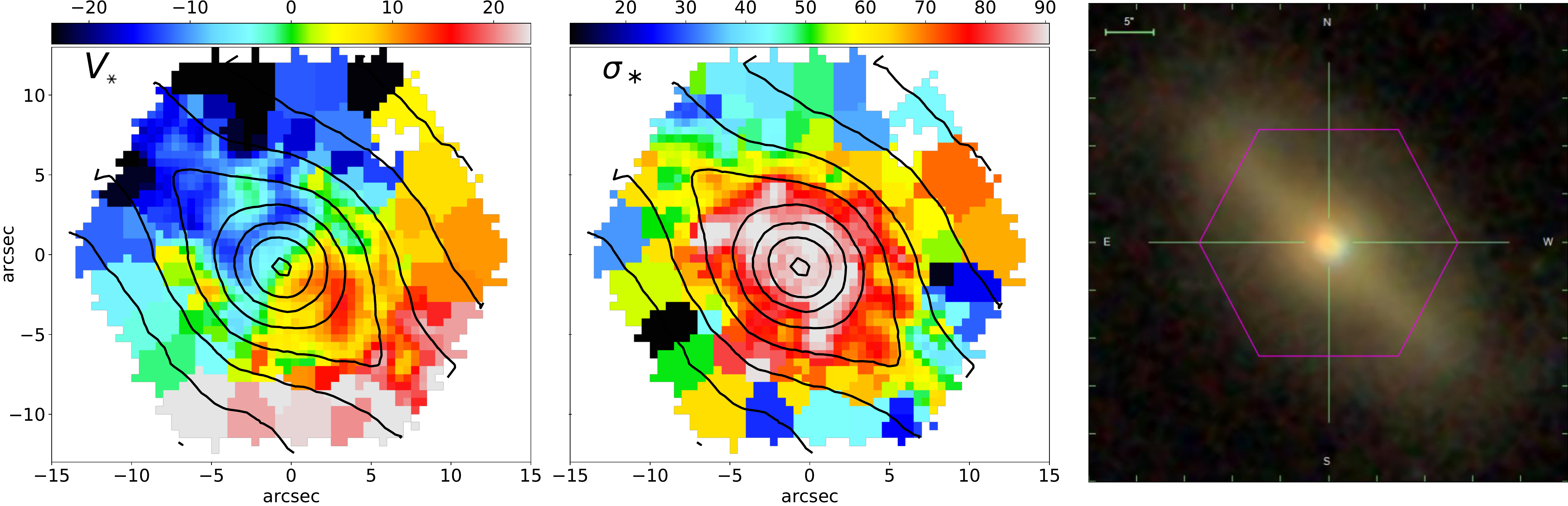}\\
\includegraphics[width=\columnwidth]{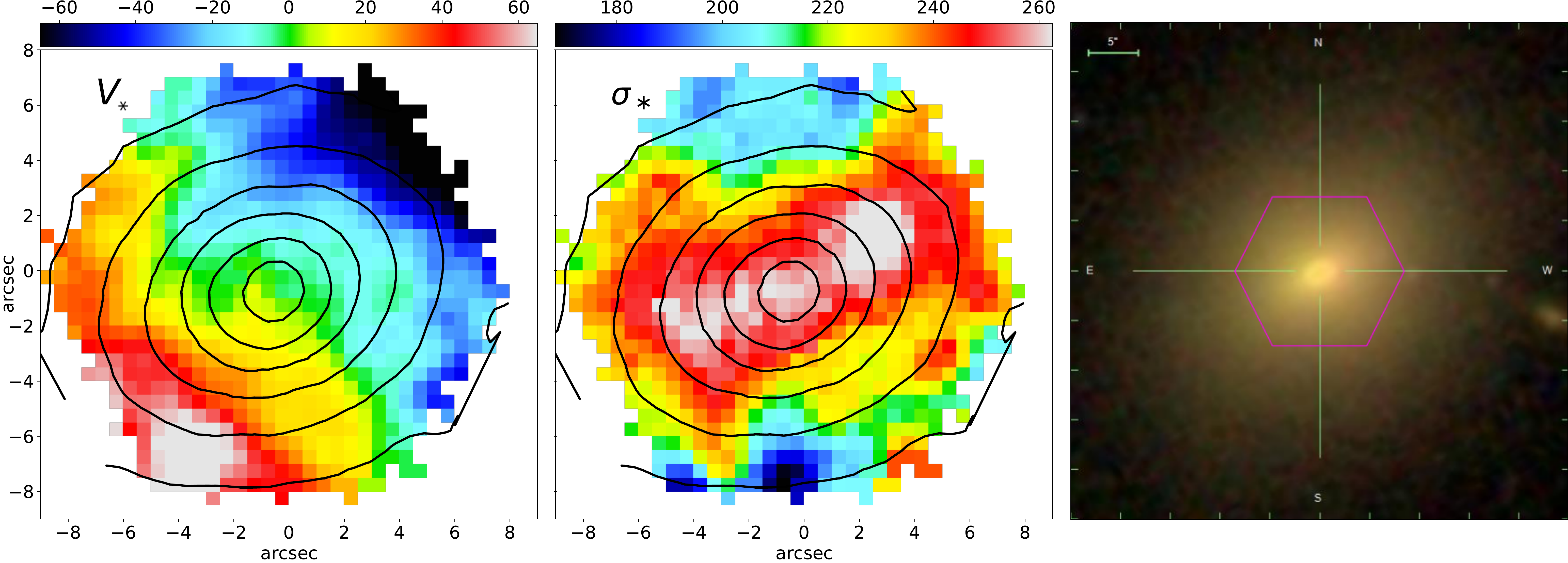}\\
\caption{Velocity map, velocity dispersion map and SDSS image of three galaxies (top to down MaNGA IDs: 1-166739, 1-135524, 1-593972) with kinematic features resembling those of CRD while not truly being so. Colorbars are in km s$^{-1}$. The same values are assigned to all spaxels of a Voronoi bin. Black lines are the flux contours. \textit{First row:} in the velocity map an inversion of rotation is seen in the northern region, but from the SDSS image it is evident that kinematic maps are defaced by the presence of external objects. \textit{Second and third rows:} the dispersion maps present two elongated peaks; the presence of a bar is revealed by the almond-shaped contours in the kinematic maps, and by the SDSS image, in the second row, and by the S-shaped zero-velocity (green) region in the velocity map, in the third row.}
\label{fig:caveat}
\end{figure}

This basic visual selection can however lead to misclassification, for some galaxies that are not truly counter-rotators can exhibit kinematic maps resembling those of CRDs. Fortunately, some peculiarities help us discard `fake CRDs'. Therefore, we always inspected also the SDSS image, and, for a proper classification, we made the following selection checks:
\begin{itemize}
\item \textbf{Presence of external objects.} In general, galaxies in the SDSS image often show neighbouring objects; in some cases, these influence significantly the fitted kinematics. For example, in the first row of Figure \ref{fig:caveat}, the velocity map appears to exhibit an inversion of the rotation in the northern region. However, the SDSS image, as well as the flux contours, clearly reveal the presence of many external objects; we concluded that the observed kinematic maps are influenced by these objects, which was also supported by the highest velocity dispersion values measured in those same bins where the velocity appears inverted. Therefore, we excluded galaxies with clear evidence of an external influence in the kinematic maps.

\item \textbf{Barred galaxies:} The presence of a bar can influence the kinematics in such a way that the dispersion map appears elongated over the extension of the bar, as in the second row of Figure \ref{fig:caveat}. From the SDSS image, we can though spot the presence of a bar, as well as from the characteristic almond-shaped flux contours \citep{Bureau_2005} in the kinematic maps, as in the fifth contour from the center outwards; visually, almond-shaped contours differ from disky contours because of the sharpening near the major axis. In another case (third row of Figure \ref{fig:caveat}), we can see the peaks in the velocity dispersion map, but no almond-shaped contours or evidences of bars in the SDSS image; however, in the velocity map the zero-velocity bins (in green) form an S-shape pattern, which arises when a bar is present (e.g. Figure 9 of \citealt{Cappellari_2003}); thus we excluded such galaxies from our sample.

\item \textbf{KDC:} Galaxies with a kinematic decoupled core (KDC) may have kinematic maps resembling those of a CRD (e.g., \citealt{kdc}). The clearest evidence to identify a KDC is the presence of a kinematic or a photometric twist \citep[e.g.][hereafter C16]{C16}. The former is recognisable from a change of the rotation axis in the velocity map, as well as from asymmetric $\sigma_\ast$ peaks, and from misalignment of the rotation or the $\sigma_\ast$ peaks with respect to the photometric major axis; the latter, instead, corresponds to a twist of the flux contours. Another peculiarity of KDC is that the kinematic features are confined in the very inner regions, while those of CRDs typically arise at $\gtrsim$ R$_e$.
\end{itemize}

Of course, even with accurate checks, no classification is perfect and some misclassified cases can remain. In particular, it is not always easy to distinguish a KDC from a true CRD, because their kinematic maps can be very similar. Aiming at building a large sample, we include borderline cases. For example, our sample includes few galaxies showing both counter-rotation and the two $\sigma_\ast$ peaks, even though they are confined in the very inner region of the galaxy. 

\begin{figure}
\includegraphics[width=\columnwidth]{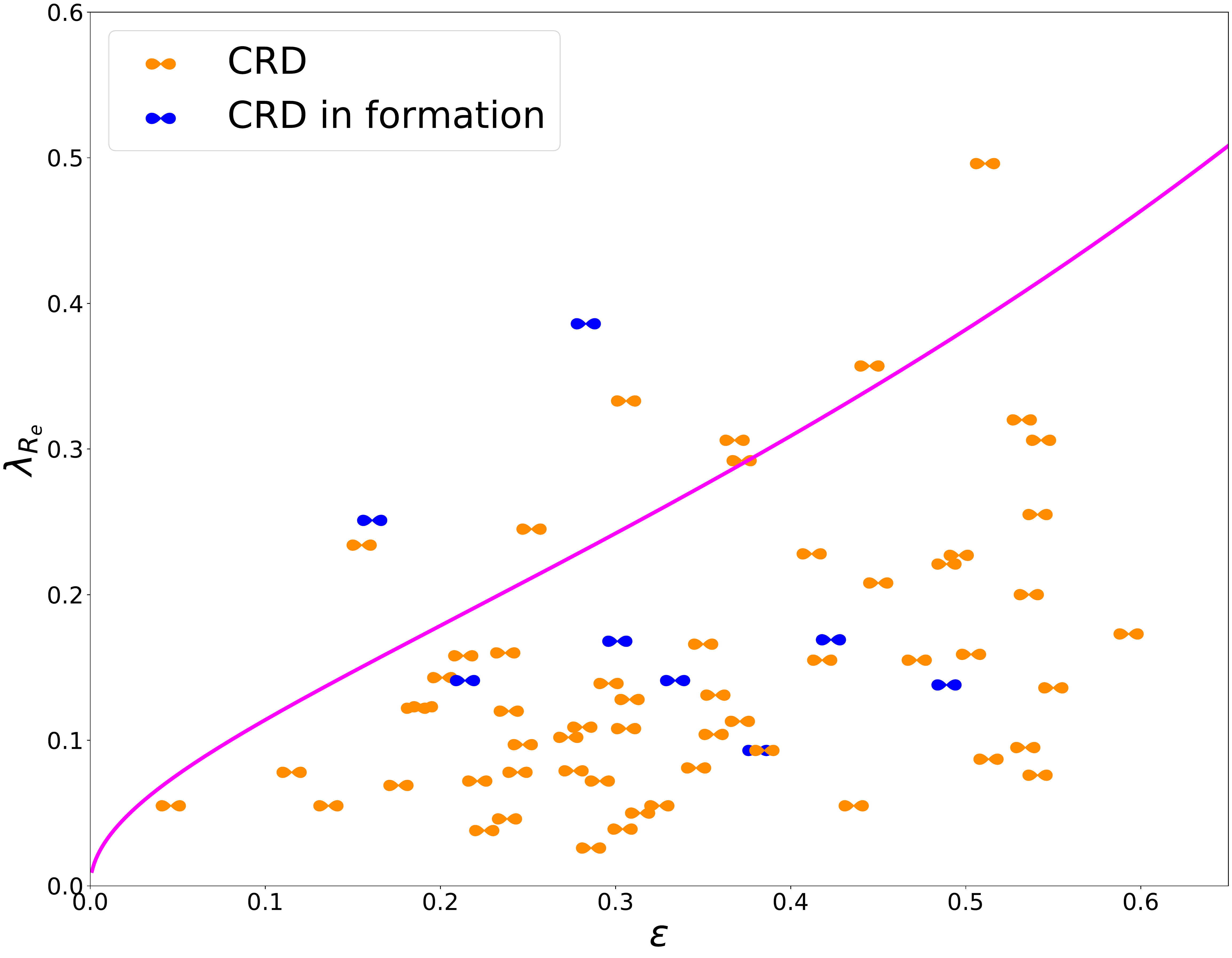}
\caption{Final sample of CRDs plotted on the ($\lambda_{R_e},\varepsilon$) diagram. The magenta line is the same of Figure \ref{fig:lamelgraham}. Galaxies labelled as `CRD in formation' have a bluish or irregular shape in the SDSS image and disturbed kinematics.}
\label{fig:finalsample}
\end{figure}

The final sample counts 64 best CRD candidates, which is the largest to date for this kinematic class; the list is given in Table \ref{tab:crd} together with the main properties of CRDs used here. The sample is mostly composed of ETGs, of which 38 are ellipticals (E) and 23 are lenticulars (S0); additionally, the sample includes a spiral galaxy (S) and two galaxies with uncertain morphologies. In Figure \ref{fig:finalsample} we plot CRDs on the ($\lambda_{R_e},\varepsilon$) diagram. It is clear that the large majority of the CRDs lie below the magenta line. This can be explained as follows: normal fast rotators, without significant counter-rotation, tend to have a velocity ellipsoid close to oblate (see fig.~11 of C16). They are expected to lie above the magenta line due to the lack of physical axisymmetric equilibrium solutions at high anisotropy when the velocity ellipsoid is close to oblate \citep{Wang21}. However, the presence of the counter-rotation lowers the global observed velocity (and thus $\lambda_{R_e}$). This tends to bring the CRDs below the magenta line. Finally, it is interesting to notice that the sample includes 8 galaxies that appear to be formed just recently, that we labelled as `CRD in formation', based on the bluish colors or irregular shapes in the SDSS image, and disturbed kinematics.

\subsection{Detectability of CRDs: method and limitations}\label{sect:detectability}
To study the detectability of CRDs in the MaNGA survey, we build dynamical models using the Jeans Anisotropic Modelling method and software JAM\footnote{Available from \url{https://pypi.org/project/jampy/}} \citep{JAM2008,JAM2020}; in particular, we used the cylindrically aligned version JAM$_{\rm cyl}$. Briefly, JAM$_{\rm cyl}$ models are solutions for the steady-state axisymmetric Jeans equations under the assumption of constant mass-to-light ratio (M/L) and cylindrically-aligned velocity ellipsoid: $\beta = 1-(\sigma_z/\sigma_{\mbox{\scriptsize{R}}})^2$ and $\gamma = 1-(\sigma_\phi/ \sigma_{\mbox{\scriptsize{R}}})^2$.

Although fast rotators are well described by oblate velocity ellipsoid ($\sigma_\phi \approx \sigma_{\mbox{\scriptsize{R}}} \gtrsim \sigma_z$) (e.g., section 3.4.3 of C16) CRDs present strong tangential anisotropy due to the counter-rotation; however, their kinematics is well reproduced when modeling the two disks separately, as two fast rotators counter-rotating in the same total gravitational potential, and then sum their contributions (fig.~12 of C16). 

We then produce our models by following this lead. In particular, we build each CRD model by considering three components: a bulge, a disk corotating with the bulge (`+'), and a counter-rotating disk (`$-$'); for all these components, we fix the two anisotropies to characteristic values for fast rotators $\beta = 0.2$ and $\gamma = 0$ (see fig.~9 of C16). The surface brightness profiles used in JAM are the Multi-Gaussian Expansion (MGE) fit of the radial profiles of the three components, obtained using the method described in \cite{mge_2002}\footnote{Available from \url{https://pypi.org/project/mgefit/}}. We assume that the brightness profile for each of the three components follows the S\'{e}rsic law:
\begin{equation}
y \, (r) = \exp\left\{ -b_n \left[ \left(\frac{r}{R_e}\right)^\frac{1}{n} -1  \right] \right\} ,
\end{equation}
\noindent where $R_e$ is the effective (projected half-light) radius, $n$ is the S\'{e}rsic index and $b_n$ is a parameter that depends on the choice of $n$, and is equal to $b_n = 2 n - 1/3 + 4/(405  n) + 46/(25 515  n^2 )$, given in \citet{Ciotti_Bertin}. 

We now give the physical properties we used to build our CRD models and discuss these choices and their limitations afterward.
We use the same S\'{e}rsic index, $n=1$, for both disks, resulting in exponential profiles with scale lengths $R_{+} \equiv R_{e,+} / b_1$ and $R_{-} \equiv R_{e,-} / b_1$, where $b_1 = 1.678$; for the bulge we adopt the median Sersic index measured for bulges by \citet{atlas_xvii}, $n = 1.7$ (then, $b_{1.7} = 3.135$). We fix the ratio $R_{+}/R_{\rm e,bulge} = 3.1$, which is the median value measured for fast rotators by \citet{atlas_xvii}, and corresponding to $R_{+} = 2.5$ kpc and $R_{\rm e,bulge} = 0.8$ kpc; \citet{Bluck_2014} showed that the great majority of fast rotators in the local universe have a bulge-to-total mass ratio lower than 0.2, so we fix $M_{\rm bulge}/M_{+} = 0.25$ (thus assuming that the total mass is the sum of the bulge and the disk masses). We fix the intrinsic axial ratios $q_{+} = q_{-} = 0.25$ for the two disks, and $q_{b} = 0.6$ for the bulge, which are the peak values of the distributions of axial ratios for fast and slow rotators, respectively, in the ATLAS$^{\mbox{\scriptsize{3D}}}$ survey \citep{atlas_xxiv}. We also include a central black hole of $10^8 M_\odot$. We then impose that the gravitational potential is due to the combination of the MGEs describing the three stellar components (a dark matter halo is not included), and corresponding to a total mass equal to the median stellar mass of our CRDs, namely $\log_{10}(M/M_\odot$) = 10.575. Finally, we adopt a distance of 154.8 Mpc, corresponding to the angular size distance at the median redshift of MaNGA, $z=0.0376$. To take into account the spatial resolution, we convolve the JAM models with the full width at half maximum, ${\rm FWHM} = 2.5''$ characteristics of the MaNGA data; we also use the spatial sampling and the corresponding signal-to-noise of a real CRD (MaNGA ID: 1-94773)\footnote{This CRD was selected because its mass and redshift are similar to the values chosen for our model.}, as produced by the DAP, and a spatial coverage of 1.5$R_e$; further, we use the formal error of the kinematic spectral fitting (see section \ref{sect:singlecompfit}) on each spatial bin as the noise for the final velocity and velocity dispersion model maps.

\begin{figure}
\includegraphics[width=\columnwidth]{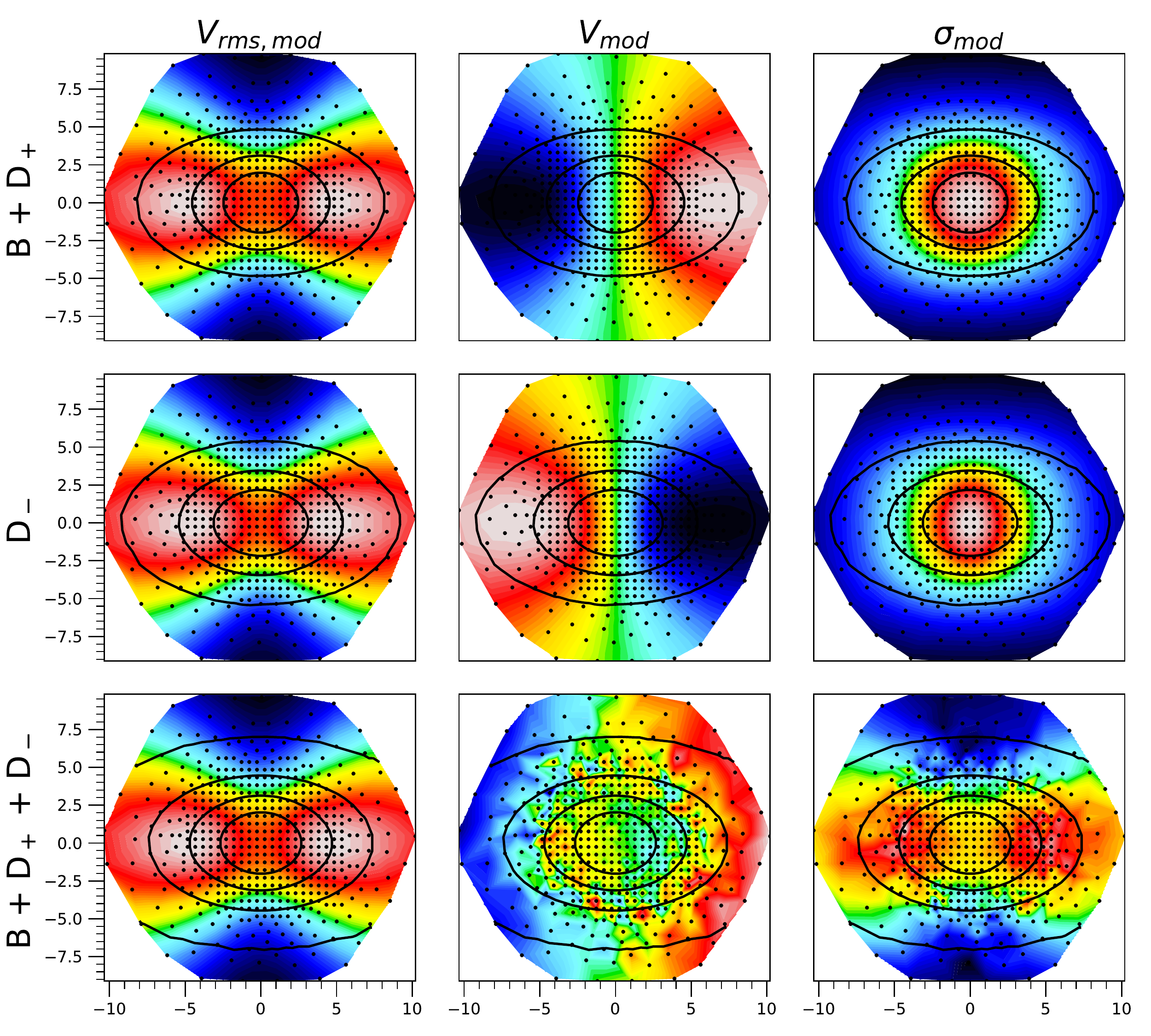}
\caption{Root mean square velocity ($V_{\rm rms,mod}$), velocity ($V_{\rm mod}$) and velocity dispersion ($\sigma_{\rm mod}$) models of the bulge combined with the disk corotating with it (B + D$_{\mbox{\scriptsize{+}}}$) (first row), of the counter-rotating disk (D$_{-}$) (second row), and of the luminosity weighted sums of the two components (third row). The small dots indicate the location of the centroids of each Voronoi bin. These models are produced using $M_{-}/M_{+} = 1$ , $R_{-}/R_{+} = 3/4$ and $i = 60^\circ$. The maps of $V_{\rm mod}$ and $\sigma_{\rm mod}$ in the bottom row include a realistic noise level, like that we used for our assessment of the CRDs detectability.}
\label{fig:model_example}
\end{figure}

We build the models by varying: (i) the mass ratio of the two disks, $M_{-}/M_{+}$, ranging from 0.2 to 1.4 at step of 0.2; (ii) the scale length ratio of the two disks, $R_{-}/R_{+}$, ranging from 0.5 to 2 at step of 0.25; (iii) the inclination $i$ by sampling 10 values uniformly distributed on the sphere of the viewing angles, namely $i = \arccos{p}$ with $p$ uniformly distributed in the open interval $[0,1]$. This results in $i = 18^\circ, 32^\circ, 41^\circ, 49^\circ, 57^\circ, 63^\circ,  70^\circ, 76^\circ, 81^\circ, 87^\circ$  ($i=90^\circ$ being edge-on). The total number of models considered is thus 420.
Each CRD model was obtained as follows. We first produced two root-mean-square velocity ($V_{\rm rms,mod}$) and two mean velocity ($V_{\rm mod}$) models for the bulge with the co-rotating disk, and for the counter-rotating disks, separately (see the first and second rows of Figure \ref{fig:model_example}). They are both computed by using in JAM$_{\rm cyl}$ their surface brightness as tracer in the same total gravitational potential. To obtain the final kinematic maps, we weigh the contribution of the two counter-rotating components with the surface brightness $\Sigma$ of the MGE models, using the following expressions
\begin{align}
    &\Sigma_{\rm tot} V_{\rm mod}  = \Sigma_{+} V_{\rm mod +} + \Sigma_{-} V_{\rm mod -}\\
    &\Sigma_{\rm tot} V^2_{\rm rms, mod}  = \Sigma_{+} V^2_{\rm rms, mod +} + \Sigma_{-} V^2_{\rm rms, mod -}\\
    &\sigma_{\rm mod}^2 = V^2_{\rm rms, mod} - V^2_{\rm mod}
\end{align}
Finally, we add the noise to $V_{\rm{mod}}$ and $\sigma_{\rm{mod}}$, assuming it to be Gaussian (last row of Figure \ref{fig:model_example}), and for the velocity dispersion map, we only display values greater than the instrumental correction (see section \ref{sect:binning}). 

\begin{figure*}
\includegraphics[width=\textwidth]{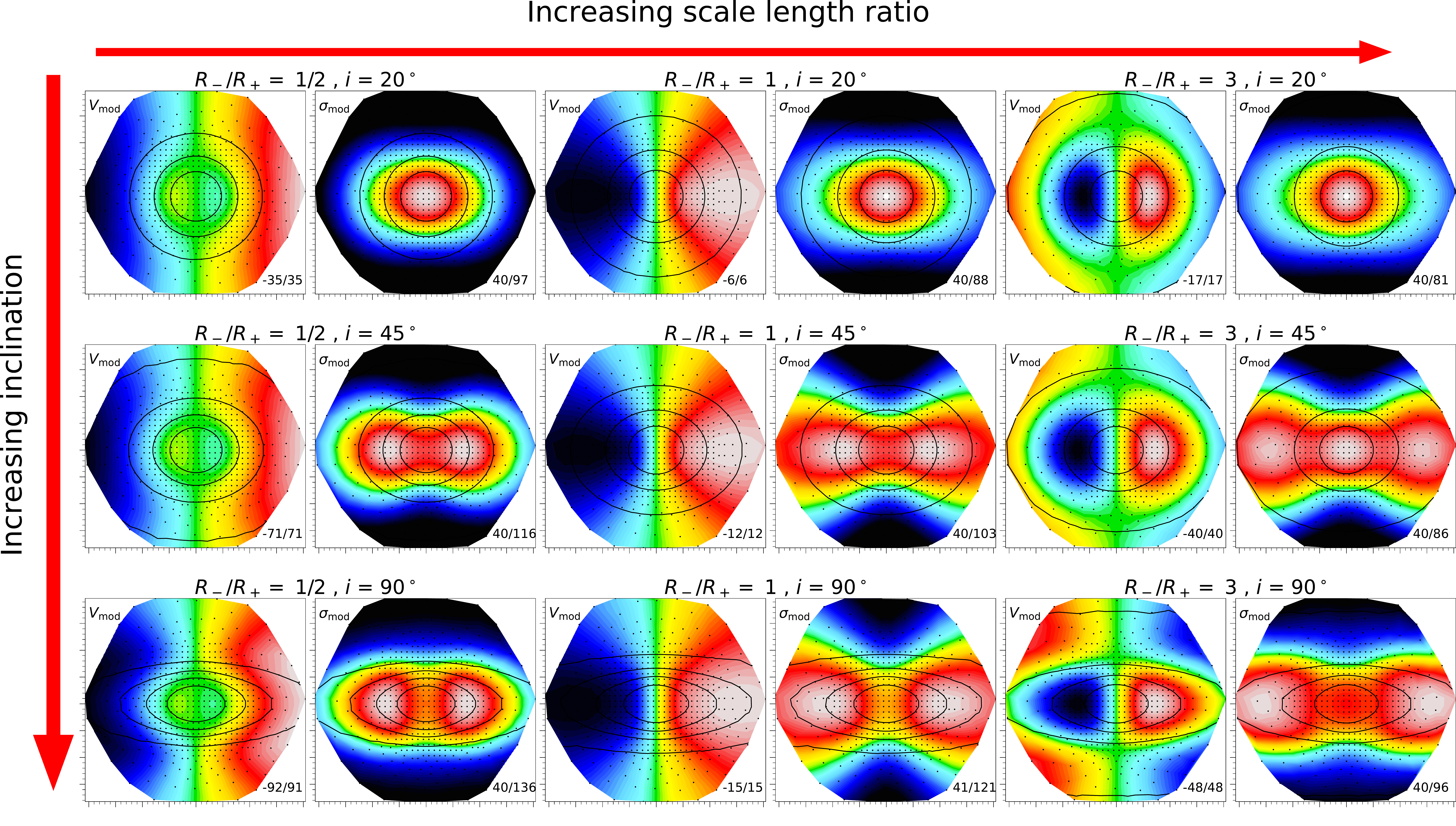}
\caption{Models with varying $R_{-}/R_{+}$ and $i$ at fixed $M_-/M_+ =3/4$; in particular, these models are made with $R_{-}/R_{+}=1/2, 1, 3$ and $i = 20^\circ, 45^\circ, 90^\circ$. The first, third and fifth columns are $V_{\rm mod}$ maps, while second, fourth and sixth columns are $\sigma_{\rm mod}$ maps. Colormaps vary from blue (lower values) to red (highest values), with green being the zero-velocity; the min/max values (in km s$^{-1}$) are showed in the bottom right corner of each panel. }
\label{fig:mfix}
\end{figure*}

\begin{figure*}
\includegraphics[width=\textwidth]{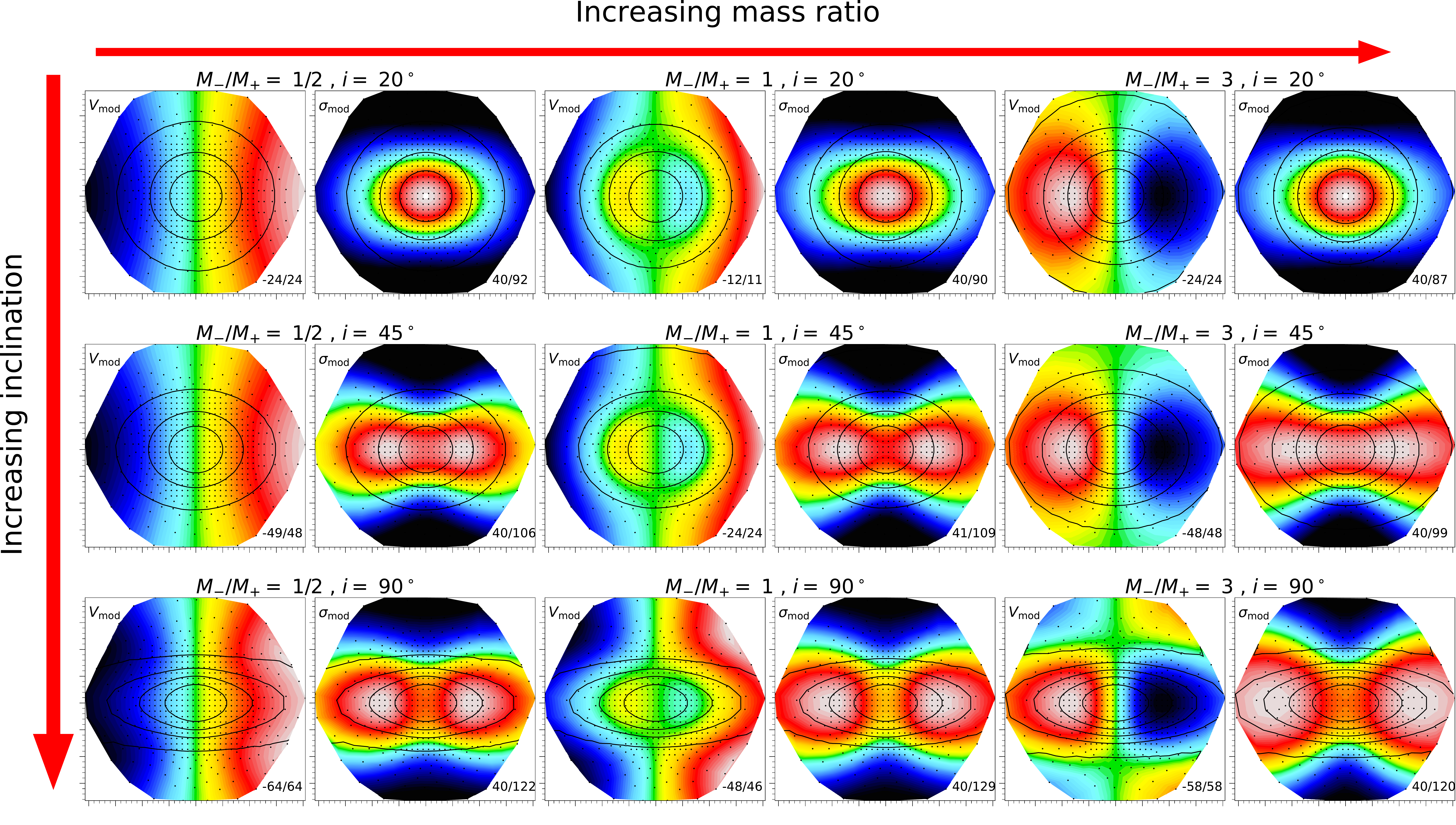}
\caption{Models with varying $M_-/M_+$ and $i$ at fixed $R_-/R_+ = 3/4$; in particular, these models are made with $M_{-}/M_{+} = 1/2, 1, 3$ and $i = 20^\circ, 45^\circ, 90^\circ$. The first, third and fifth columns are the $V_{\rm mod}$ maps, while the second, fourth and sixth columns are the $\sigma_{\rm mod}$ maps.  Colormaps vary from blue (lower values) to red (highest values), with green being the zero-velocity; the min/max values (in km s$^{-1}$) are showed in the bottom right corner of each panel.}
\label{fig:rfix}
\end{figure*}

\begin{figure*}
\includegraphics[width=\textwidth]{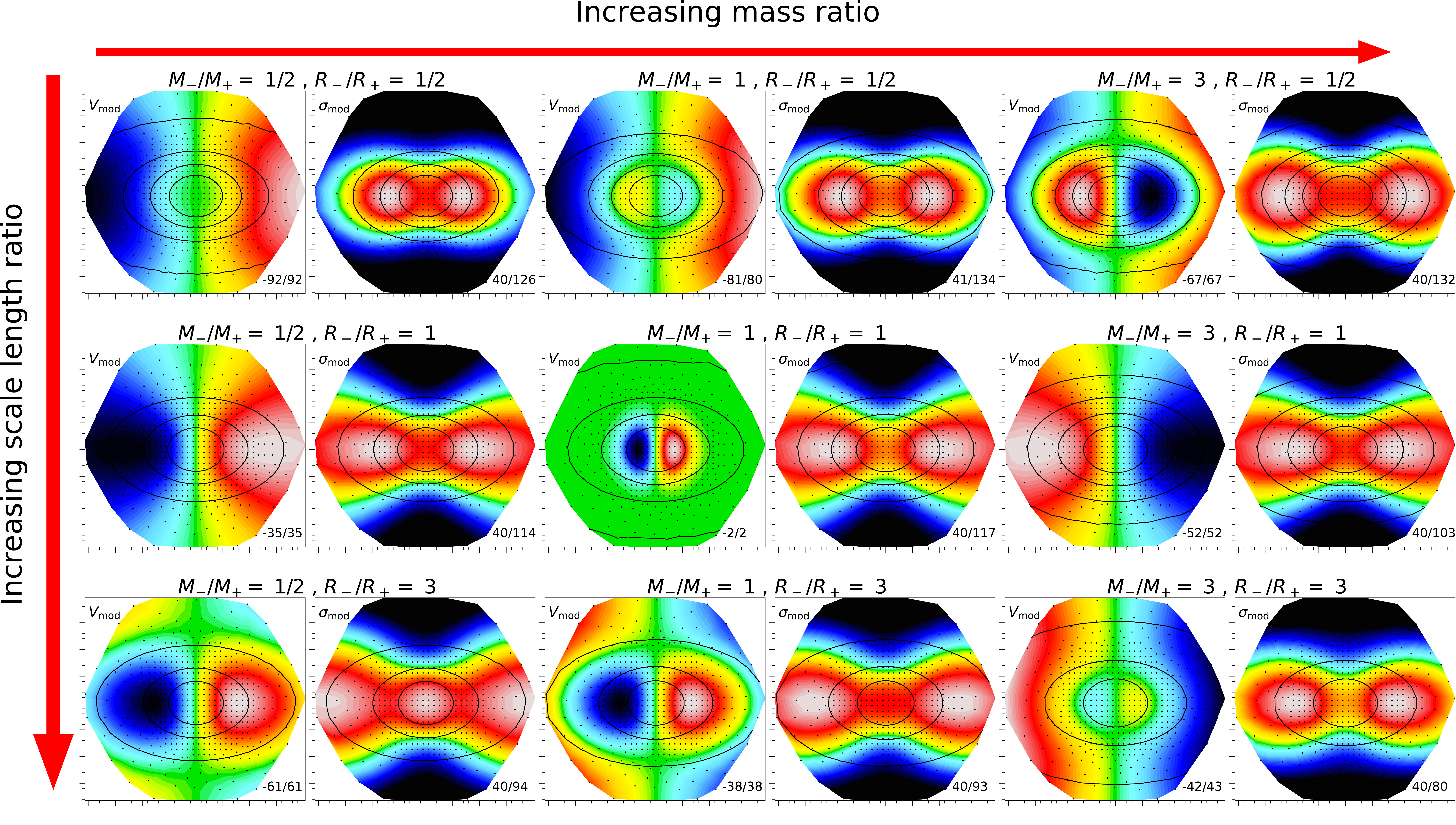}
\caption{Models with varying $M_{-}/M_{+}$ and $R_{-}/R_{+}$ at fixed $i = 60^\circ$; in particular, these models are made with $M_{-}/M_{+} = 1/2, 1, 3$ and $R_{-}/R_{+}=1/2, 1, 3$. The first, third and fifth columns are the $V_{\rm mod}$ maps, while the second, fourth and sixth columns are the $\sigma_{\rm mod}$ maps. Colormaps vary from blue (lower values) to red (highest values), with green being the zero-velocity; the min/max values (in km s$^{-1}$) are showed in the bottom right corner of each panel.}
\label{fig:ifix}
\end{figure*}

Figures \ref{fig:mfix}, \ref{fig:rfix} and \ref{fig:ifix} show how the final $V_{\rm mod}$ and $\sigma_{\rm mod}$ maps change at different couples of $M_{-}/M_+ , R_-/R_+ \mbox{ and } i$, while keeping the third parameter fixed. In these figures we show models produced using values different from those in (i), (ii) and (iii) above, and without any noise, for illustrative purposes of how the maps change with parameters.
We visually classified the CRDs of all the models, as we did for the real observations, and we found that of 420 models, 341 show at least one of the two characteristic CRD features; thus, we estimate a detectability of 81\%. As one would expect, the great majority of models with undetected CRD features are those with the lowest mass ratio, scale length ratio, and inclination.

Due to the low statistic of CRDs in the literature and to the variety of observed properties for the two disks (e.g. mass fraction, luminosity fraction, extension, thickness, etc.), the choice of the structural parameters to be used in models is not straightforward. Here, we discuss some major concerns:
\begin{itemize}
\item The bulge in the CRDs studied so far has always been found to be negligible (e.g. NGC 3593, NGC4550 \citealt{Coccato_2012} and IC 719 \citealt{katkov}). Nonetheless, we tested how the presence of the bulge influences the observed kinematics by changing its fixed parameters and found that it mainly influences the central regions of the velocity dispersion map, resulting in a central peak which is more (less) prominent when increasing (decreasing) its mass, but almost independent of its scale length; further, increasing the S\'{e}rsic index produces slightly better maps. The central velocity dispersion peak produced by the bulge prevails in the map only when the mass ratio of the two disks is low (particularly, when $M_{-}/M_{+} \leq 0.4$), and when the viewing angle is low ($i < 40^\circ$). The counter-rotation feature in the velocity map, instead, is not particularly affected by the presence of the bulge; for this reason, it does not affect importantly the general estimate of the detection rate of CRDs.\\
\item The spatial resolution of MaNGA depends on the radial coverage, on the mass, and on the redshift of the galaxy considered. While we adopted parameters characteristic of MaNGA galaxies, a low mass CRD observed at high redshift with a 2.5$R_e$ coverage could be missed by our classification. However, among the $\sim 4000$ galaxies we examined only 156 galaxies have a radial coverage reaching 2.5$R_e$; of these, 88 have mass lower than $\log_{10}(M/M_\odot) = 10.575$, and 67 of these are at redshift higher than 0.0376. Therefore, given the intrinsic rarity of CRDs among galaxies (a few percent), we believe that only a few cases could have been missed because of our choices on spatial resolution.\\
\item The values of $M_-/M_+$ and $R_-/R_+$ considered are somehow arbitrary and some cases could be unrealistic: for example, models calculated at highest mass ratio with lowest scale length ratio, e.g. $M_-/M_+ = 1.4$ and $R_-/R_+ = 0.5$, imply that a galaxy formed a very massive (more than the pre-existing) counter-rotating disk within a small region, which seems unlikely. Such extreme cases may affect our statistics, but only of few percent; indeed, all the other cases seem reasonable, since we generally expect that CRDs host a variety of mass and scale length ratios.
\end{itemize}
Other minor concerns (about, e.g., the $R_{+}$ and $R_{\rm e,bulge}$ values, the mass of the central black hole, and the exclusion of a dark matter halo) could be considered, but they are marginal issues since they do not particularly affect the detectability. In any case, a detailed modeling of CRDs is beyond the scope of this paper. We also point out that the statistics of CRDs we discuss in section \ref{sect:statistics} are broadly consistent with previous studies, even assuming a detection rate much lower than the 81\% estimated by this~analysis.

\section{Stellar and gas kinematics}\label{sect:kinem}

\subsection{Templates library, resolution matching and binning criterion}\label{sect:binning}

To extract the stellar kinematics of galaxies, a library of stellar templates of different spectral classes is generally fitted to the observed spectra. Libraries with a larger number of templates require larger computation time: for example, for \texttt{pPXF} the execution time is typically $\sim O(N_{\rm tpl})$ for $N_{\rm tpl}$ templates. Thus, we selected a subsample of spectra, representative of the whole library, following the method and software used in \cite{Westfall_2019} (hereafter, W19) for the DAP: starting from the MILES stellar template library \citep{Sanchez_Blazquez_2006, Falcon_Barroso_2011}, which counts 985 stellar spectra, we applied a hierarchical-clustering algorithm\footnote{Available from \url{https://github.com/micappe/speclus}} to produce a desired subsample of $\sim$50 representative spectra. In brief, the reduced library is obtained by fitting each template with every other one in the MILES library, and, for each couple of templates, measuring the `distance', in terms of residuals of the fit between the two spectra, calculated using equation (1) of W19; finally, spectra with distances lower than a fixed value $d_{\mbox{\scriptsize{max}}}$\footnote{This corresponds to the parameter \texttt{threshold} in the \texttt{scipy} function \texttt{cluster.hierarchy.fcluster}, with \texttt{criterion = `distance'}, used to perform the clustering of the spectra.} are clustered together.

With maximum distance of $d_{\mbox{\scriptsize{max}}}$ = 0.065, we obtained 55 clusters. We then normalized each MILES spectrum to a mean of unity, and averaged all spectra in each cluster, in order to have a single representative spectrum per cluster; these averaged spectra are the templates for the actual kinematic fits. We finally removed from our library three spectra with prominent emission lines, resulting in a final set of 52 templates. W19 showed that using a subsample instead of the whole library leads to marginal biases in the extracted kinematics, while speeding up the computation time very significantly. Like W19, we refer to our distilled library as `MILES-HC'.

In general, when measuring the stellar kinematics, one wants the templates' spectral resolution to match that of the observed spectra. The implications of ignoring the requirement of resolution matching, when fitting MaNGA spectra with the MILES library, are discussed in detail in section 7.4.3 of W19. They found that using spectra with higher spectral resolution than MaNGA data (the median instrumental resolution for MaNGA is $\sim$16\% larger than that of the MILES library) is actually convenient, particularly when the intrinsic stellar velocity dispersion is lower than the resolution of MaNGA. For this reason, we perform the fit of the mean velocity and the velocity dispersion by keeping both template and galaxies spectra at their native resolution.

To account for the mismatch, after the fit we correct the extracted velocity dispersion as follows:
\begin{equation}
\sigma_\ast = \sqrt{\sigma_{\mathtt{ppxf}}^2 - \sigma_{\mbox{\scriptsize{inst}}}^2}
\label{eq:sigma}
\end{equation}
where $\sigma_\ast$ is the intrinsic stellar velocity dispersion, $\sigma_{\mathtt{ppxf}}$ is the kinematic solution of the fit, and $\sigma_{\mbox{\scriptsize{inst}}}$ is the instrumental correction. The DAP calculated a median value of $\sigma_{\mbox{\scriptsize{inst}}} \approx$ 40 km s$^{-1}$, that we choose as the fixed instrumental correction in equation \eqref{eq:sigma} for all the dispersion maps. Notice that when $\sigma_{\mathtt{ppxf}}$ < $\sigma_{\mbox{\scriptsize{inst}}}$ we get an unphysical value for $\sigma_\ast$; however, \cite{ppxf} showed that it is not due to a failure of the fit, but it is the intrinsic stellar dispersion to be lower than the instrumental resolution, instead. In our maps, we set the unphysical values to the minimum positive $\sigma_\ast$ of the remaining Voronoi bins.

W19 showed that for MaNGA data an average value of S/N $\geq$ 10 (in $g$-band) per log-rebinned spectral pixel of 69 km s$^{-1}$ (see section \ref{sect:singlecompfit}) is sufficient for reliable measurements of the first two moments of the LOSVD, and they used this value for the DAP. However, when fitting two kinematic components, it is fundamental to have sufficiently high S/N spectra to make the two components spectroscopically more distinguishable. On the other hand, binning data at very high S/N results in a significant loss of spatial resolution, and, in particular, CRDs can lose their peculiar kinematic features. To compromise between the request of a relatively high S/N, in order to better distinguish the two components, and the need for a relatively low S/N, in order to preserve enough spatial resolution, we chose to bin galaxies either with (S/N)$_{\mbox{\scriptsize{min}}}$ = 25 or (S/N)$_{\mbox{\scriptsize{min}}}$ = 15, depending on the resulting number of spatial bins: in practice, after excluding all spaxels with S/N $\leq$ 1, we first bin all galaxies to (S/N)$_{\mbox{\scriptsize{min}}}$ = 25; if the resulting number of bins is < 100, we rebin to (S/N)$_{\mbox{\scriptsize{min}}}$= 15. The binning procedure has been performed using the Voronoi binning method and software\footnote{Available from \url{https://pypi.org/project/vorbin/}}, described in \cite{Cappellari_2003}. 

Another difference with W19, is that we do not take the covariance between spaxels into account for the production of the maps to avoid too large Voronoi bins at large galactocentric radii, at the expense of a slightly lower S/N.

\subsection{Single component fits}\label{sect:singlecompfit}

For the extraction of the stellar kinematics, we fitted the datacubes provided by the MaNGA Data Reduction Pipeline (DRP) \citep{Law_2016}, using the penalized pixel fitting method and software\footnote{Available from \url{https://pypi.org/project/ppxf/}} (\texttt{pPXF}) of \cite{ppxf}; in particular, we only fitted the first two moments of the LOSVD. Before the fit starts, we set some parameters. First, as the spectra provided by the DRP are logarithmically rebinned in wavelength, we logarithmically rebinned the template spectra. Secondly, we set a constant velocity scale (which in \texttt{pPXF} corresponds to the wavelength sampling in pixel space) $V_{\mbox{\scriptsize{scale}}} = c\Delta\ln \lambda \approx 69$ km s$^{-1}$ , where c is the speed of light, as the wavelength sampling of the MaNGA spectra is constant, and defined exactly as $\Delta\ln(\lambda)=\ln(10)\times10^{-4}$, chosen to have a step of $10^{-4}$ in log$_{10}$($\lambda$). To measure the kinematics with respect to the galaxy rest-frame, we de-redshift the observed spectra. We restrict the fitting spectral range, which is smaller for MILES than for MaNGA, to $\sim 3700-7400$ \AA . Finally, we masked regions of gas emission.

The kinematics solutions were obtained by performing two fits, both including degree-eight additive Legendre polynomials, which change the strength of spectral lines, to correct the continuum. We first considered as input spectrum the sum of the spectra of all spatial bins, normalized to their median value, and fitted it with all the 52 stellar templates of MILES-HC, setting the input noise to be constant with wavelength. From this first fit, we made a robust estimation of the standard deviation $\sigma_{\mbox{\scriptsize{std}}}$ of the residuals, and masked all regions of the spectrum deviating > 3$\sigma_{\mbox{\scriptsize{std}}}$. We then performed a second fit, this time on every spatial bin of the galaxy. As template we only used the best-fit spectrum of the first fit, constructed as the weighted sum of the template spectra of MILES-HC; as the input constant noise we use $\sigma_{\mbox{\scriptsize{std}}}$. Thus, we derived the two kinematic solutions for each spatial bin. The resulting rest-frame stellar velocity ($V_\ast$) and (corrected) velocity dispersion ($\sigma_\ast$) maps obtained for the whole sample are presented in an electronic appendix attached to this article.

\subsection{Two-component fits and $\chi^2$ maps}\label{sect:twocompfit}

To find the intrinsic velocity and velocity dispersion of the two stellar disks separately, one has to fit two kinematic components. However, even when the presence of two counter-rotating disks is obvious from the kinematic maps, the spectroscopical distinction of the two kinematic components is not possible if the velocity difference of the two stellar components is $\Delta V < \sigma_{\mbox{\scriptsize{inst}}}$, as this implies that the two Gaussians are spectrally unresolved. Similarly, if the velocity dispersions of the two stellar disks are intrinsically large, the LOSVD of the two components will be hardly resolvable; furthermore, the recovery of the two components becomes more difficult when the S/N is low. 

\begin{figure}
\includegraphics[width=.49\columnwidth]{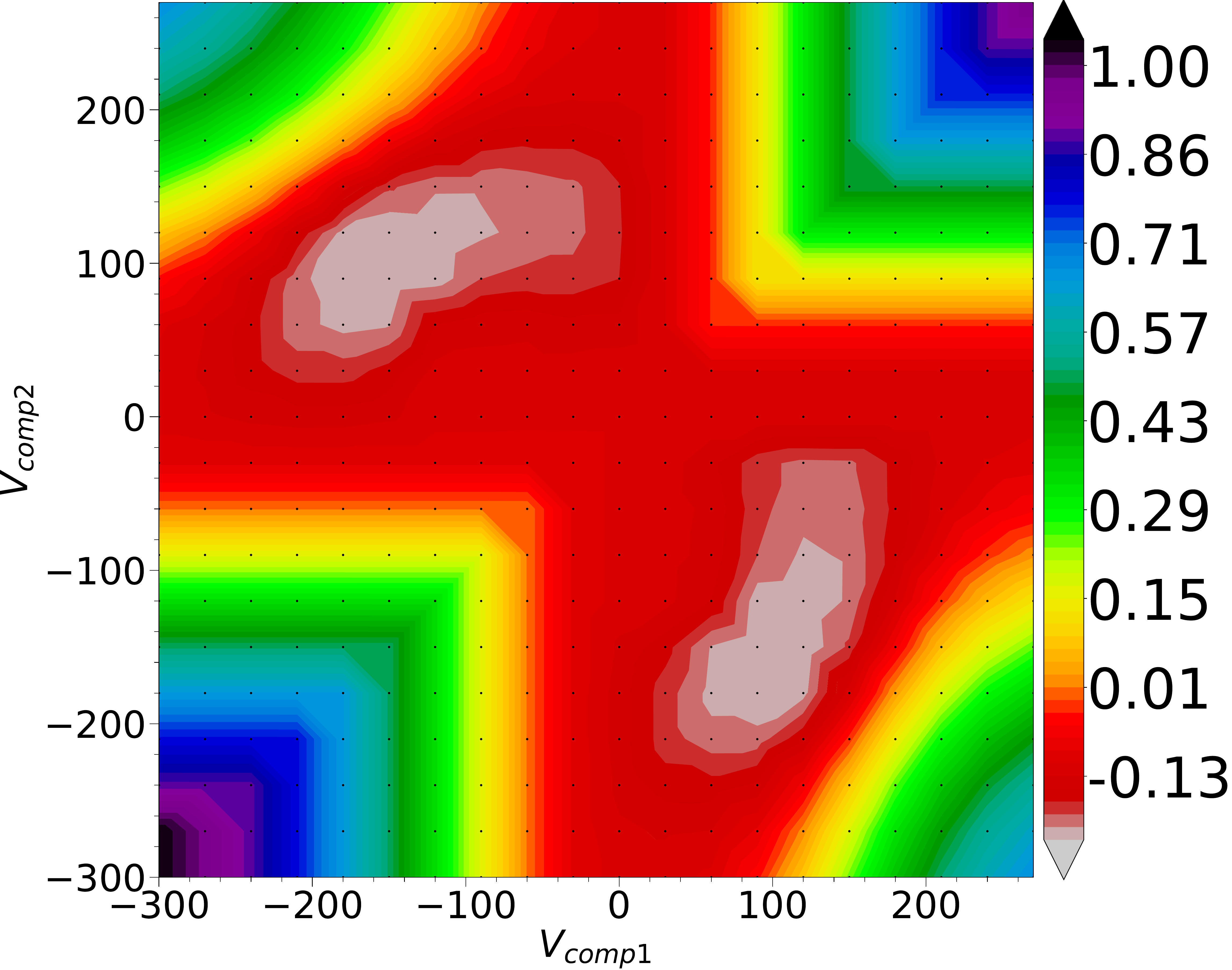}
\includegraphics[width=.49\columnwidth]{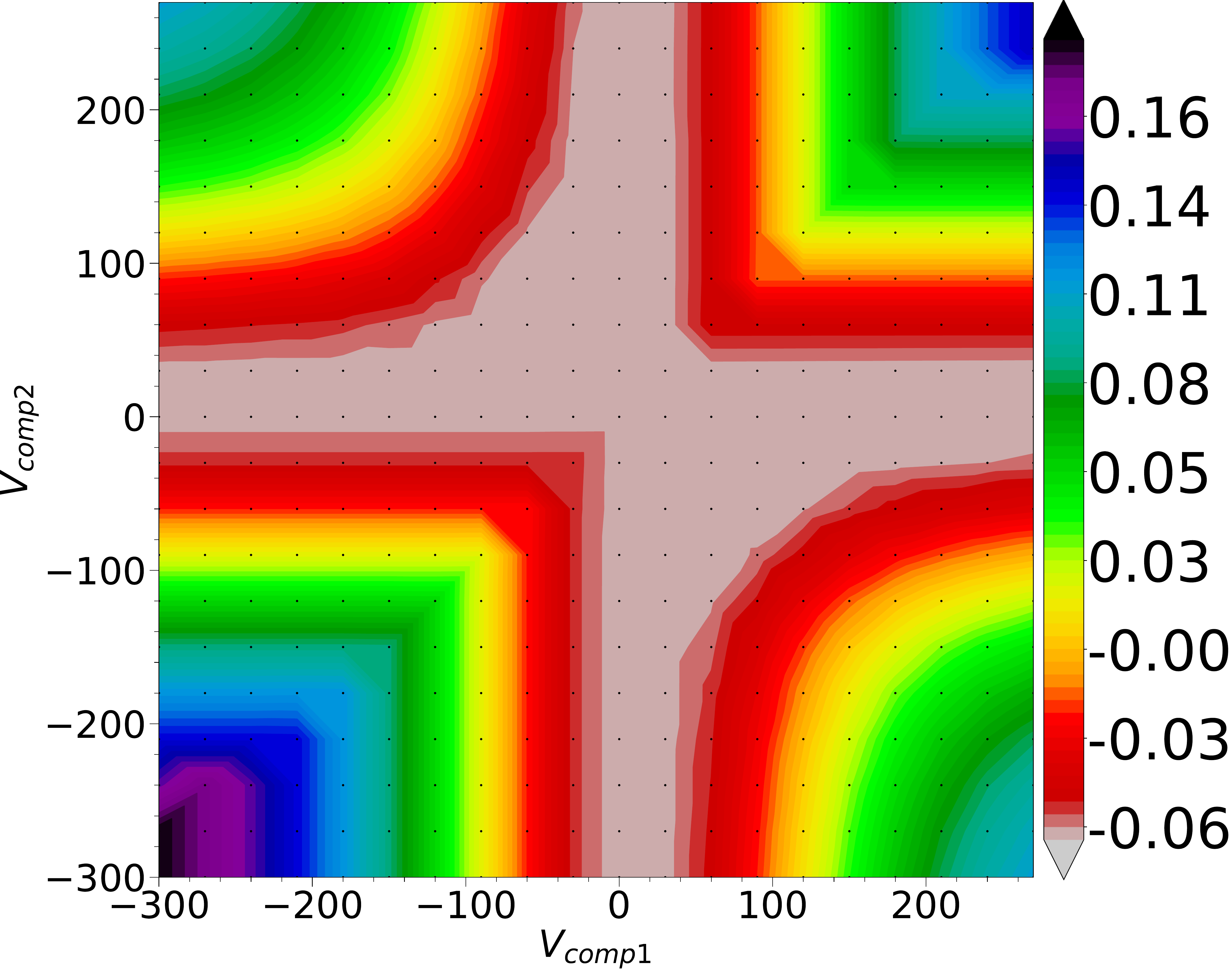}
\caption{$\chi^2$ maps of galaxies with MaNGA IDs 1-38543 (left) and 1-167555 (right). The x and y axes are the fitted velocities. The black dots are the couples of velocities at which the local minimum $\chi^2$ is estimated. The colorbars are the log$_{10}(\chi^2$/DOF). \textit{Left panel:} the global minimum $\chi^2$ is found in two distinct regions (in white) of the map; this indicates that the two counter-rotating disks are spectroscopically distinguishable. \textit{Right panel:} the global minimum $\chi^2$ is found in a single, cross-shaped (white) region; in this case, the two disks are not spectroscopically distinguishable. See the electronic Appendix B for details.}
\label{fig:exchi}
\end{figure}

The recovery of the two kinematic components presents an additional complication when performing fits with a program designed to find a local minimum, like \texttt{pPXF}, because it requires a precise estimate of the input starting velocities of the two components. In fact, the best fit solutions are defined as the ones that minimize the $\chi^2$; since \texttt{pPXF} uses a local minimization algorithm, the global minimum will be missed if the input starting velocities are not accurate. To overcome this problem, we looked for the global minimum by creating a grid of $\chi^2$ values, obtained by fitting each of the two components with a set of input velocities; for each couple of velocities for the two components, we calculated the reduced $\chi^2$ (namely the $\chi^2$/DOF, where DOF are the degrees of freedom), and plotted it on a map. This method has already been used in \cite{mitzkus} and in \cite{Tabor_2017}. In the electronic appendix (B) to this paper we give details on the method we used to produce $\chi^2$ maps, and the tests we made to improve it. Here, we only state that when the two counter-rotating disks are spectroscopically distinguishable, the $\chi^2$ map will have two distinct minima, like those shown in the left panel of Figure \ref{fig:exchi}; on the contrary, when the two disks are not distinguishable the global minimum $\chi^2$ will be a cross, as in the right panel of Figure \ref{fig:exchi}. 

\subsection{Gas and stars position angles, and kinematic misalignment}\label{sect:stargaspa}
To compare gas and stellar rotation, we calculate the global kinematic position angle (PA) of the major axis, measured from north to east, of the velocity maps using the \verb|fit_kinematic_pa| routine\footnote{Available from \url{https://pypi.org/project/pafit/}}, described in Appendix C of \cite{Krajnovic_2006}. The stellar kinematics are extracted as described in section \ref{sect:singlecompfit}; for the ionized gas, instead, we simply use the kinematics of the H$\alpha$ emission as extracted by the DAP (W19; \citealt{Belfiore_2019}). We excluded from the sample galaxies whose gas or stellar kinematics do not show a clear rotation pattern, or have many pixels with non-detection, making hard a reliable estimate of the PA, resulting in a sample of 42 CRDs.

To obtain meaningful values of the PA, we took care of some issues to avoid bad fitting with \verb|fit_kinematic_pa|. First, we dealt with high-velocity bins by performing a robust estimate of the standard deviation $\sigma_{\mbox{\scriptsize{std}}}$ of the velocities of bins, and then excluded all bins deviating by more than 3$\sigma_{\mbox{\scriptsize{std}}}$. Secondly, even though foreground stars are flagged during the fit, the IFU footprint can be disturbed by the presence of background/neighbouring galaxies; thus, when needed, we masked all the spatial bins outside a certain elliptical radius, depending on the region of influence of the foreground/background galaxy. We adopted the same masking also for few galaxies where the routine failed in fitting the PA due to untrustworthy external bins, whose velocity deviate from the main velocity field, resulting in an incorrect PA which could be clearly determined by eye.\\
\begin{figure}
\includegraphics[width = \columnwidth]{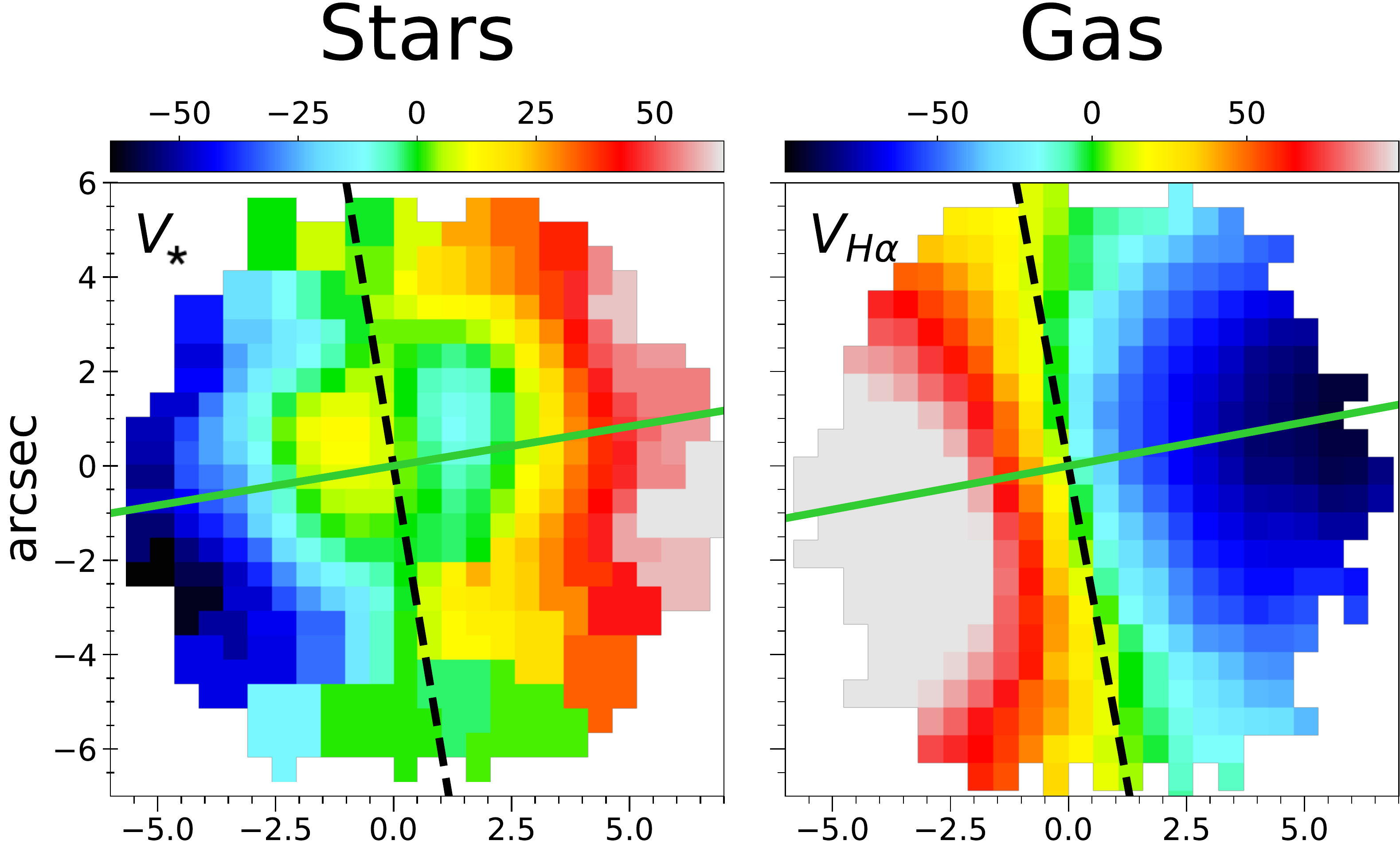}
\includegraphics[width = \columnwidth]{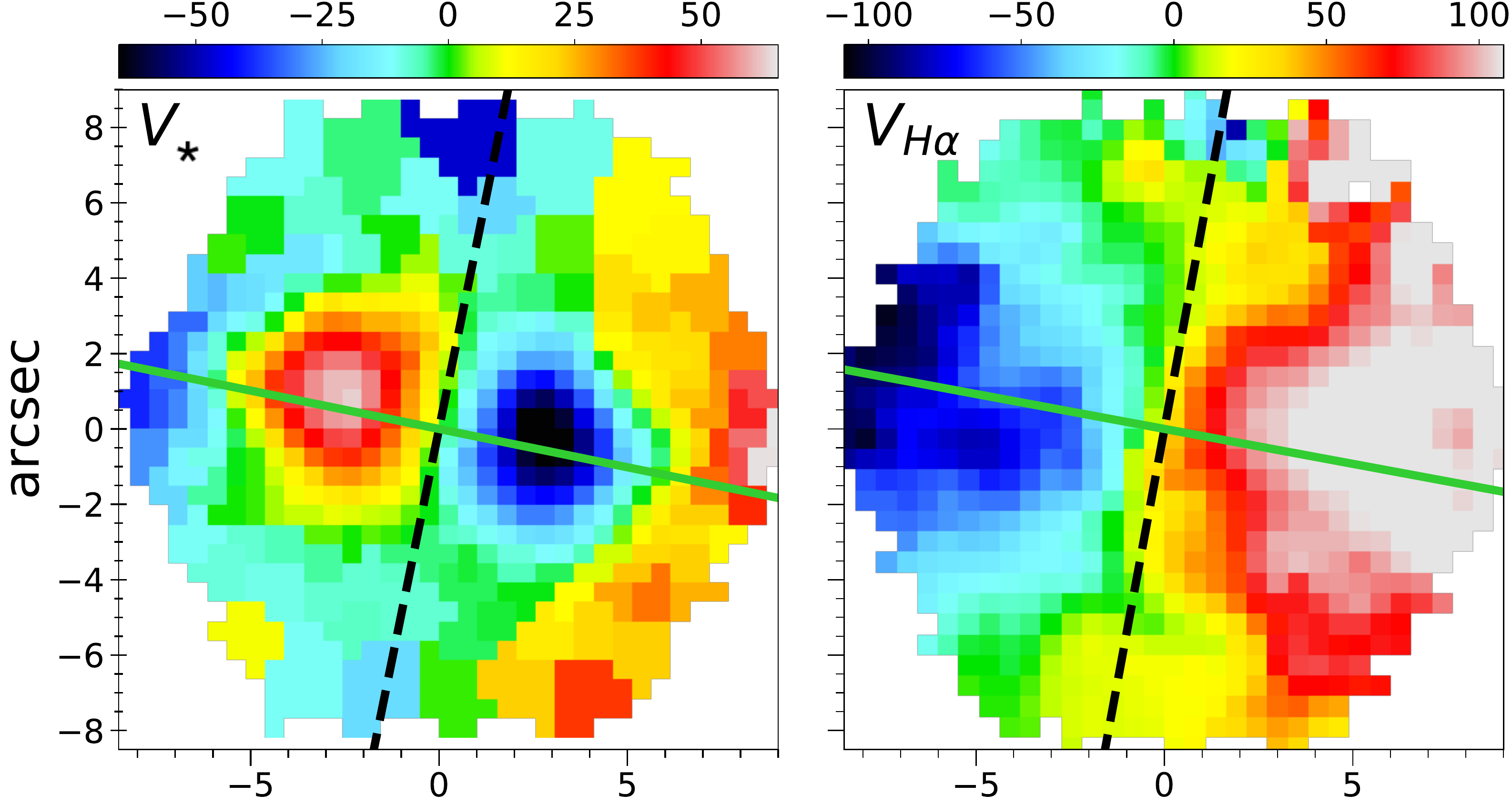}
\includegraphics[width = \columnwidth]{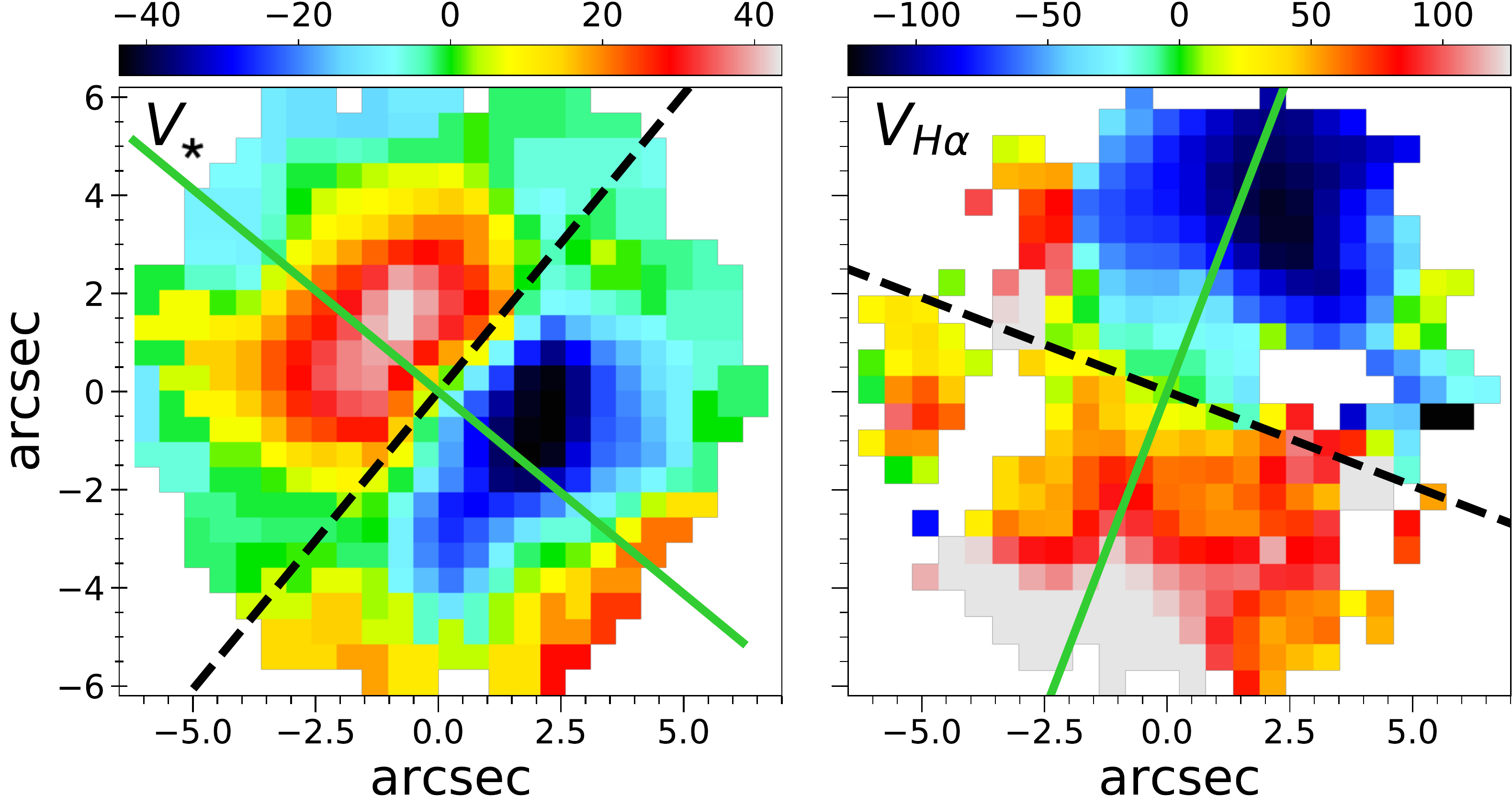}
\caption{Three representative examples of stars (left columns) and gas (right columns) velocity fields of galaxies with the gaseous disk corotating with the inner disk (upper panels, MaNGA ID: 1-179561), with the outer disk (middle panels, MaNGA ID: 1-136248) and misaligned (lower panels, MaNGA ID: 1-163594). The straight lines in green are the major axis PAs, and the black dashed lines mark the rotation axes (calculated as PA$+90^\circ$). Colorbars are the velocity ranges in km s$^{-1}$.}
\label{fig:stargas_ex}
\end{figure}
 We compared the gas and stellar kinematics by considering the difference of their position angles:
\begin{equation}
\Delta \mbox{PA} = | \mbox{PA}_{\mbox{\scriptsize{stars}}} - \mbox{PA}_{\mbox{\scriptsize{gas}}} | \; .
\label{eq:pa}
\end{equation}

When fitting the stellar PA, \verb|fit_kinematic_pa| does not distinguish between the sense of rotation of the two counter-rotating disks; however, since they share the same rotation axis, we can define the misalignment between gas and stars using the stellar rotation axis as reference, without worrying about the sense of rotation. Furthermore, since the routine returns a PA between 0$^\circ$ and 180$^\circ$, and does not discriminate between the redshifted and blueshifted sides, we considered misaligned those galaxies with 30$^\circ \leq \Delta \mbox{PA} \leq$ 150$^\circ$, namely those whose stellar and gas rotation axes have at least a 30$^\circ$ misalignment, regardless of their direction of rotation. The choice of 30$^\circ$ is arbitrary, and we choose it to allow for a comparison with previous studies (e.g. \cite{Davis_2011}, \cite{Bryant_19}).

Errors calculated by the routine are the 3$\sigma$ confidence limit of the fit, and include the error on the $\chi^2$. However, the latter, even when the output PA is evidently good, often yields large errors that are not statistically meaningful. This is because the routine compares the observational data to a symmetrized model, and even little deviations from symmetry can yield large errors. This happens commonly when fitting the gas PA, since it often presents asymmetries or clumps. Even larger are the errors on the stellar PA, because of the inversion of the velocity field due to the counter-rotation. For these reasons, we do not provide errors calculated by the routine, and we assume a fixed representative uncertainty of $5^\circ$ for all fitted PAs, that is the maximum error estimated by \cite{Duckworth_19}, and in agreement with the median uncertainty of $5.2^\circ$ for the ATLAS$^{\rm 3D}$ sample quoted by \cite{Krajnovic_2011}. The values of $\Delta$PA are given in Table \ref{tab:crd}.

For those galaxies who have their gas and stellar PA aligned, we determine whether the gas corotates with the `inner' disk, namely the one whose kinematics prevail in the central regions of the galaxy, or the `outer' disk, i.e. the one prevailing outwards, when the rotation is inverted. To establish which of the two disks the gas corotates with, we checked the kinematic maps and visually determined it. We show examples of alignment with the inner disk, alignment with the outer disk, and misalignment in Figure \ref{fig:stargas_ex}. For those CRDs lacking the counter-rotation feature in the velocity map, we assume that the only disk that appears is the outer one; for example, if the gas counter-rotates with the only disk that is detected, we assume it is corotating with the inner disk.

\section{Stellar Population Fitting}\label{sect:fitpop}
\subsection{Stellar population maps}\label{sect:popfit}
Since the strength of the spectral lines is rather sensible to the S/N, we tested if any substantial variation occurs in the fitting of the stellar population properties by increasing the target (S/N)$_{\mbox{\scriptsize{min}}}$ of the spatial bins. We found that those galaxies binned with (S/N)$_{\mbox{\scriptsize{min}}}$ = 25 do not show any noticable change when binned up to (S/N)$_{\mbox{\scriptsize{min}}}$ = 75, while those with (S/N)$_{\mbox{\scriptsize{min}}}$=15 can be different in the external regions, but the overall maps do not change significantly. Therefore, we kept the same spatial binning adopted for the kinematics (section \ref{sect:binning}), in order to allow for a direct comparison between maps, and limited further considerations on the  stellar population properties to the central regions, where the S/N is higher.

To extract the stellar population properties, we simultaneously fitted the stellar and the gas kinematics with \texttt{pPXF}. To fit the stellar component, we use the MIUSCAT SSP models \citep{Vazdekis_2010}, based on the Padova isochrones \citep{Girardi_00}, with a Salpeter initial mass function (IMF), and we restricted to the safe range of parameters described in \cite{Vazdekis_2012}. The final library consists of 150 SSP models of twenty-five different ages, ranging between 0.06 Gyr and 15.84 Gyr at logarithmic steps, and six different metallicities [M/H] $= -1.71, -1.31, -0.71, -0.4, 0.0, +0.22$ dex. The templates for the gas emission lines are Gaussians; in particular, we fit the lines of the Balmer series H$\alpha$, H$\beta$, H$\gamma$ and H$\delta$, for which we fixed the flux ratios, the [SII] doublet at $\lambda$ = 6717, 6731 \AA , [OIII] at $\lambda$ = 5007 \AA \, and [NII] at $\lambda$ = 6583 \AA . Before performing the fits, we de-redshifted the galaxy spectrum to the rest-frame, normalized the SSP templates to their median value and logarithmically rebinned them; further, we applied a mask for the night sky emission lines of the [OI] at $\lambda$ = 5578, 6301 \AA \, and NaI at $\lambda$ = 5890, 5896 \AA , for these lines can be important in the MaNGA spectra. When fitting the stellar population, we are interested in the true strength of the spectral lines, and therefore we do not fit any additive polynomial; on the other hand, we use degree-eight multiplicative Legendre polynomials.

We extracted the stellar population properties performing two fits on each spatial bin. From the first fit, we obtained the two kinematic solutions for stars and gas, and the $\sigma_{\mbox{\scriptsize{std}}}$ of the residuals. Then, the second fit is performed setting the kinematic solutions of the first fit as the starting guesses for the stellar and gas components, and the $\sigma_{\rm std}$ as the input noise. From this second fit the average ages and metallicities are calculated as the $r$-band luminosity weighted sums of the individual stellar population values. For this, we normalized the SSP spectral templates to the same flux in the $r$-band before fitting with pPXF. We used the weights $w_i$ returned by the pPXF fits and the corresponding ages and metallicities of each SSP template as follows to compute the luminosity-weighted quantities:
\begin{equation}
\langle \log_{10}\mbox{Age} \rangle = \frac{\sum_i{w_i \log_{10}\mbox{Age}_i}}{\sum_i w_i}
\label{eq:age}
\end{equation}
\begin{equation}
\langle \mbox{[M/H]} \rangle = \frac{\sum_i{w_i \mbox{[M/H]}_i}}{\sum_i {w_i}}
\label{eq:met}
\end{equation}
where $w_i$ is the weight of the $i$-th stellar populations, and sums are performed over the whole MIUSCAT populations. Errors on age and metallicities are estimated as the rms of the residuals between a smoothed model of the data, computed with the implementation of \cite{atlas_xx}\footnote{Available from \url{https://pypi.org/project/loess/}.} of the LOESS algorithm \citep{Cleveland_78}, and the actual data.

The values of $\langle$log$_{10}$Age$\rangle$ and $\langle$[M/H]$\rangle$ of each bin are luminosity-weighted to obtain their mean within one effective radius (more precisely, the elliptical Petrosian 50\% light radius in SDSS r-band provided by the DRP, and given in Table \ref{tab:crd}) that represents our ``global'' population properties log$_{10}$Age and [M/H] (also given in Table \ref{tab:crd}). To evaluate gradients, we first construct the radial profiles of these properties by taking the median values inside elliptical annuli of logarithmically-spaced radii, varying from R$_e$/8 to R$_e$; then, we take as gradients the slopes of linear fits performed on the logarithmic profiles, using the \verb|lts_linefit|\footnote{Available from \url{https://pypi.org/project/ltsfit/}} routine described in \cite{atlas_xv} and defined as
\begin{equation}
\nabla \mbox{Age} =  \Delta \langle \mbox{log}_{10}\mbox{Age} \rangle / \Delta \mbox{log}_{10}R \; ,
\label{eq:gradage}
\end{equation}
\begin{equation}
\nabla \mbox{[M/H]} =  \Delta \langle \mbox{[M/H]} \rangle / \Delta \mbox{log}_{10}R \; .
\label{eq:gradmet}
\end{equation}
We consider a gradient to be positive if the corresponding age or metallicity increases from the center outwards. The values of $\nabla$Age and $\nabla$[M/H] are given in Table \ref{tab:crd}. Note that the radial gradients can vary depending on the radial coordinate used, like linear or logarithmic in radius (e.g., \citealt{Zhuang19}). However, population gradients are better estimated using a logarithmic radial coordinate, because many quantities inside galaxies are well approximated by power-law functions of radius. This is the case, for instance, for surface brightness profiles (e.g., \citealt{Lauer95}), stellar mass-to-light ratios (e.g., \citealt{Ge21}), stellar population (e.g., \citealt{SauronXVII}, \citealt{Li_18}), or colour (e.g., \citealt{Carollo97}). The logarithmic gradient of a power law is a constant, and can be quantified by a single number; the linear gradient of a power law, instead, varies with radius. It cannot be quantified by a single number and becomes meaningless for a whole galaxy. \\

\begin{figure}
\centering
\includegraphics[width=\columnwidth]{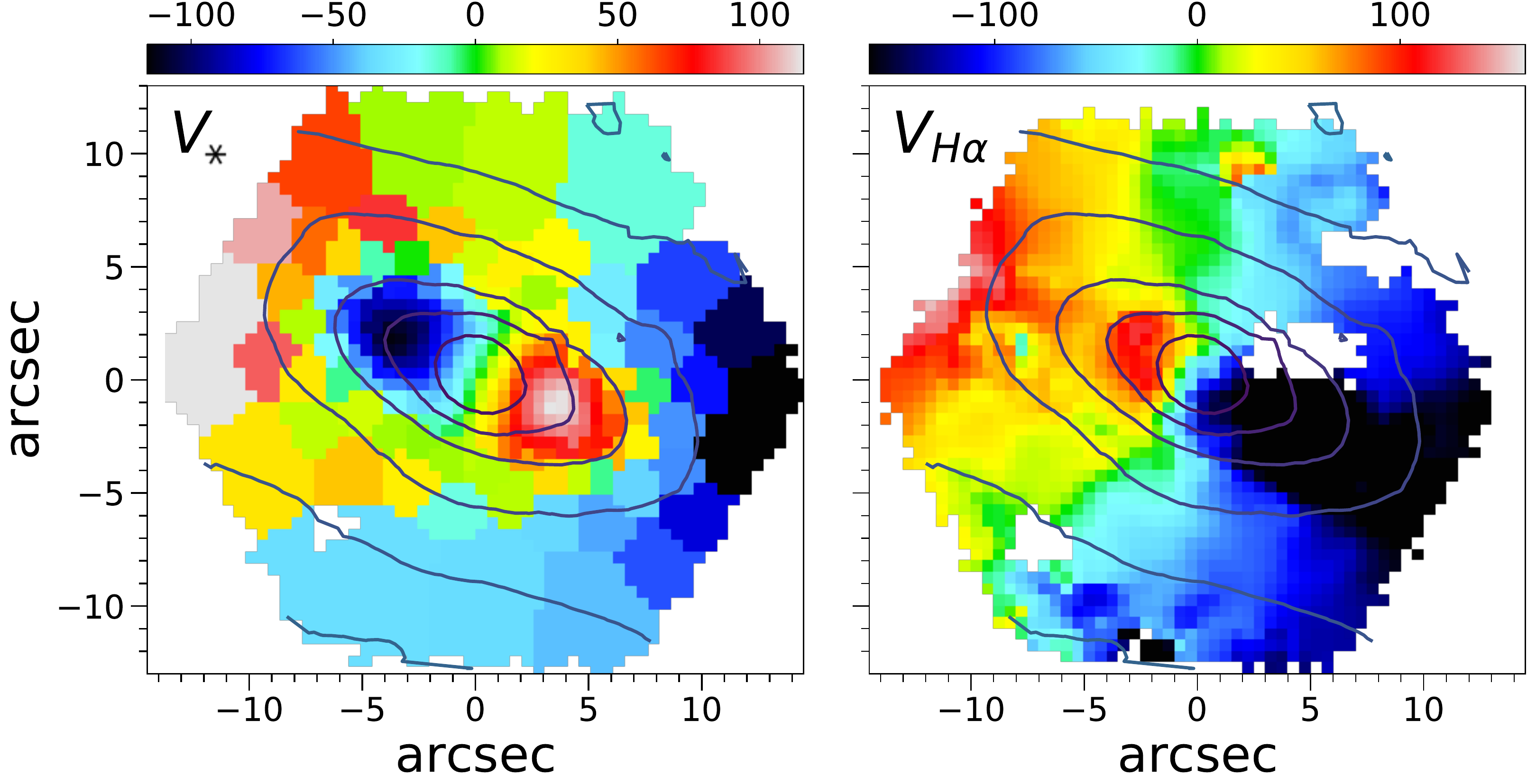}\\

\includegraphics[width=.55\columnwidth]{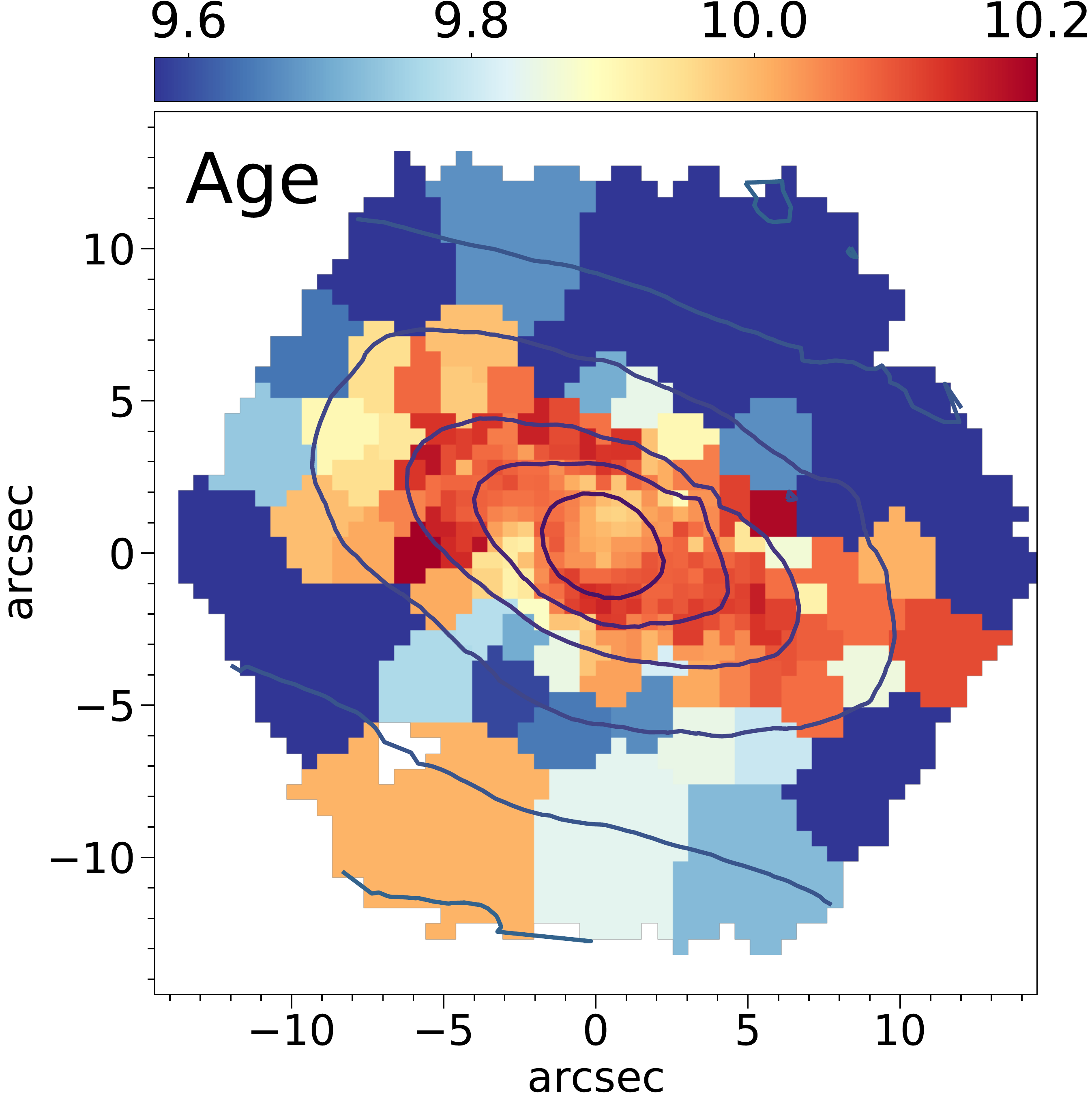}
\caption{Stellar and gas velocity maps (upper panels) compared to the logarithmic age map (lower panel) of a CRD (MaNGA ID: 1-174947). Velocities are in km s$^{-1}$. Age values are calculated using equation \eqref{eq:age}. In this example, we can see an abrupt change in age, coinciding with the inversion of the stellar rotation. The stellar rotation is inverted halfway from the third to the fourth isophote, starting from the center.The outer disk, which is corotating with the ionised gas, is also the younger one.}
\label{fig:ex_gaspop}
\end{figure}

\subsection{Gas-stars corotation and age maps}\label{sect:gaspop}

For those galaxies with the gas and stellar rotation aligned, we investigated whether the gaseous disk corotates with the younger or the older disk by comparing the velocity maps, both of H$\alpha$ and stars, with the age maps. Age maps are often noisy at large radii, and since the outer disk typically arises at radii larger than R$_e$, in general it is not straightforward to determine which disk is the older and which the younger. Further, the age difference from the inner to the outer regions could be due to a genuine variation of the stellar population, rather than a difference in age of the two disks. 

To take into account these considerations, we assigned a distinct age to the disk the gas is corotating with only when the following three conditions are met: (1) the change in the stellar population coincides with the inversion of the stellar rotation; (2) the age difference of neighbouring spaxels in the region of inversion is significant; (3) most spatial bins associated with the inner or the outer disk have the same age within $\sim$ 0.1 dex. For example, Figure \ref{fig:ex_gaspop} shows how most bins within the region of inversion have log$_{10}$Age$ \gtrsim 10$, while outside it most bins have log$_{10}$Age $\lesssim 9.75$; we can also see the abrupt change in age of neighbouring bins up to $\sim 0.6$ dex in the region of the stellar rotation inversion. Here, the gas clearly corotates with the outer stellar disk, which is also the younger one.

\begin{figure}
\includegraphics[width=\columnwidth]{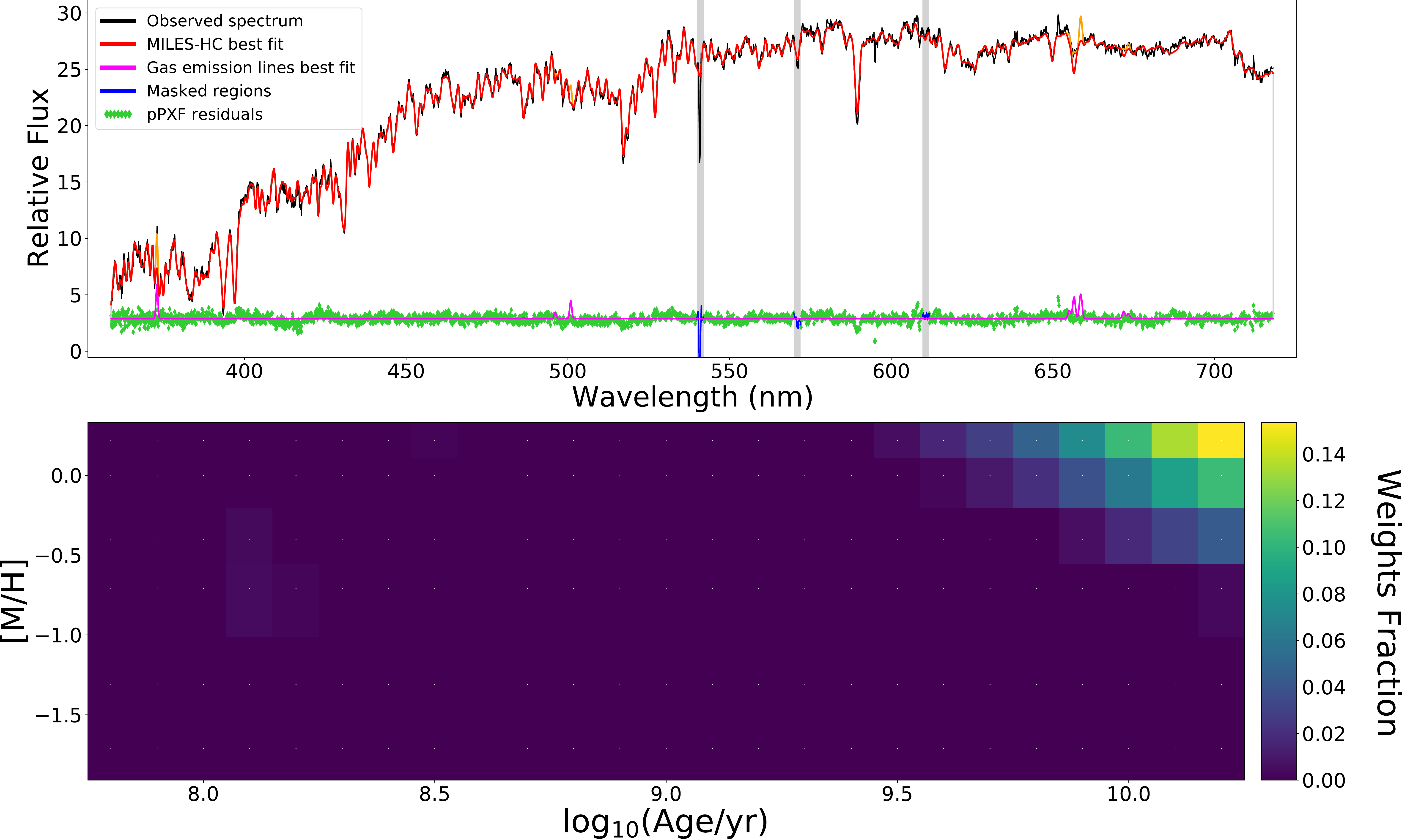}
\includegraphics[width=\columnwidth]{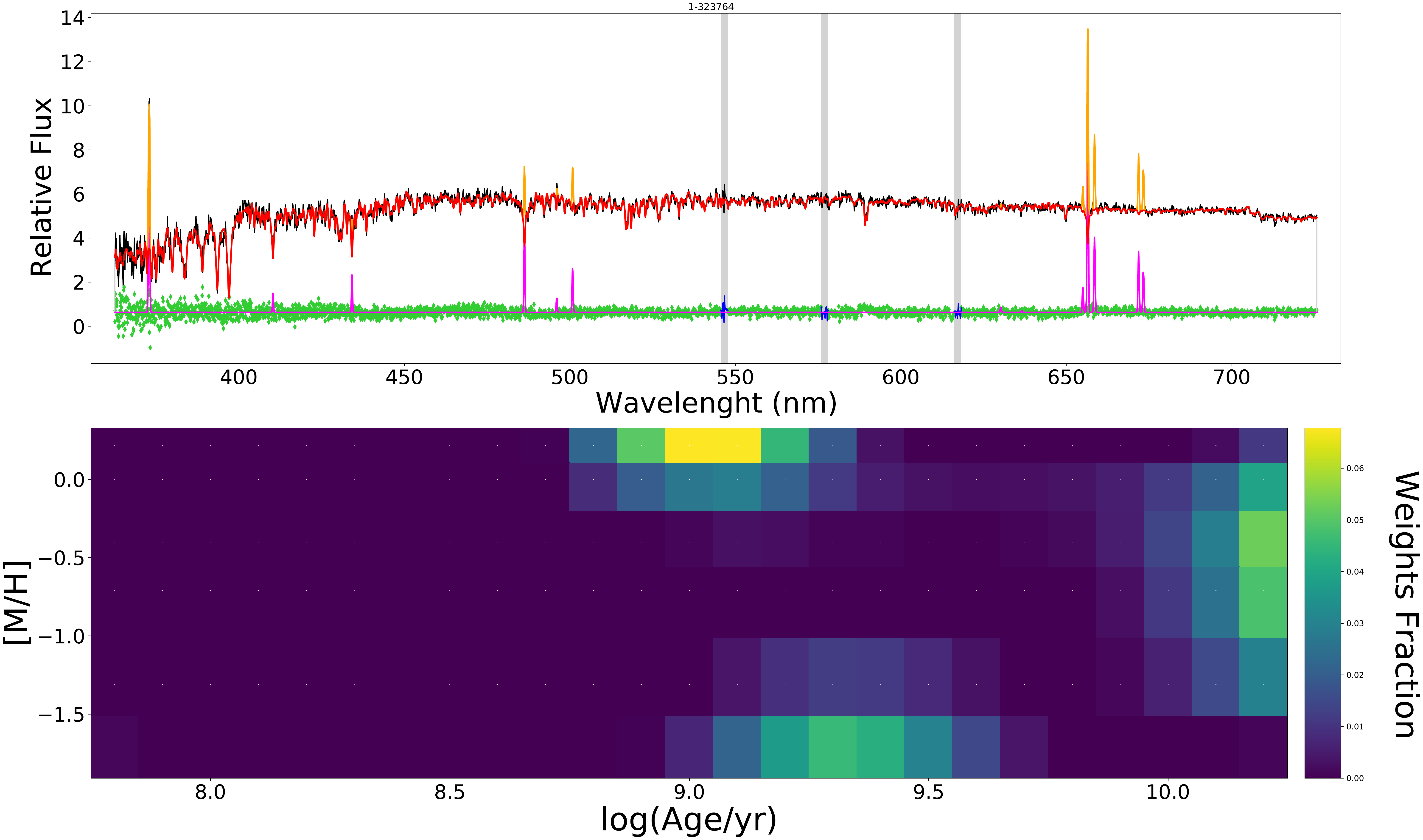}
\includegraphics[width=\columnwidth]{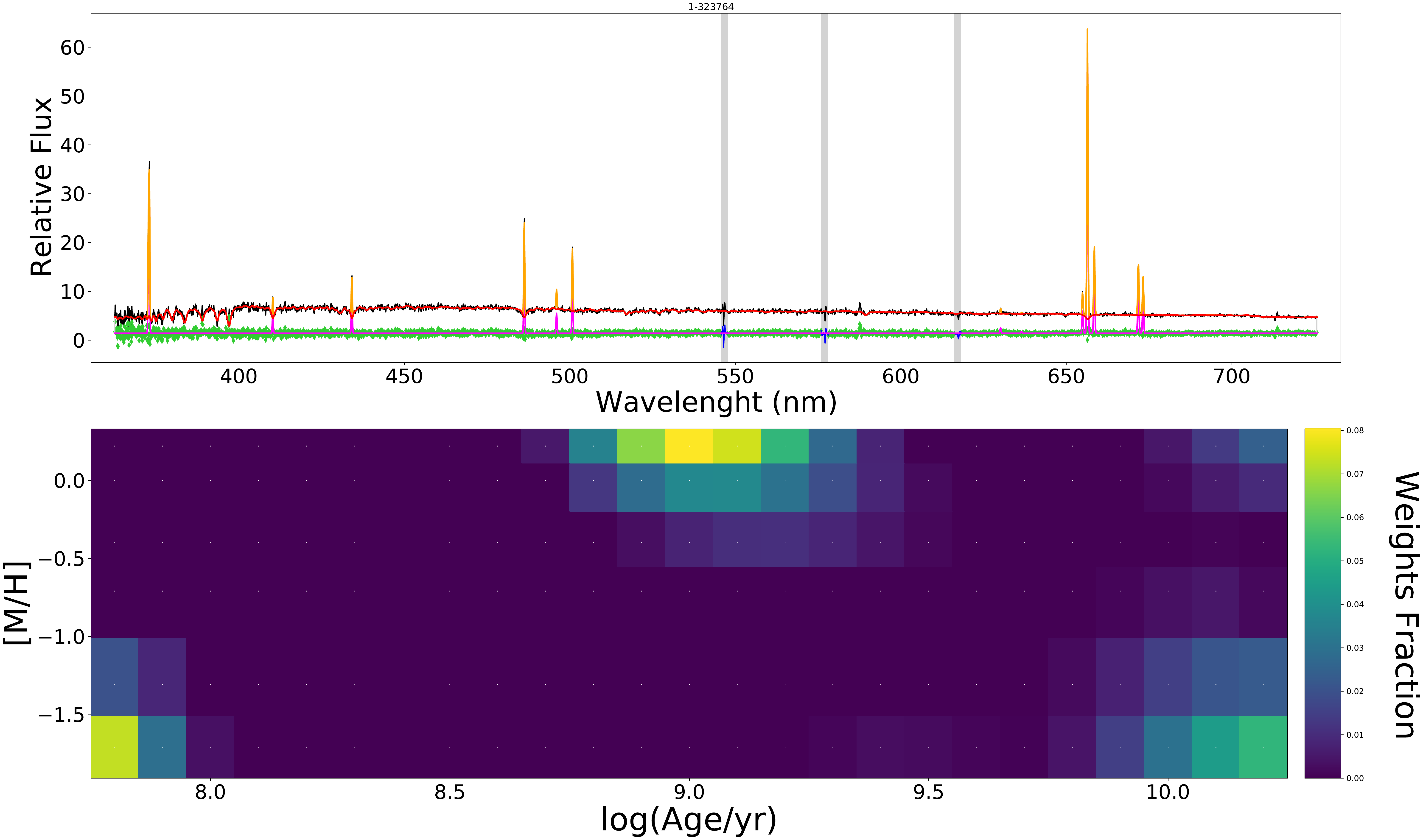}
\caption{Examples of three stellar population regularised fits with pPXF and relative weights fraction maps of the fitted SSPs. \textit{Upper panels:} spectral fits of a central spatial bin of a galaxy (MaNGA ID: 1-248869) showing only one peak in the weights map; the same peak appears in all the other spatial bins considered, thus this galaxy is classified as `unimodal'. \textit{Middle and lower panels:} spectral fits of two different spatial bins of a galaxy (MaNGA ID: 1-323764) showing multiple peaks in the weights maps. In this case, the peaks change at different spatial bins, and this galaxy is thus classified as `star-forming'. Instead, galaxies showing the same multiple peaks at different spatial bins are classified as `multimodal'.}
\label{fig:regul_ex}
\end{figure}

\begin{table*}
\caption{Properties of CRDs.}
\label{tab:crd}
\begin{tabularx}{\textwidth}{*{11}{c}}
\hline
& & & & & & & & & &\\
MaNGA ID & Morphology & log$_{10}$(M$_\ast$/M $_\odot$) & R$_e$ [ kpc ] & $\Delta$PA [ $^\circ$ ] & log$_{10}$Age [ yr ] & [M/H] [ dex ] & $\nabla$Age & $\nabla$ [M/H] & Modality\\
& & & & & & & & & &\\
(1) & (2) & (3) & (4) & (5) & (6) & (7) & (8) & (9) & (10) \\
& & & & & & & & & &\\
\hline
& & & & & & & & & &\\
12-84617${\color{blue}\ddagger}$ & E & $9.49$ & $1.46$ & $22.5$ ($\blacktriangle$) & $9.25$ & $-1.05$ & $+0.03$ & $-0.04$ & sf\\ 
1-113520 & E & $10.03$ & $1.07$ & $135.0$ ($\bigstar$) & $9.88$ & $-0.15$ & $+0.24$ & $-0.22$ & multi\\ 
1-115097$^\dagger$ & S0 & $10.54$ & $2.72$ & $11.0$ ($\blacktriangle$) & $9.92$ & $-0.16$ & $+0.10$ & $-0.35$ & multi\\ 
1-38347 & E & $11.11$ & $2.30$ & $ - $ & $10.06$ & $-0.01$ & $+0.02$ & $-0.19$ & multi\\ 
1-339061$^\dagger$ & E & $10.67$ & $3.41$ & $29.0$ ($\color{cyan}\bullet$) & $9.79$ & $-0.09$ & $-0.18$ & $-0.53$ & multi\\ 
1-44047$^\dagger$ & E & $10.55$ & $1.67$ & $17.0$ ($\bullet$) & $10.04$ & $-0.15$ & $-0.05$ & $+0.01$ & multi\\ 
1-44483$^\dagger$ & S0 & $10.49$ & $2.33$ & $8.0$ ($\color{cyan}\bullet$) & $9.95$ & $-0.08$ & $+0.17$ & $-0.36$ & multi\\ 
1-556514 & E & $11.36$ & $4.06$ & $ - $ & $10.04$ & $-0.01$ & $+0.01$ & $-0.23$ & uni\\ 
1-37494 & E & $10.62$ & $2.07$ & $10.0$ ($\color{cyan}\bullet$) & $10.06$ & $-0.33$ & $-0.01$ & $-0.07$ & $-$\\ 
1-37155 & E & $11.32$ & $4.46$ & $24.0$ ($\bullet$) & $9.75$ & $+0.07$ & $+0.29$ & $-0.52$ & multi\\ 
1-38543$^\dagger$ & E & $10.73$ & $1.76$ & $0.5$ ($\color{red}\blacktriangle$) & $9.80$ & $+0.09$ & $-0.04$ & $-0.14$ & multi\\ 
1-47248$^\dagger$ & E & $11.56$ & $6.06$ & $ - $ & $10.03$ & $+0.08$ & $+0.21$ & $-0.24$ & uni\\ 
1-137890 & S0 & $10.39$ & $1.67$ & $0.5$ ($\color{cyan}\blacktriangle$) & $9.31$ & $-0.19$ & $+0.08$ & $-0.31$ & sf\\ 
1-255220 & S0 & $10.23$ & $1.20$ & $30.0$ ($\bigstar$) & $9.88$ & $-0.26$ & $+0.02$ & $-0.13$ & multi\\ 
1-282035${\color{blue}\ddagger}$ & S0 & $10.13$ & $1.72$ & $120.0$ ($\bigstar$) & $8.61$ & $-0.52$ & $+0.70$ & $-0.89$ & sf\\ 
1-251783 & E & $10.55$ & $1.37$ & $ - $ & $9.75$ & $-0.06$ & $+0.02$ & $-0.16$ & multi\\ 
1-419257$^\dagger$ & S0 & $10.65$ & $2.61$ & $1.0$ ($\color{cyan}\blacktriangle$) & $9.67$ & $+0.00$ & $-0.20$ & $-0.31$ & sf\\ 
1-418023 & E & $10.36$ & $0.99$ & $41.0$ ($\bigstar$) & $9.97$ & $-0.74$ & $-0.17$ & $+0.24$ & $-$\\ 
1-167555 & E & $9.90$ & $2.24$ & $85.5$ ($\bigstar$) & $9.95$ & $-0.53$ & $+0.36$ & $-0.33$ & multi\\ 
1-274440 & S0 & $10.0$ & $1.20$ & $17.5$ ($\color{cyan}\blacktriangle$) & $9.22$ & $-0.46$ & $+0.00$ & $+0.06$ & sf\\ 
1-274545${\color{blue}\ddagger}$ & U & $10.16$ & $2.12$ & $4.0$ ($\blacktriangle$) & $9.27$ & $-0.59$ & $-0.08$ & $-0.05$ & sf\\ 
1-275185 & E & $11.51$ & $6.41$ & $ - $ & $10.04$ & $+0.05$ & $-0.01$ & $-0.18$ & uni\\ 
1-167044 & E & $11.58$ & $5.83$ & $ - $ & $10.07$ & $+0.13$ & $+0.10$ & $-0.25$ & uni\\ 
1-166613 & E & $11.53$ & $4.57$ & $ - $ & $9.86$ & $+0.13$ & $+0.06$ & $-0.27$ & multi\\ 
1-246175 & S0 & $10.35$ & $2.81$ & $8.5$ ($\color{cyan}\bullet$) & $9.85$ & $-0.58$ & $-0.16$ & $+0.11$ & multi\\ 
1-210728 & E & $10.77$ & $3.08$ & $ - $ & $10.04$ & $-0.17$ & $-0.06$ & $-0.23$ & multi\\ 
1-248869$^\dagger$ & E & $11.24$ & $3.31$ & $26.5$ ($\bullet$) & $10.02$ & $+0.12$ & $+0.03$ & $-0.24$ & uni\\ 
1-136248$^\dagger$ & S0 & $10.84$ & $4.04$ & $1.0$ ($\color{cyan}\bullet$) & $9.93$ & $-0.03$ & $-0.10$ & $-0.33$ & multi\\ 
1-179561 & E & $10.48$ & $1.53$ & $1.5$ ($\color{cyan}\blacktriangle$) & $9.39$ & $-0.56$ & $+0.18$ & $-0.19$ & sf\\ 
1-635590$^\dagger$ & E & $11.64$ & $4.94$ & $ - $ & $10.07$ & $+0.14$ & $+0.04$ & $-0.37$ & uni\\ 
1-113698${\color{blue}\ddagger}$ & S0 & $9.38$ & $1.10$ & $10.5$ ($\color{cyan}\blacktriangle$) & $9.20$ & $-0.81$ & $-0.10$ & $+0.12$ & multi\\ 
1-44722 & E & $9.88$ & $4.68$ & $ - $ & $10.05$ & $-0.47$ & $-0.12$ & $-0.35$ & $-$\\ 
1-45016 & E & $11.67$ & $6.27$ & $ - $ & $10.04$ & $+0.04$ & $+0.06$ & $-0.14$ & uni\\ 
1-163594 & E & $11.53$ & $2.89$ & $108.5$ ($\bigstar$) & $9.88$ & $+0.07$ & $+0.10$ & $-0.33$ & multi\\ 
1-248410 & S0 & $9.94$ & $1.31$ & $2.5$ ($\color{red}\bullet$) & $9.61$ & $-0.50$ & $+0.17$ & $-0.12$ & sf\\ 
1-235983 & E & $10.04$ & $0.75$ & $95.5$ ($\bigstar$) & $10.01$ & $-0.14$ & $+0.04$ & $-0.18$ & multi\\ 
1-236144 & S & $10.92$ & $5.77$ & $19.0$ ($\bullet$) & $9.92$ & $-0.35$ & $-0.01$ & $-0.30$ & multi\\ 
1-174947$^\dagger$ & S0 & $11.0$ & $5.64$ & $2.0$ ($\color{cyan}\bullet$) & $10.05$ & $-0.06$ & $-0.02$ & $-0.55$ & $-$\\ 
1-278079 & E & $11.96$ & $7.53$ & $ - $ & $9.98$ & $+0.16$ & $+0.06$ & $-0.09$ & uni\\ 
1-188530 & E & $11.0$ & $3.69$ & $164.0$ ($\blacktriangle$) & $10.19$ & $-0.34$ & $-0.04$ & $-0.07$ & uni\\ 
1-149172${\color{blue}\ddagger}$ & S0 & $9.49$ & $1.12$ & $38.0$ ($\bigstar$) & $9.26$ & $-0.70$ & $-0.12$ & $+0.02$ & sf\\ 
1-94773 & E & $10.56$ & $1.6$ & $ - $ & $9.72$ & $-0.22$ & $+0.02$ & $-0.01$ & multi\\ 
1-94690$^\dagger$ & S0 & $10.56$ & $2.63$ & $6.0$ ($\color{cyan}\blacktriangle$) & $9.84$ & $-0.28$ & $-0.03$ & $-0.20$ & multi\\ 
1-323766 & E & $10.16$ & $1.04$ & $0.5$ ($\blacktriangle$) & $9.82$ & $-0.32$ & $+0.16$ & $-0.24$ & multi\\ 
1-323764${\color{blue}\ddagger}$ & S0 & $9.94$ & $2.18$ & $7.5$ ($\color{cyan}\bullet$) & $9.42$ & $-0.52$ & $-0.36$ & $-0.41$ & sf\\ 
1-135244$^\dagger$ & E & $11.19$ & $2.53$ & $ - $ & $9.99$ & $+0.06$ & $-0.04$ & $-0.19$ & multi\\ 
1-549076$^\dagger$ & U & $11.4$ & $5.21$ & $ - $ & $10.01$ & $+0.09$ & $-0.05$ & $-0.12$ & uni\\ 
1-314719 & E & $11.91$ & $15.51$ & $ - $ & $9.96$ & $+0.12$ & $+0.13$ & $-0.46$ & uni\\ 
1-299176 & E & $11.66$ & $6.79$ & $ - $ & $9.99$ & $+0.14$ & $-0.01$ & $-0.26$ & uni\\ 
1-298940 & E & $10.63$ & $2.74$ & $ - $ & $9.94$ & $-0.24$ & $-0.11$ & $-0.18$ & multi\\ 
1-593328$^\dagger$ & E & $11.71$ & $6.99$ & $0.0$ ($\blacktriangle$) & $10.02$ & $+0.11$ & $-0.14$ & $-0.27$ & uni\\ 
1-322291 & S0 & $10.2$ & $2.56$ & $ - $ & $9.76$ & $-0.22$ & $+0.07$ & $-0.37$ & multi\\ 
1-633000 & S0 & $10.27$ & $1.62$ & $31.5$ ($\bigstar$) & $9.37$ & $-0.47$ & $+0.18$ & $+0.20$ & sf\\ 
1-251198 & S0 & $10.64$ & $2.47$ & $17.0$ ($\color{cyan}\blacktriangle$) & $9.87$ & $-0.35$ & $+0.11$ & $+0.02$ & multi\\ 
1-266244${\color{blue}\ddagger}$ & S0 & $9.63$ & $1.18$ & $ - $ & $9.31$ & $-0.80$ & $-0.13$ & $+0.07$ & sf\\ 
1-234115 & E & $10.53$ & $2.13$ & $3.5$ ($\blacktriangle$)& $10.07$ & $-0.25$ & $-0.03$ & $-0.26$ & $-$\\ 
1-457547 & S0 & $10.24$ & $1.56$ & $13.0$ ($\color{cyan}\bullet$) & $9.66$ & $-0.31$ & $-0.14$ & $-0.24$ & sf\\ 
1-188177$^\dagger$ & E & $10.59$ & $3.17$ & $22.5$ ($\bullet$)& $9.89$ & $-0.26$ & $-0.15$ & $-0.48$ & multi\\ 
1-269227$^\dagger$ & E & $10.76$ & $1.54$ & $-$ & $10.04$ & $+0.07$ & $+0.02$ & $-0.13$ & uni\\ 

& & & & & & & & & &\\
\end{tabularx}
\end{table*}

\begin{table*}
\ContinuedFloat
\caption{\textit{(continue)}}
\begin{tabularx}{\textwidth}{*{11}{c}}
\hline
& & & & & & & & & &\\
MaNGA ID & Morphology & log$_{10}$(M$_\ast$/M $_\odot$) & R$_e$ [ kpc ] & $\Delta$PA [ $^\circ$ ] & log$_{10}$Age [ yr ] & [M/H] [ dex ] & $\nabla$Age & $\nabla$ [M/H] & Modality\\
& & & & & & & & & &\\
(1) & (2) & (3) & (4) & (5) & (6) & (7) & (8) & (9) & (10) \\
& & & & & & & & & &\\
\hline
& & & & & & & & & &\\
1-318513 & S0 & $10.24$ & $1.39$ & $3.0$($\bullet$) & $9.75$ & $-0.60$ & $+0.02$ & $-0.05$ & multi\\ 
1-42660${\color{blue}\ddagger}$ & S0 & $10.75$ & $2.84$ & $6.0$ ($\color{cyan}\bullet$) & $9.68$ & $-0.33$ & $-0.31$ & $-0.49$ & multi\\ 
1-121871 & S0 & $10.41$ & $2.72$ & $10.5$ ($\bullet$) & $9.61$ & $-0.29$ & $+0.02$ & $-0.22$ & sf\\ 
1-297822 & E & $10.9$ & $2.18$ & $-$ & $9.67$ & $-0.12$ & $+0.04$ & $-0.12$ & multi\\ 
1-386322 & E & $10.22$ & $1.56$ & $4.0$ ($\bullet$) & $9.87$ & $-0.20$ & $+0.06$ & $-0.24$ & multi\\ 
& & & & & & & & & &\\
\hline

\end{tabularx}
\begin{minipage}{\textwidth}

Columns: (1) The MaNGA ID of the galaxy. The symbol ${\color{blue}\ddagger}$ marks galaxies labelled as `CRD in formation' in section \ref{sect:selection}, while $\dagger$ marks galaxies exhibiting two minima in the $\chi^2$ map. (2) Galaxy morphology: `E' for Elliptical, `S' for Spiral, `S0' for Lenticular, `U' for unclassified morphology. (3) Stellar mass taken from \citealt{G19benchmark}. (4) Elliptical Petrosian 50\% light radius in SDSS r-band. (5) Difference of the kinematic position angles between the stellar and the gas velocity fields. Symbols in brackets are the same of Figure \ref{fig:MS_stargas}. Associated error: 7$^\circ$. (6) and (7) Luminosity weighted mean age and metallicity within 1 R$_e$. Typical errors: 0.04 dex and 0.05 dex, respectively. (8) and (9) Age and metallicity gradients. (10) Modality of the weights maps from regularised fits: `uni' for unimodal, `multi' for multimodal, `sf' for star-forming.
\end{minipage}
\end{table*}

\begin{figure*}
\includegraphics[width=\textwidth]{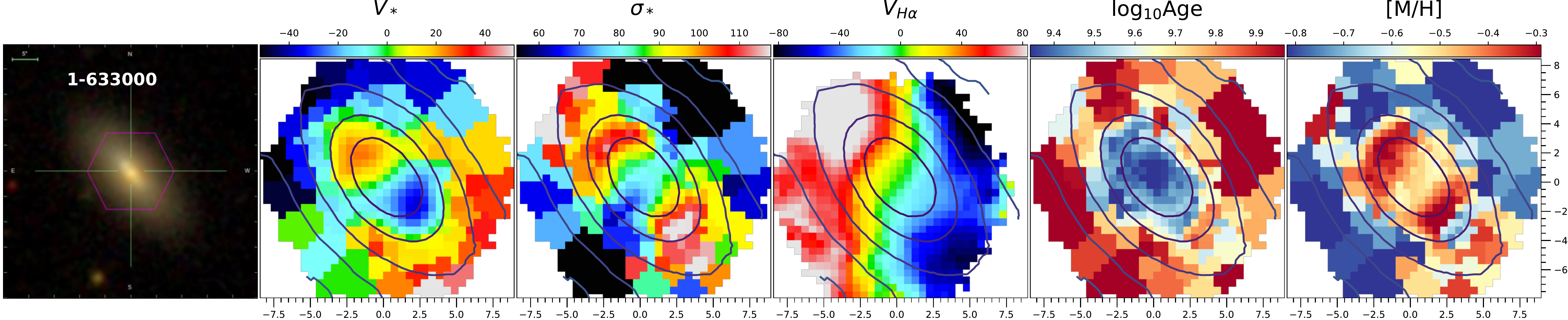}
\caption{Columns: (1) SDSS image with MaNGA ID overwritten; (2) and (3) Stellar velocity and velocity dispersion maps (in km s$^{-1}$); (4) H$\alpha$ velocity map (in km s$^{-1}$); (5) age map (log$_{10}$(Age/yr)); (6) metallicity [M/H] map (in dex). Ticks are in arcsec. The same maps for all the other CRDs are shown in the electronic appendix A.}
\label{fig:crd_example}
\end{figure*}

\subsection{Regularised fits and multiple populations}\label{sect:regul}
To investigate the presence of multiple stellar populations in the same spatial region, we performed regularised fits in every spatial bin. Regularisation is a technique to overcome the ill-posed problem of the recovery of the stellar population properties (e.g. section 3.5 of \citealt{ppxf}), and it can be considered as a method to smooth the weights of the solution of the SSP models during the fit. 

To perform regularised fits, we first perform an unregularised fit, as described in \ref{sect:popfit}, and compute the errors, $\epsilon$. Afterwards, we renormalise the errors to have a $\chi^2$/DOF = 1, which is done by setting:
\begin{equation}
\epsilon_{norm} = \epsilon \times \sqrt{\chi^2/\mbox{DOF}} \; \mbox{,}
\end{equation}
and use $\epsilon_{norm}$ as the new noise; finally, we perform the regularised fit. We kept a low regularisation, and set the \texttt{pPXF} parameter \texttt{regul=5}. Three examples of regularised fits, relative to two different galaxies, are shown in Figure \ref{fig:regul_ex}. The weights fraction maps, shown below the fits, represent the weights, entering the same equations \eqref{eq:age} and \eqref{eq:met}, of the regularised fits of the SSPs. Based on our visual assessment of the number of well isolated peaks (yellow `blobs') in these maps, we distinguish the presence of `unimodality' and `multimodality' in the stellar population, if there are one or multiple peaks, respectively. In Figure \ref{fig:regul_ex}, the first weights map is an example of unimodality, while the second and third weights maps are examples of multimodality.

Since spatial bins with low S/N provide untrustworthy weights map, we say that a galaxy exhibits unimodality or multimodality by considering only those weights maps whose spatial bins have S/N $\geq 25$. Additionally, we require the weights maps of a galaxy to be similar for all the spatial bins considered; this means that, for instance, to classify as unimodal the galaxy in the upper panels of Figure \ref{fig:regul_ex}, we require that all the spatial bins considered exhibit a blob in the upper-right corner of the weights map, and not a single blob that varies in age and metallicity at different bins. Many galaxies exhibiting multimodality also exhibit different weights maps at different (often neighbouring) spatial bins, like, for instance, the second and third maps of Figure \ref{fig:regul_ex}. We attribute this behaviour to ongoing star formation. For this reason, we further distinguish multimodal galaxies between those who exhibit the same multimodality in all the considered spatial bins, and those who instead vary significantly at different bins, and label the latter as `star-forming'.

\section{Results}\label{sect:results}

In the following, we discuss the results of our study. Table~\ref{tab:crd} lists the main properties of our sample of galaxies hosting CRDs, including those derived here. The morphology and values of M$_\ast$ are taken from \cite{G19benchmark}. R$_e$ is taken from the DRP. The $\Delta$PA values are calculated as described in section \ref{sect:stargaspa}. The global ages, metallicities and relative gradients are calculated as described in section \ref{sect:fitpop}. The modality column refers to the distinction in the weights fraction maps of regularised fits, described in section \ref{sect:regul}. In Figure \ref{fig:crd_example} we show the SDSS image, along with the extracted maps of an illustrative galaxy. The same maps for all the other CRDs are shown in the electronic appendix A to this paper.

\begin{figure*}
\includegraphics[width = .16\textwidth]{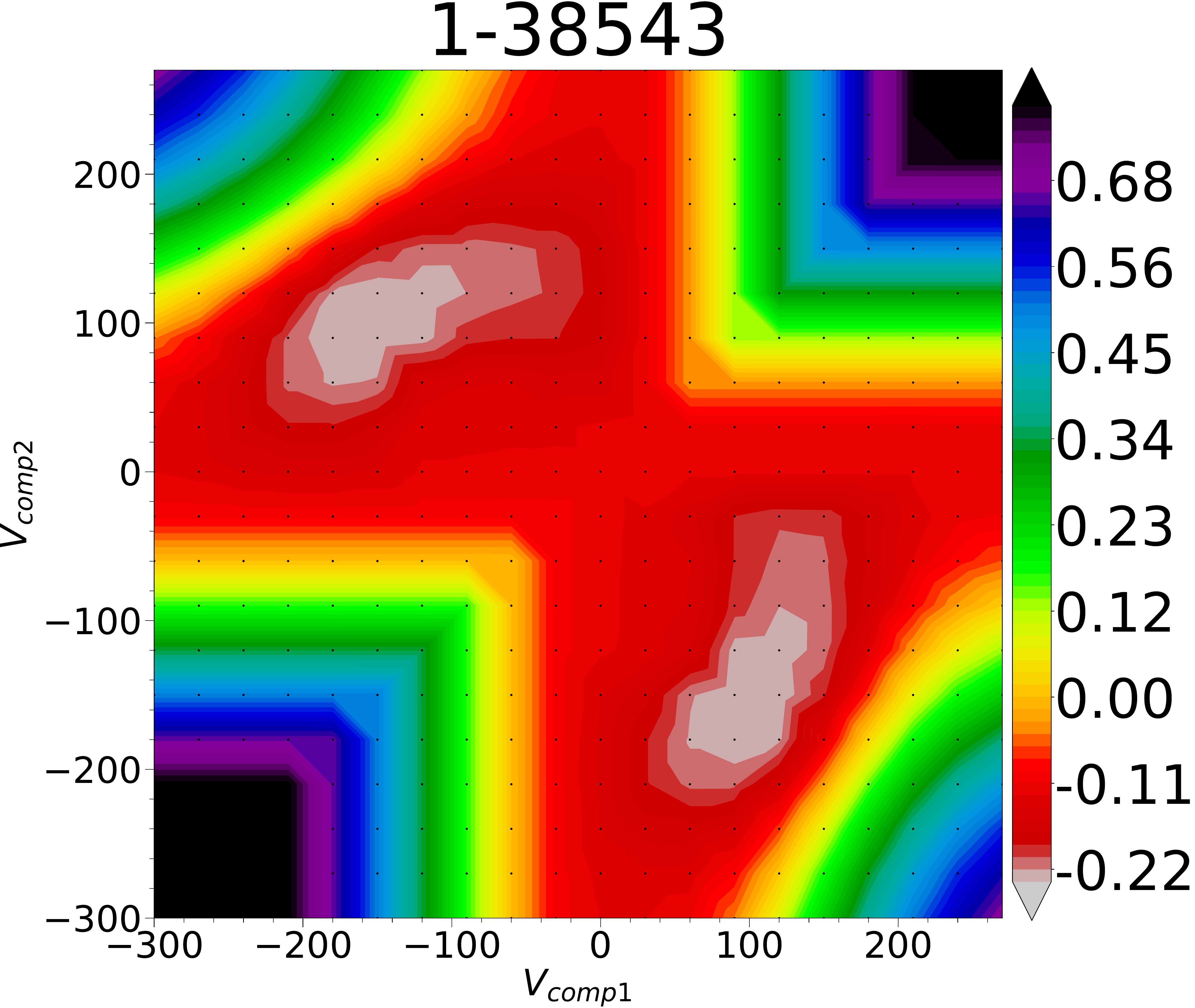}
\includegraphics[width = .16\textwidth]{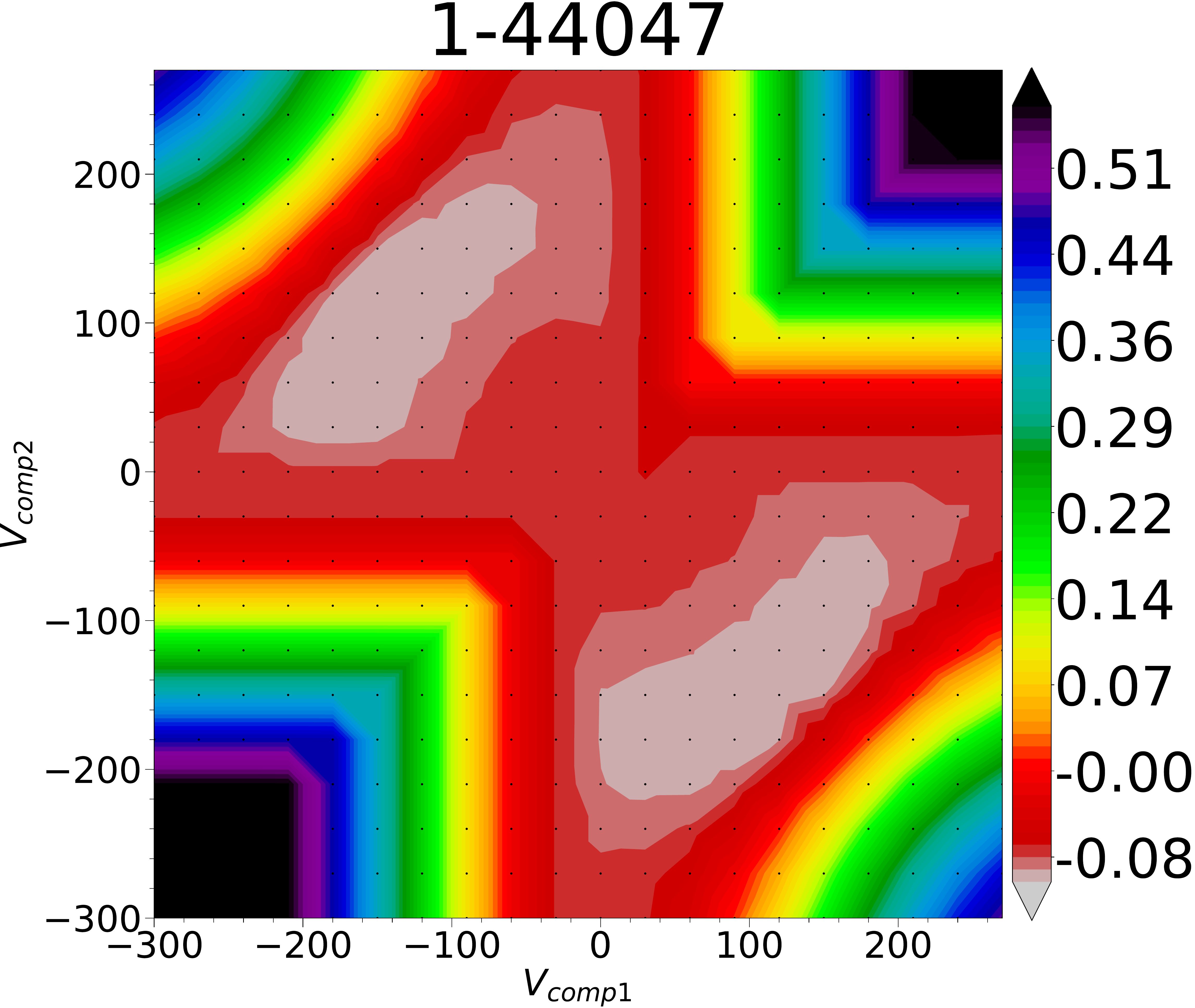}
\includegraphics[width = .16\textwidth]{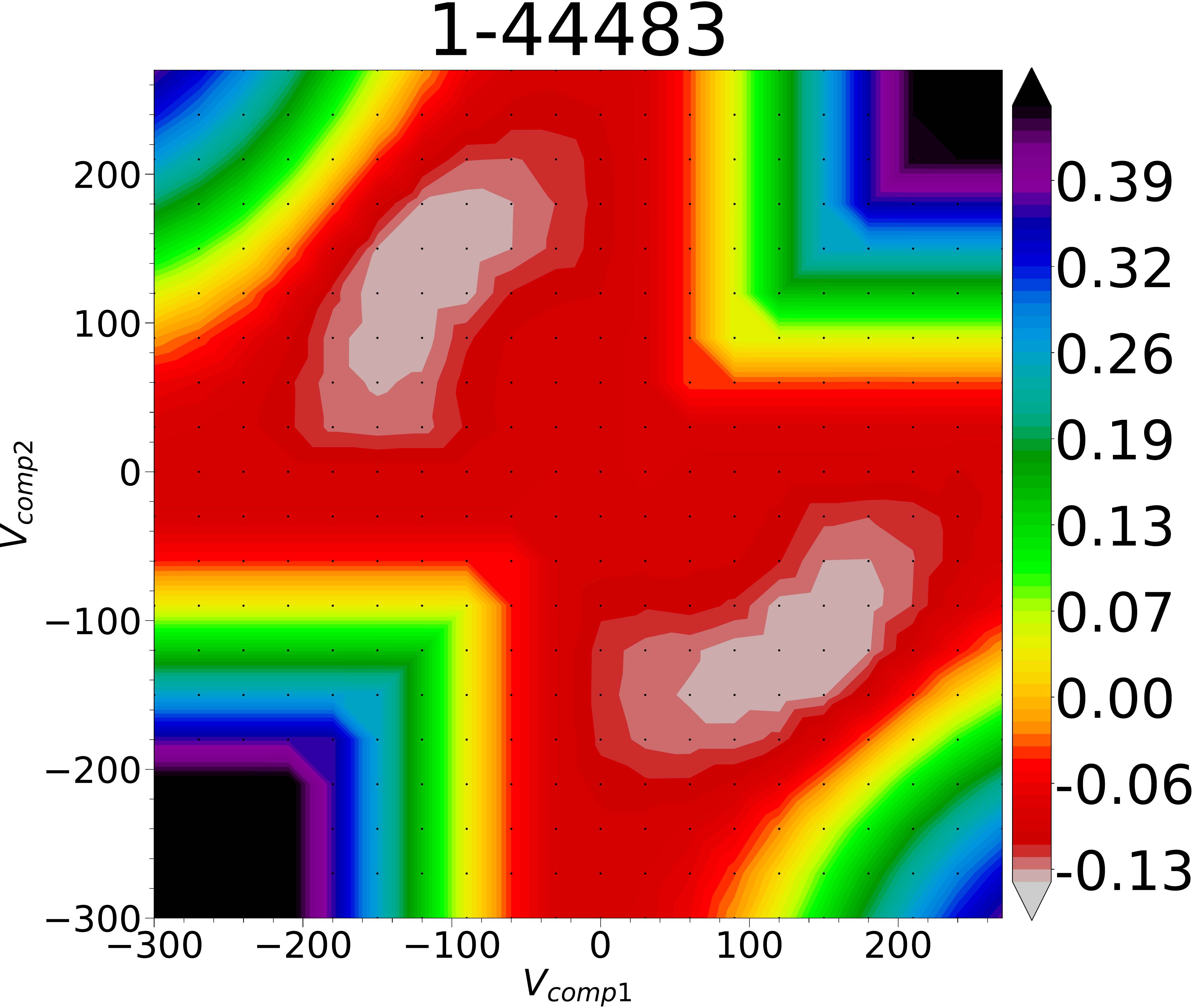}
\includegraphics[width = .16\textwidth]{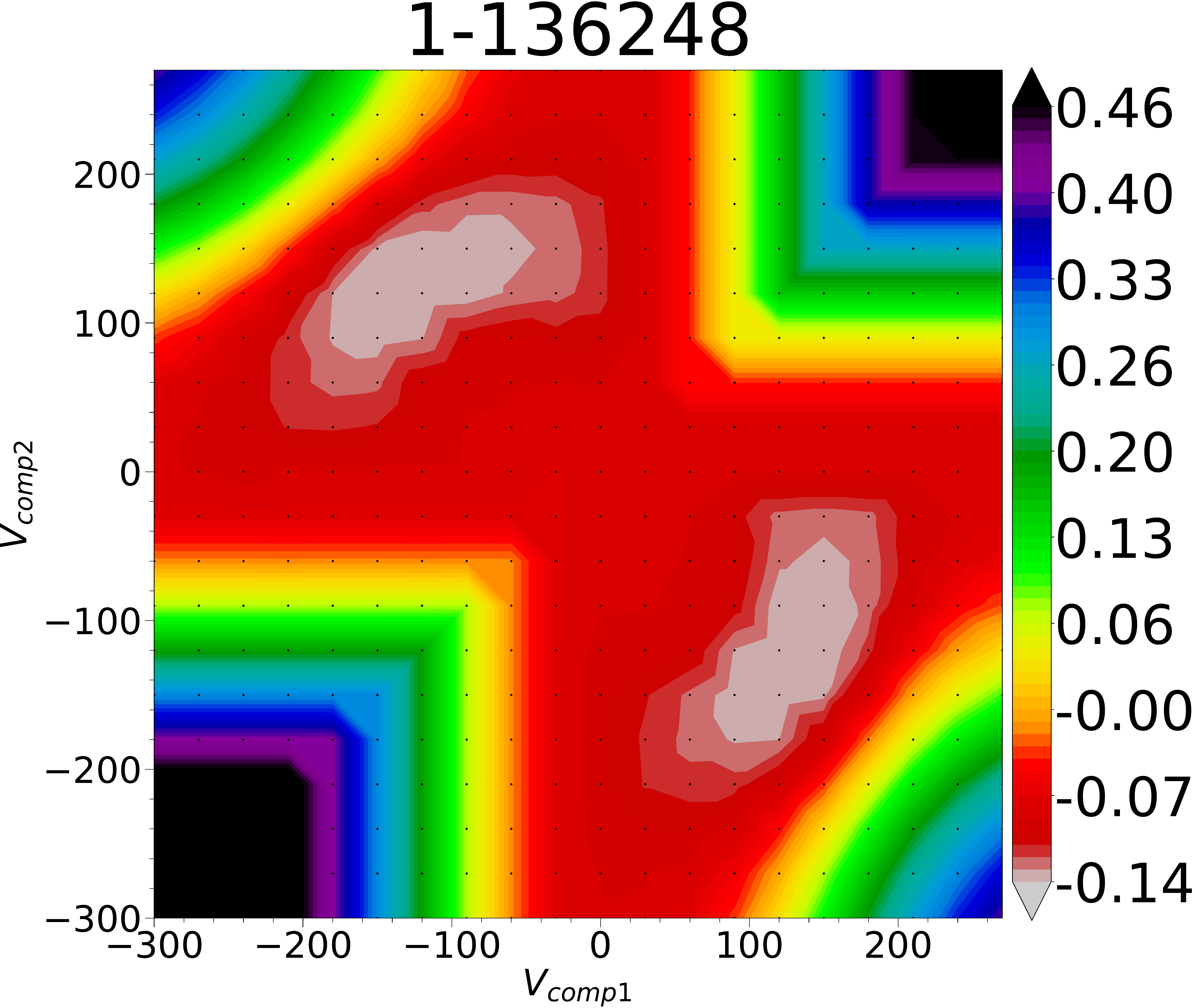}
\includegraphics[width = .16\textwidth]{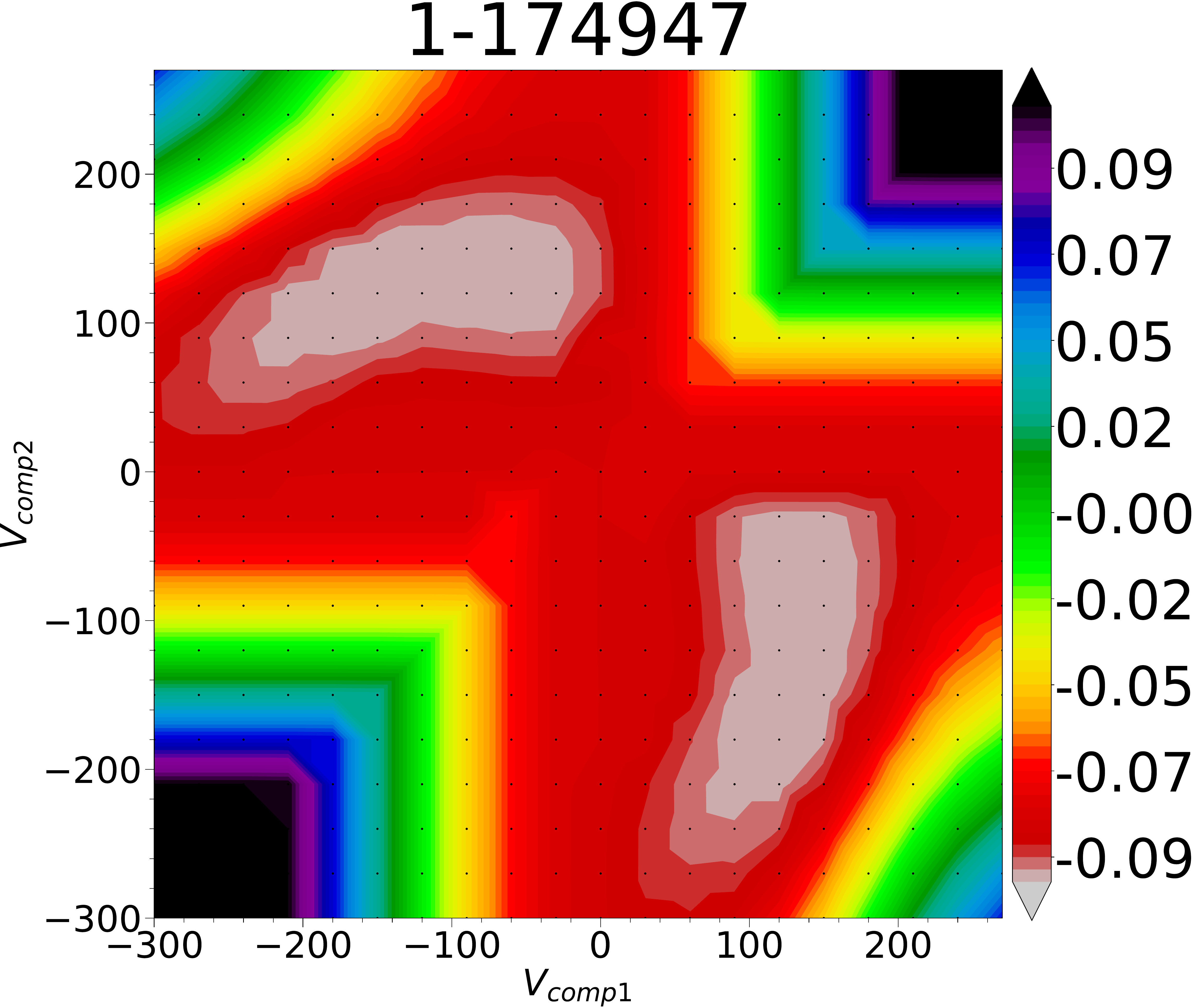}
\includegraphics[width = .16\textwidth]{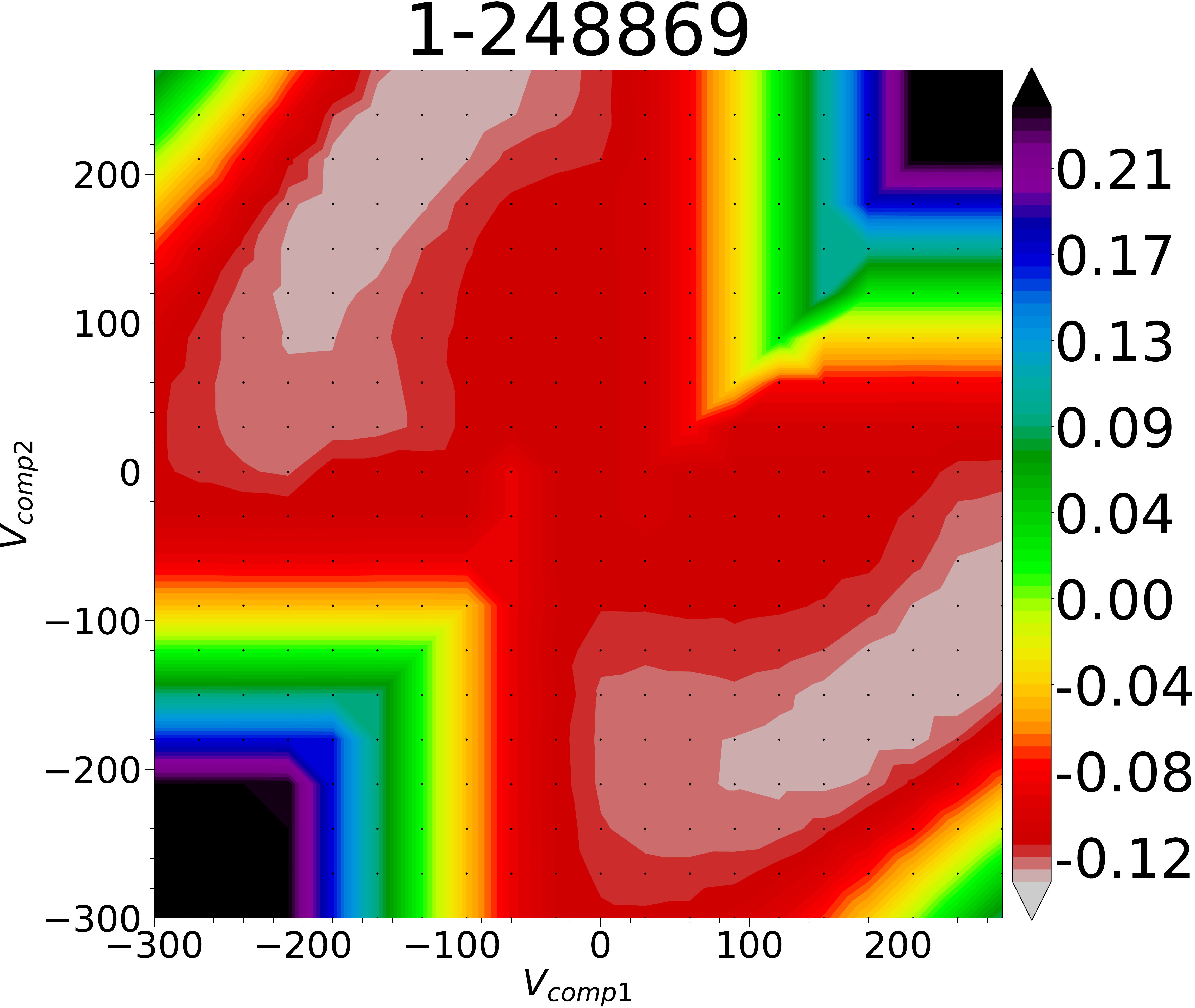}\\
\includegraphics[width = .16\textwidth]{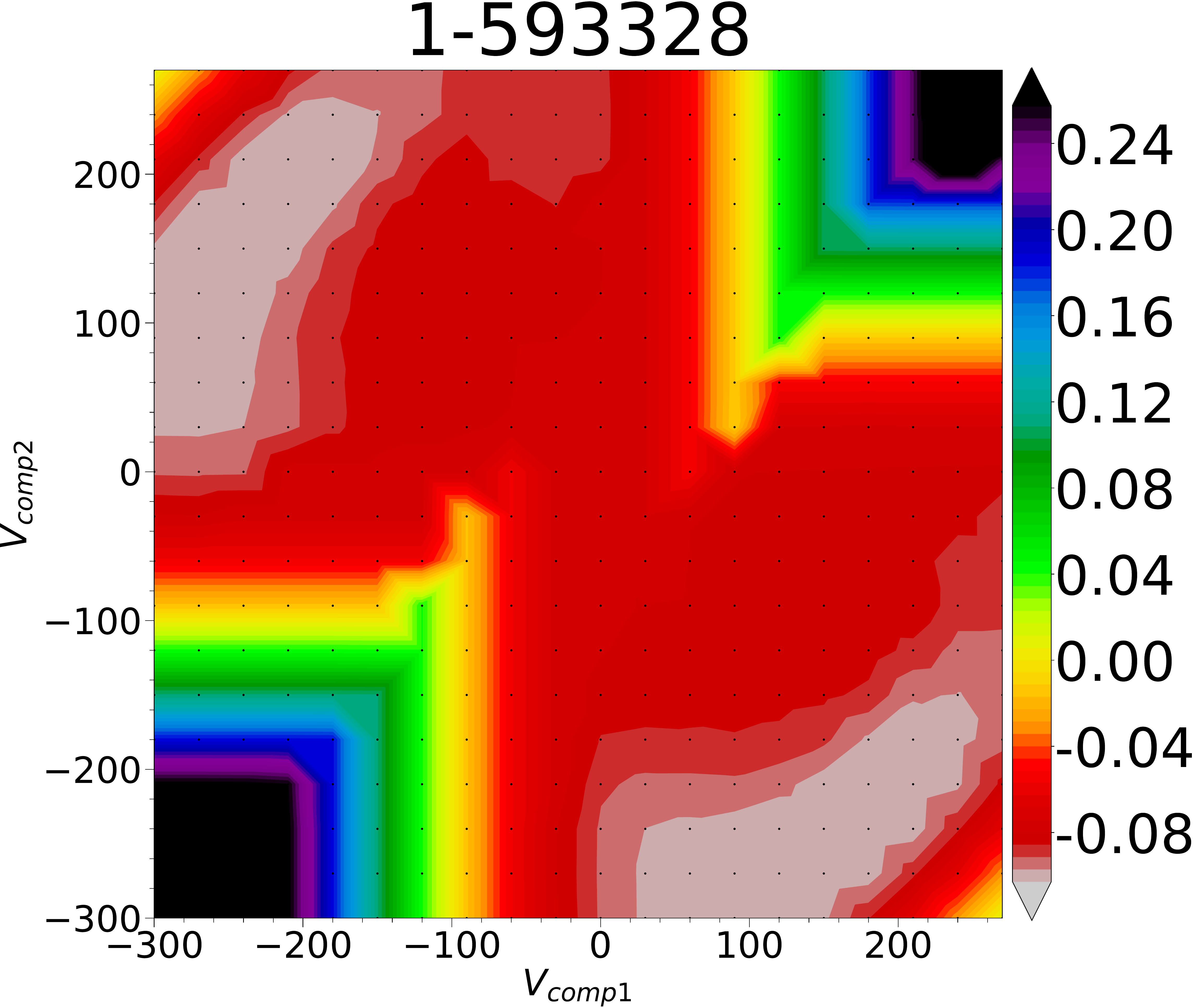}
\includegraphics[width = .16\textwidth]{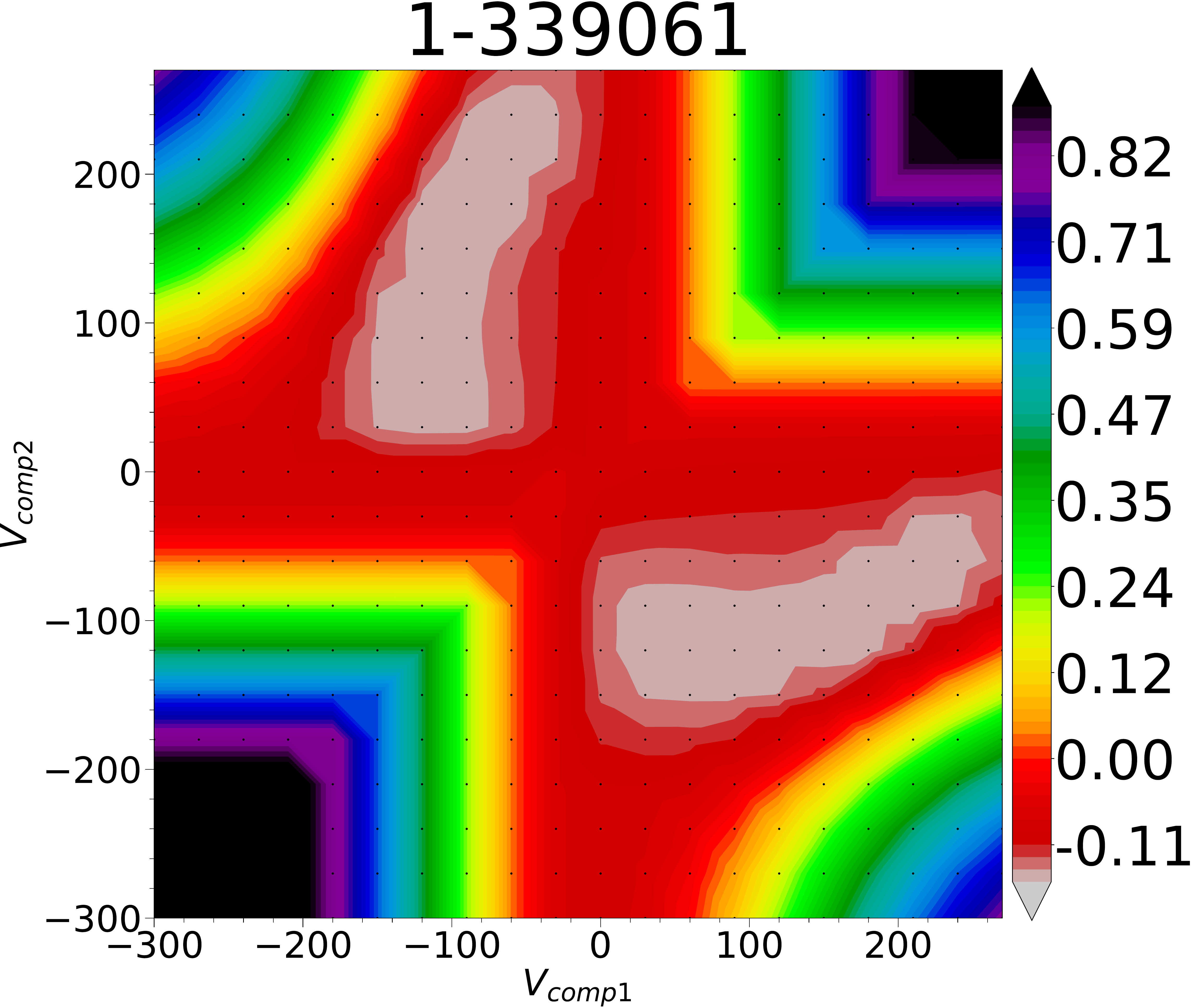}
\includegraphics[width = .16\textwidth]{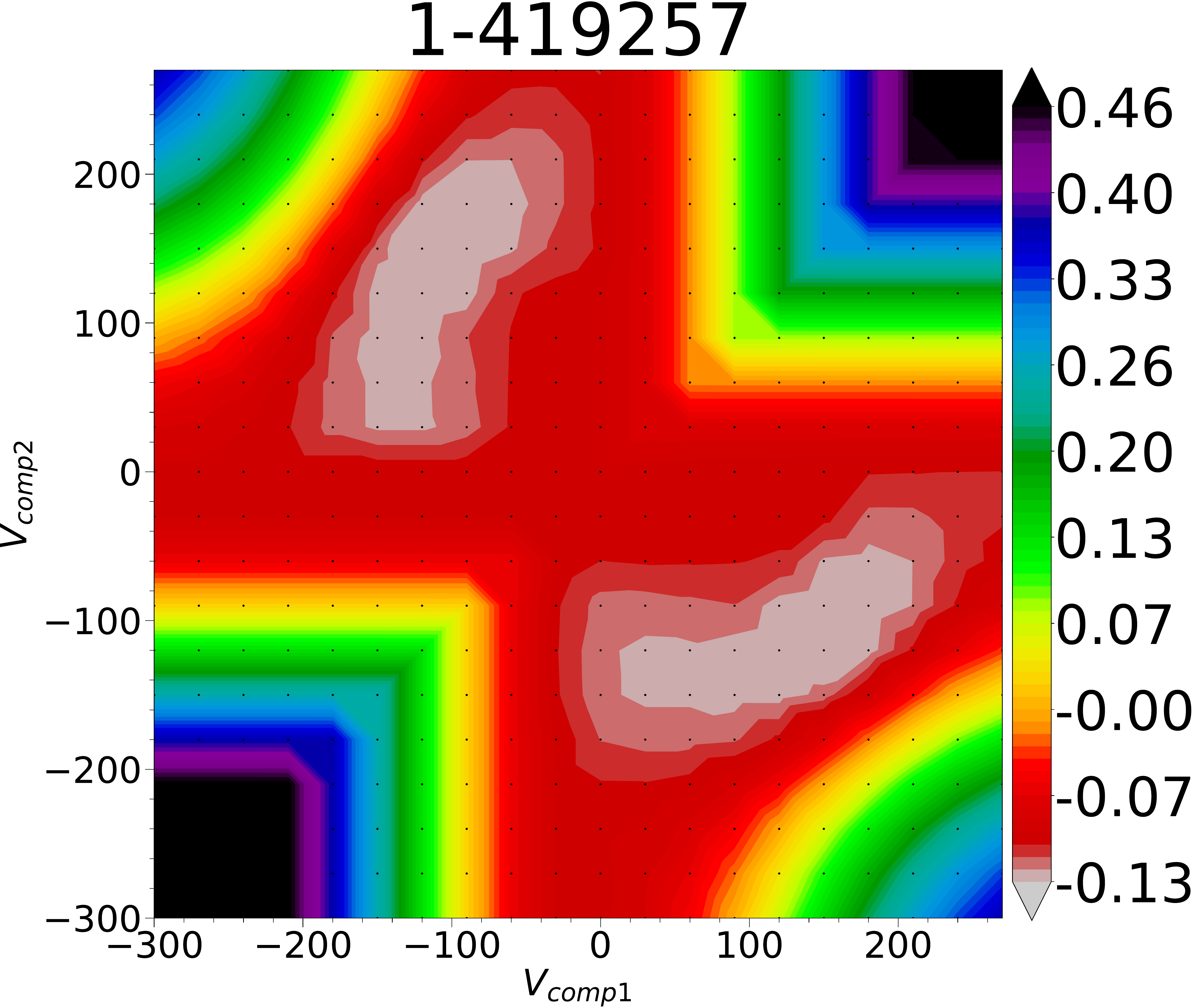}
\includegraphics[width = .16\textwidth]{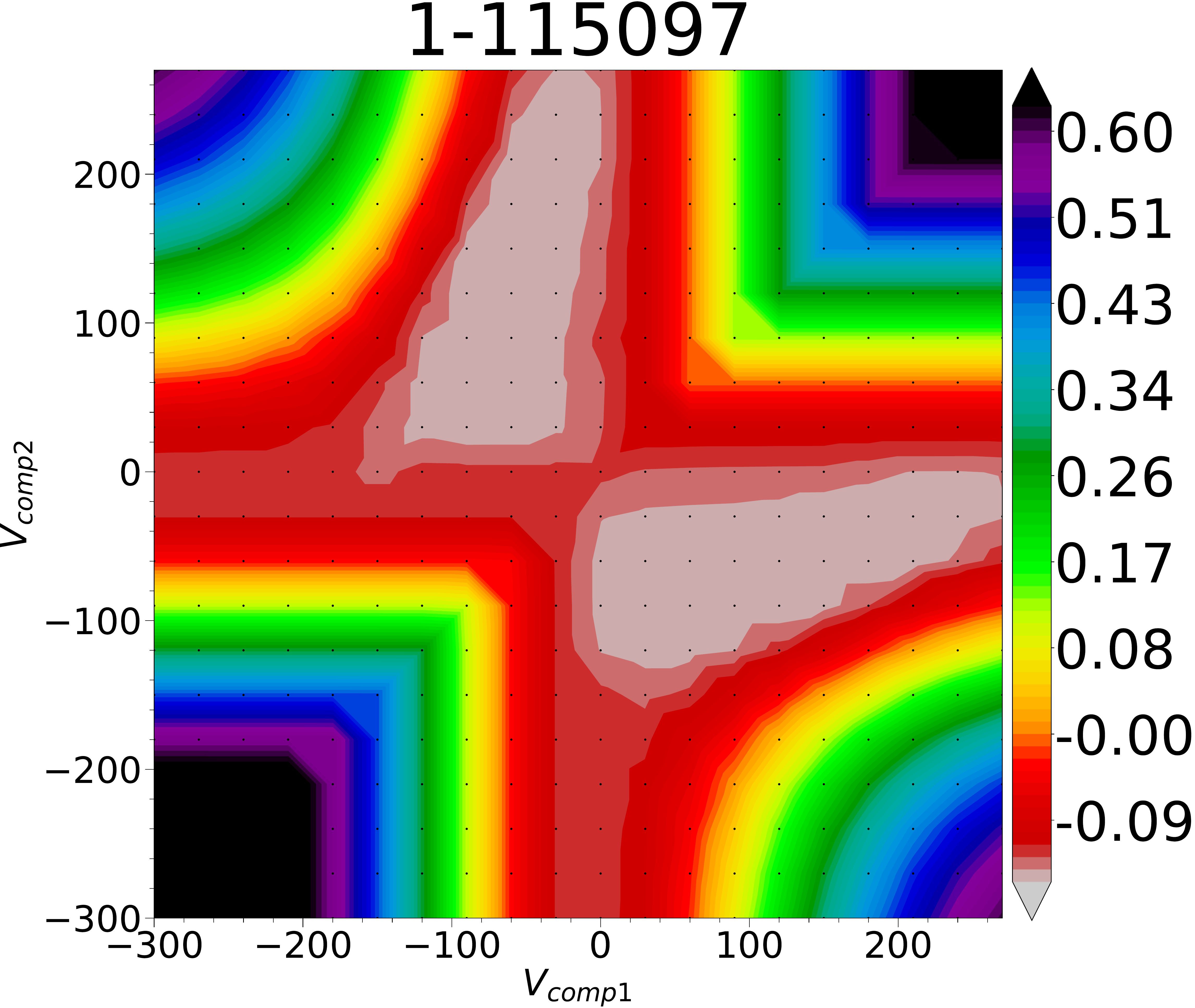}
\includegraphics[width = .16\textwidth]{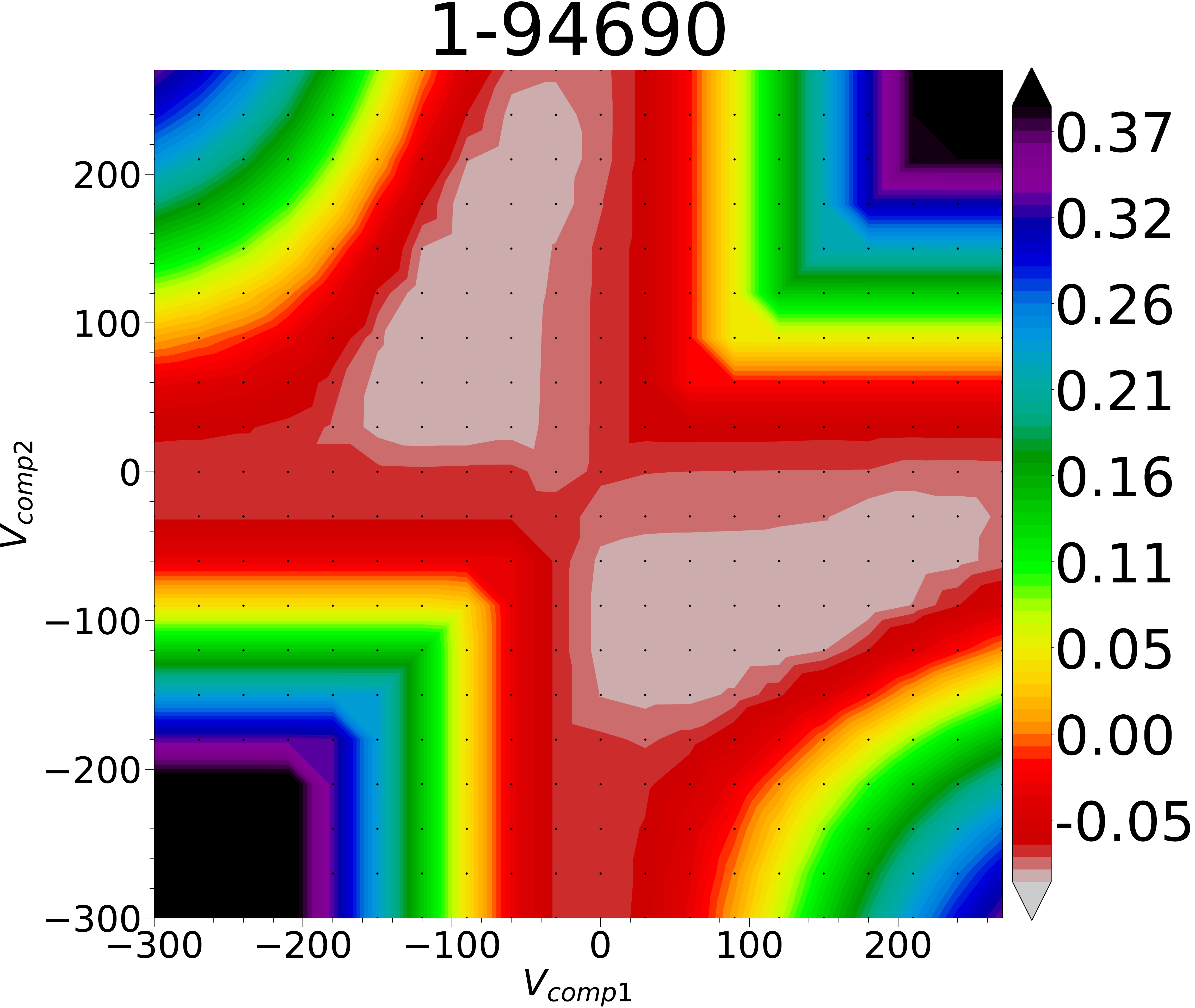}
\includegraphics[width = .16\textwidth]{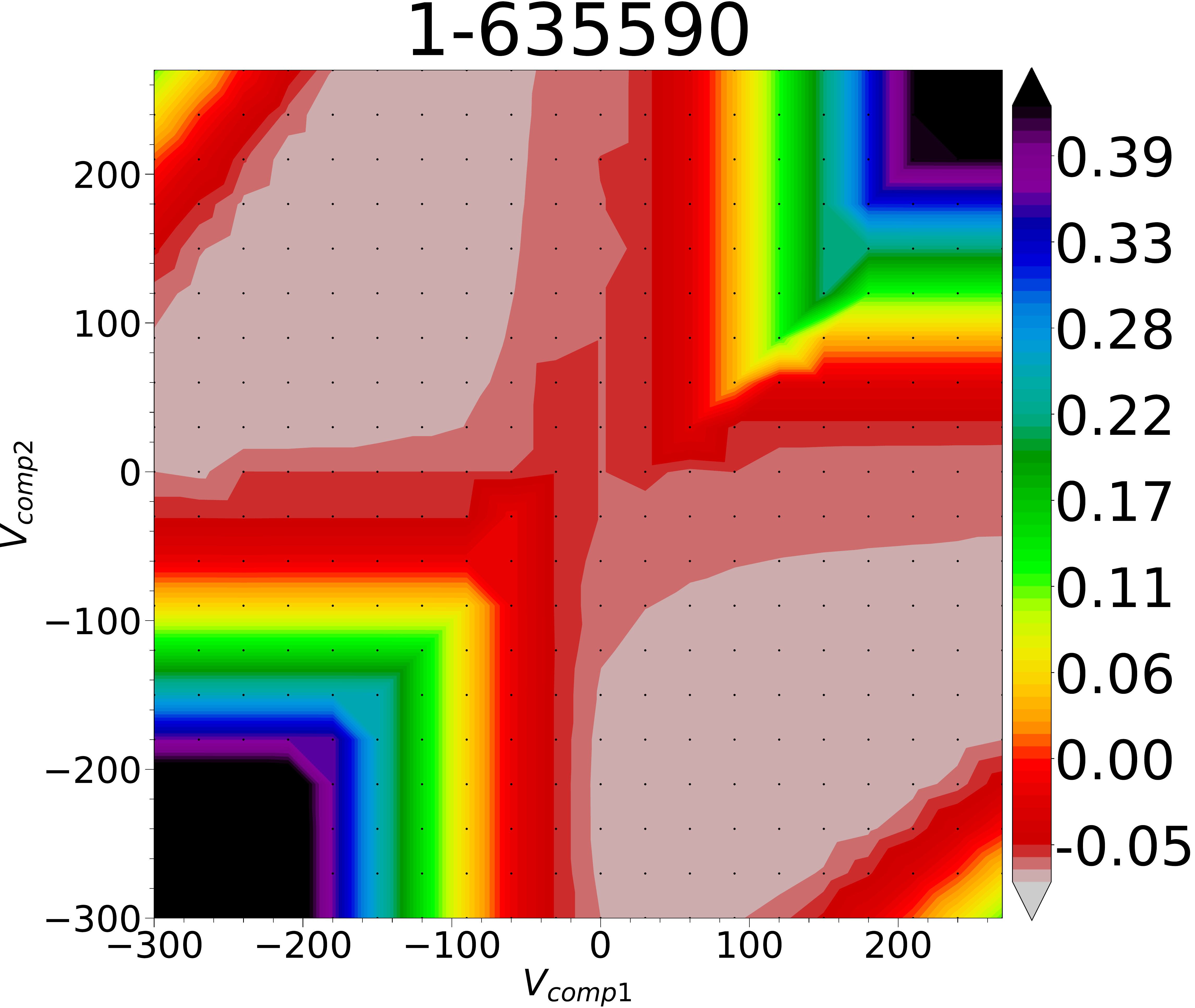}\\
\includegraphics[width = .16\textwidth]{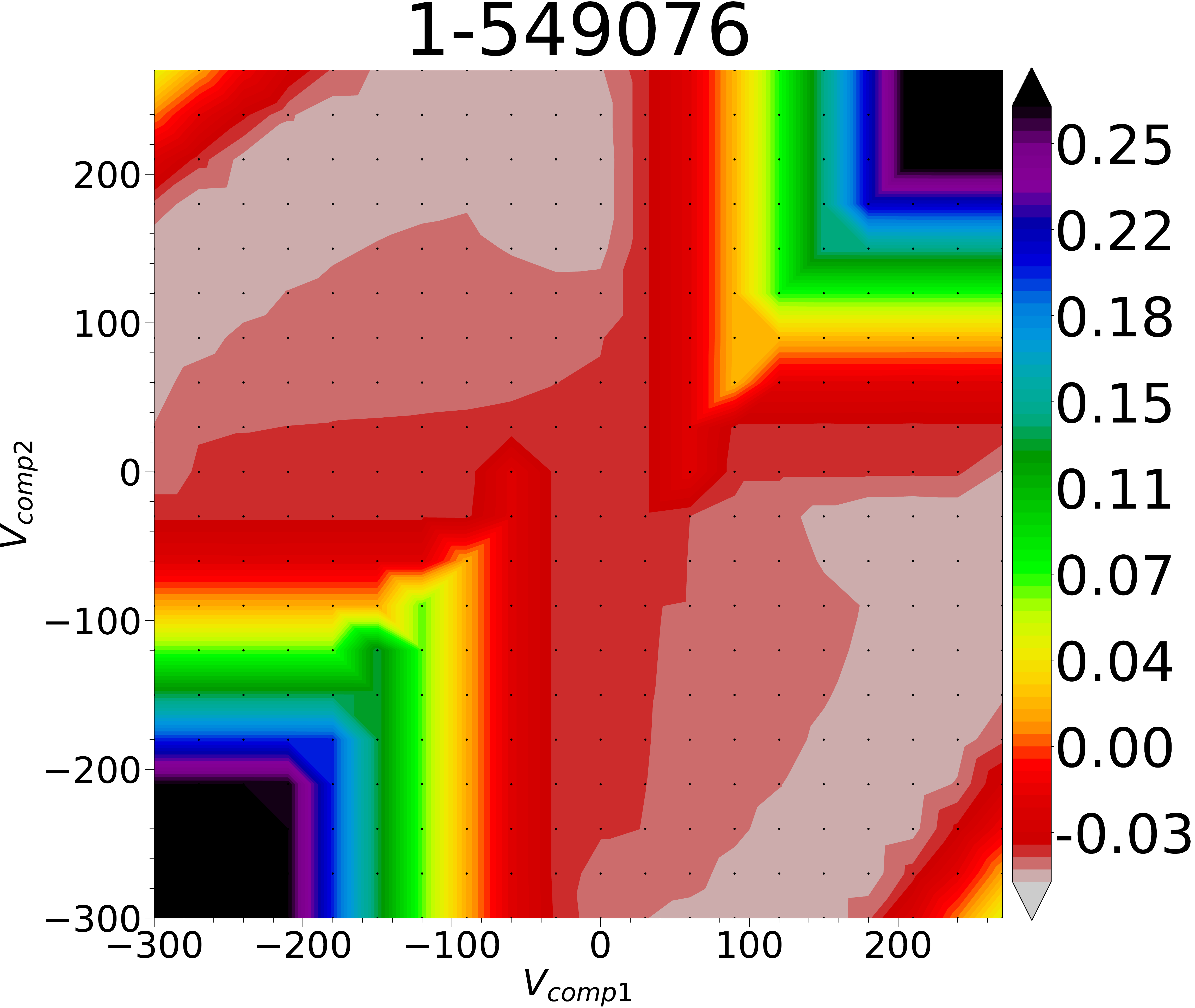}
\includegraphics[width = .16\textwidth]{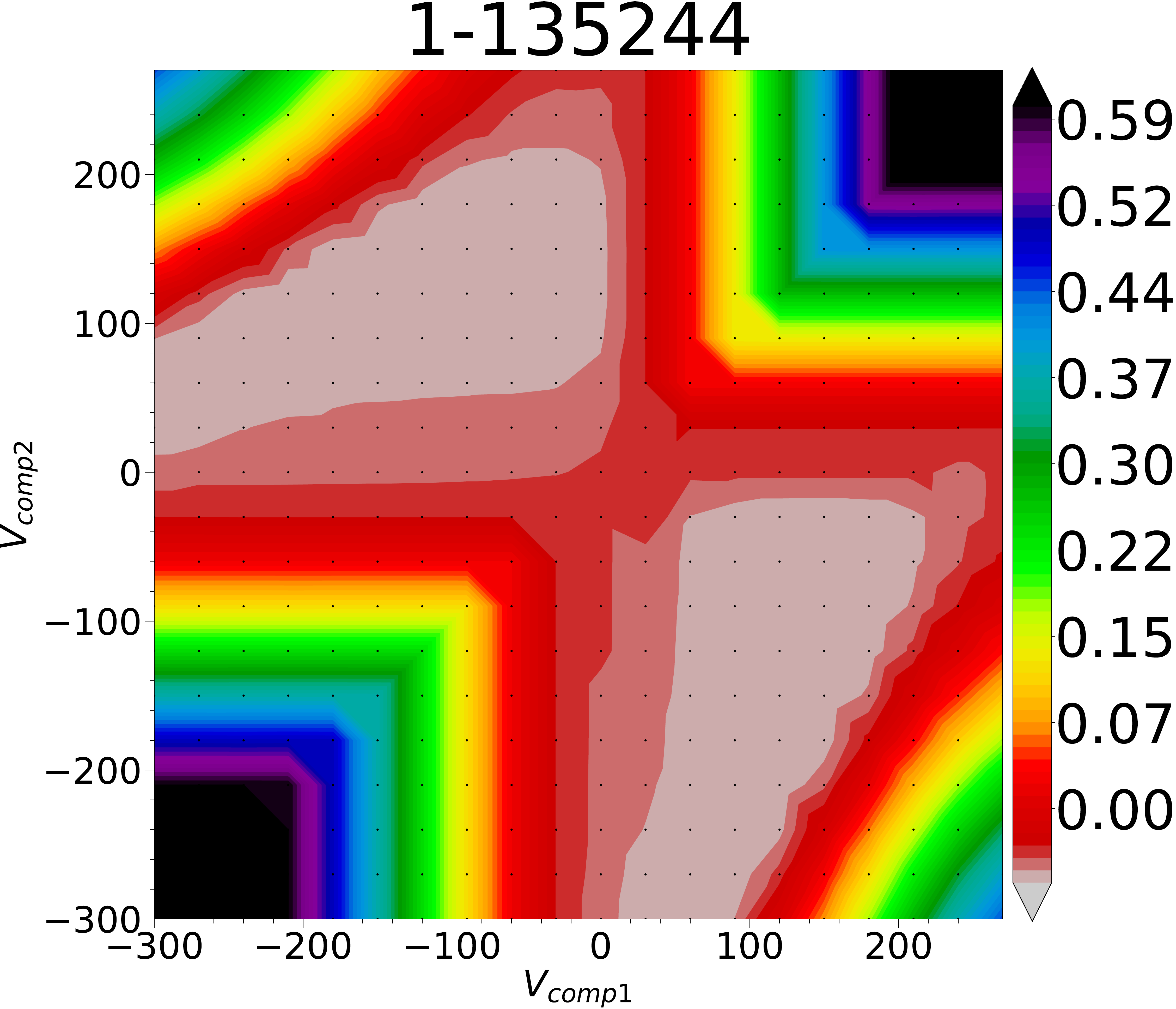}
\includegraphics[width = .16\textwidth]{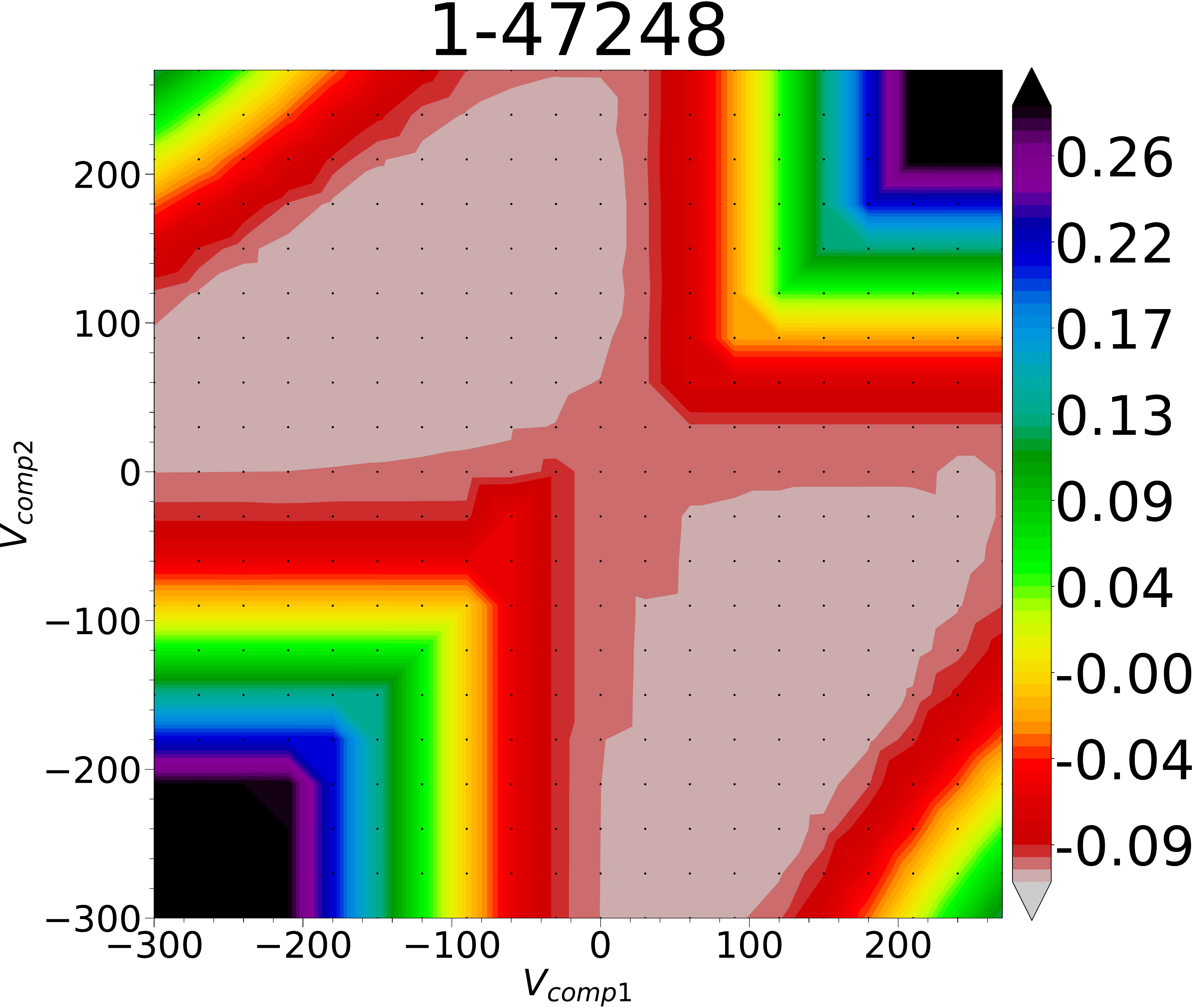}
\includegraphics[width = .16\textwidth]{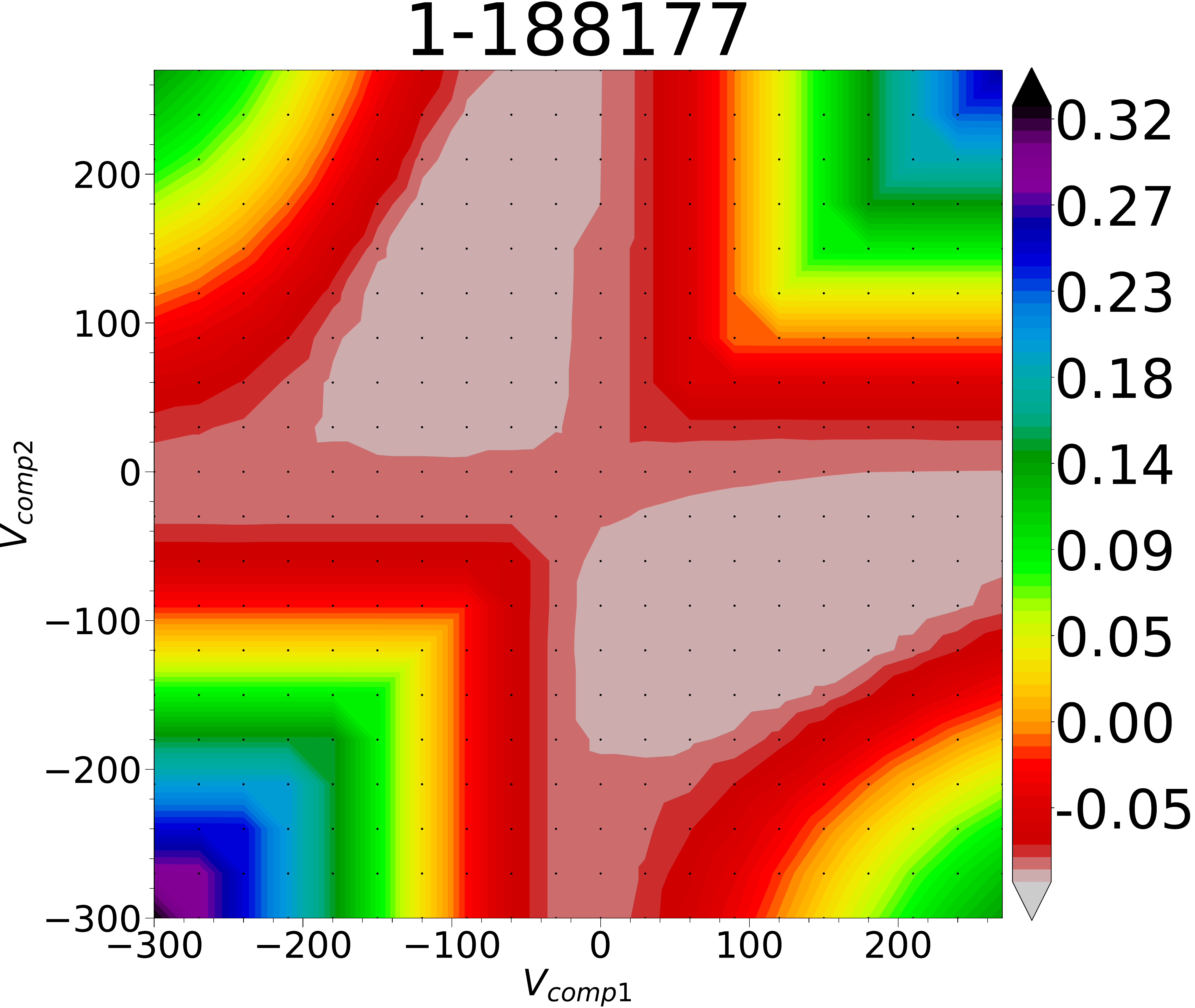}
\includegraphics[width = .16\textwidth]{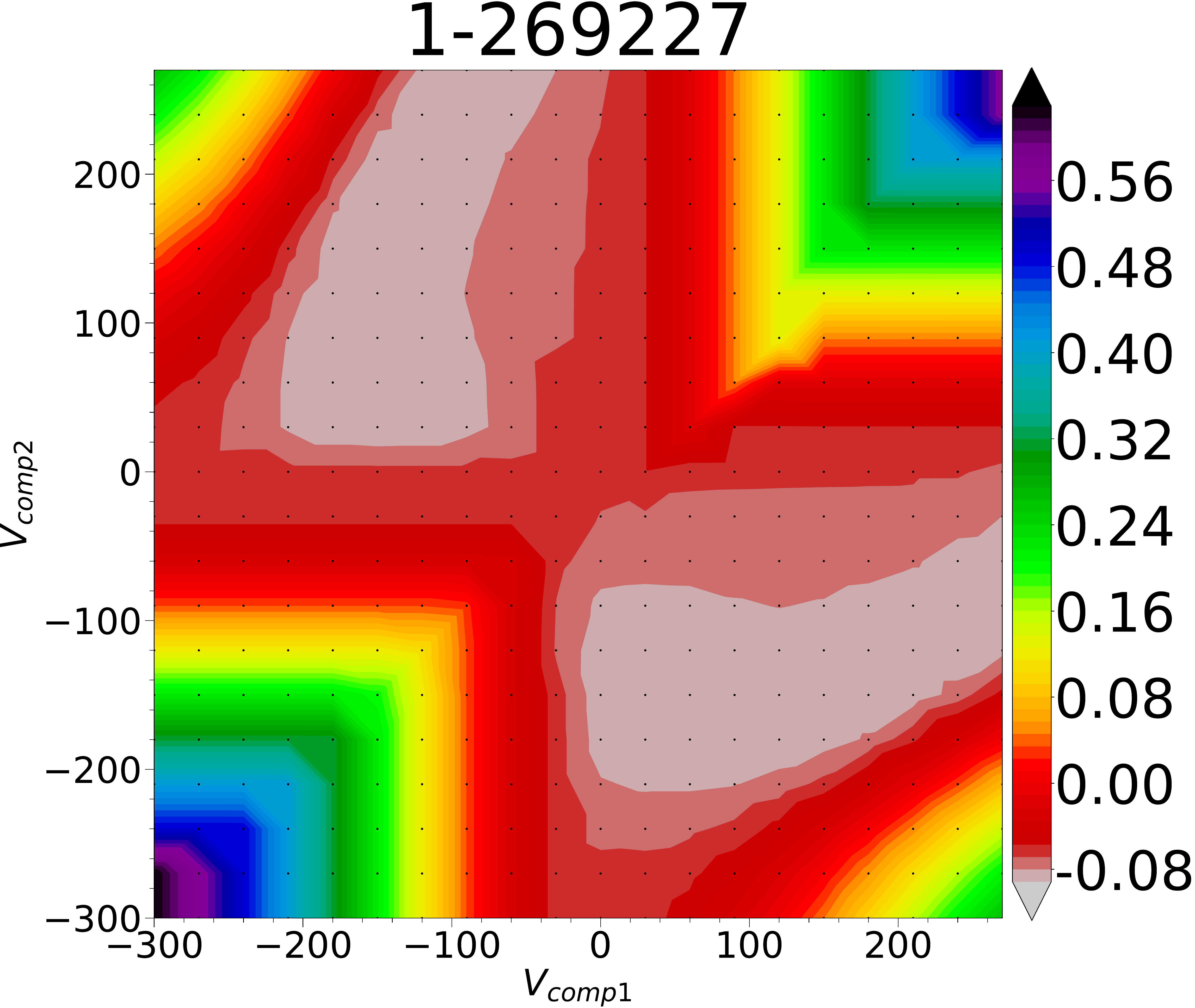}

\caption{$\chi^2$ maps of CRDs exhibiting two distinct minima, namely those where the two kinematic components are spectroscopically distinguishable. These maps were obtained by fitting a single representative spectrum per galaxy, chosen at one of the two $\sigma_\ast$ peaks, where the separation between the components is the largest (see the electronic appendix B for details). MaNGA-IDs are plotted above each panels. Colorbars values are the log$_{10}$($\chi^2$/DOF). In each panel, the x and y axes are the fitted $V_{\mbox{\scriptsize{comp1}}}$ and $V_{\mbox{\scriptsize{comp2}}}$, ranging within $\pm 300$ km s$^{-1}$ at $V_{\mbox{\scriptsize{step}}} = 30$ km s$^{-1}$. }
\label{fig:bestchi2}

\end{figure*}

\subsection{Statistics of CRDs}\label{sect:statistics}
In section \ref{sect:selection} we described how we identified 64 counter-rotators from the MaNGA sample, by requiring the appearance, in the kinematic maps, of one or both of two characteristic features: the counter-rotation and the two peaks in $\sigma_\ast$. As mentioned in the Introduction, our capability of identifying CRDs is limited by both instrumental and intrinsic issues. To estimate the limits of detectability, we built dynamical models of a representative CRD in MaNGA and checked whether the two disks can be distinguished, by taking into account the instrumental properties of the survey, and by varying the main structural properties of the two disks affecting the kinematic maps, namely their mass ratio and their scale-length radius ratio, and the galaxy inclination. Then, we examined the maps resulting from models and selected CRD based on the criteria used in section \ref{sect:selection}. The method and limitations of our modeling approach are presented and discussed in section \ref{sect:detectability}. From this analysis, we estimate a CRD detectability of 81\% in the MaNGA survey.

In our sample of 64 CRDs, 61 are ETGs, of which 38 are ellipticals and 23 lenticulars. The sample of the DR16 we considered includes 3129 ETGs, of which 1343 are ellipticals and 1786 are lenticulars. Taking into account our detectability estimate, this means that CRDs constitute $(2.3\pm0.3)\%$ of ETGs (and $<3\%$ at $95\%$ confidence\footnotemark ); in particular, $3\%$ of ellipticals ($< 5\%$ at $95\%$ confidence\footnotemark[11] ) and $1\%$ of lenticulars ($<3\%$ at $95\%$ confidence\footnotemark[11]) host a counter-rotating stellar disk (though, since ellipticals could be misclassified lenticulars such estimates should be considered carefully). The fraction of S0s is consistent with the early estimate of an upper limit of $10\%$ at $95\%$ confidence from long-slit spectroscopy \citep{Kujiken_1996}. It is also consistent with the estimate of $(1.5\pm0.8)$\% (1$\sigma$ confidence) from integral-field data by \citet{Krajnovic_2011}. Further, among the $\sim 4000$ galaxies considered, we examined the kinematic maps of 787 spirals, and found only 1 CRD, resulting in a fraction of $<1\%$ at $95\%$ confidence\footnotemark[11], consistent with the $<8\%$ estimated by \cite{Pizzella_2004}.

\footnotetext{Applying Poisson distribution statistics.}

\subsection{Spectroscopic evidence of counter-rotation}\label{sect:results_kinem}
To have a spectroscopic confirmation that these galaxies are made of two kinematic stellar components, we have performed two-components fits, and presented the method of $\chi^2$ maps. We found that 17 galaxies exhibit distinct minima in the $\chi^2$ maps, meaning that the two counter-rotating components are spectroscopically distinguishable. We show the $\chi^2$ maps of these galaxies in Figure~\ref{fig:bestchi2}.

Generally speaking, the recovery of the two components with MaNGA data has been possible for galaxies with high S/N, and with clearer CRDs' kinematic features, while for those galaxies with lower S/N and less clear kinematic maps the recovery of the two kinematic components is ineffective. However, we point out that four of the spectroscopically separable galaxies do not exhibit a clear counter-rotation (but they have very clear $\sigma_\ast$ peaks); conversely, many galaxies with evident counter-rotation or the two $\sigma_\ast$ peaks have not been found to be spectroscopically distinguishable. This is because the recovery of the two kinematic components also depends on the intrinsic stellar kinematics of the two disks, and on their contribution to the total flux.

The spectroscopical confirmation of the counter-rotating components for a larger number of CRDs requires data with higher S/N and better spatial resolution. \\

\subsection{Stellar and gaseous disks}\label{sect:results_stargas}

\begin{figure}
\centering
\includegraphics[width=\columnwidth]{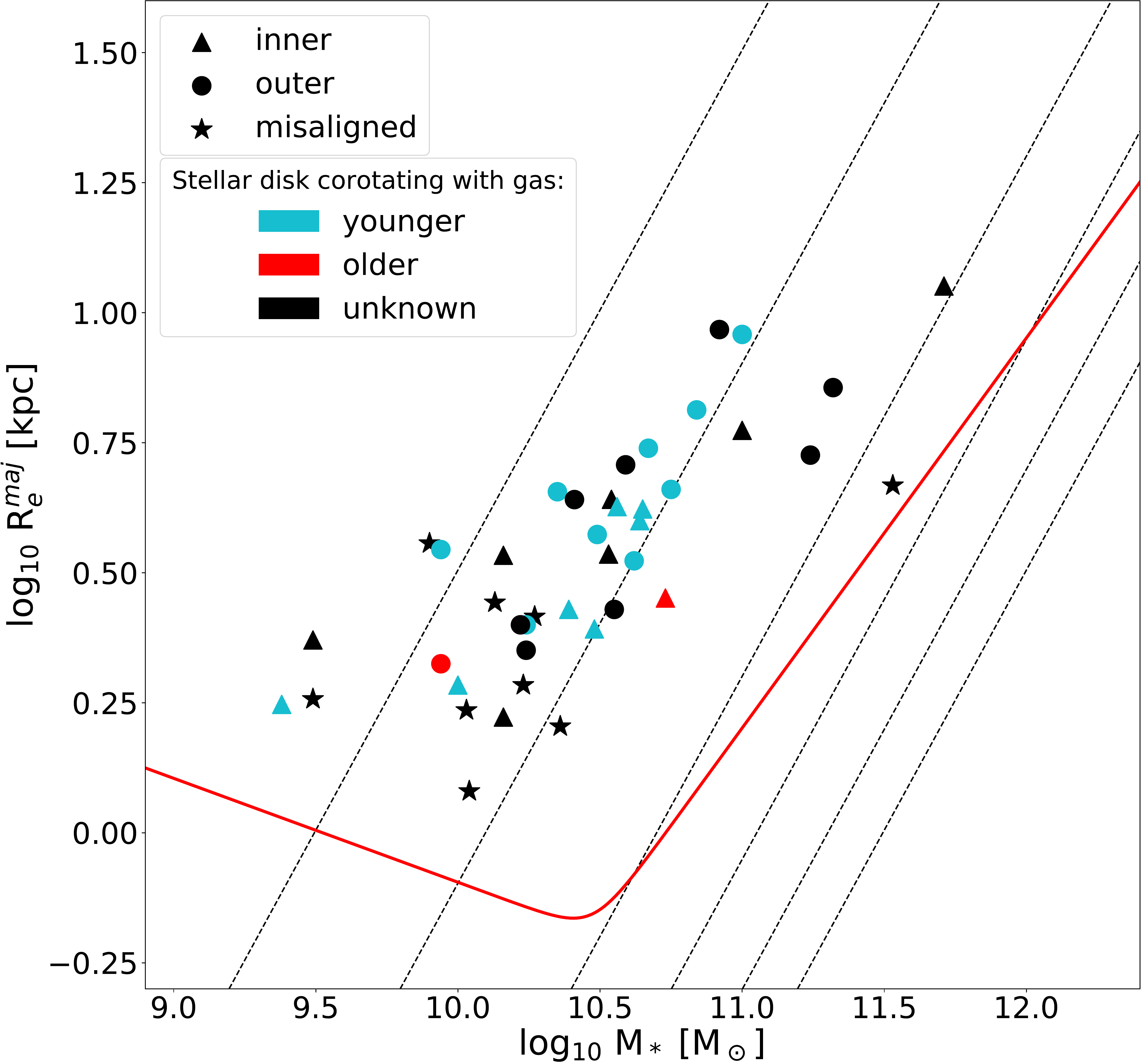}
\caption{Alignment of the ionised gas with respect to the stellar rotation on the mass-size plane of CRDs. Here, M$_\ast$ is the stellar mass, and R$_e^{maj} = 1.61 \times \mbox{R}_e$ is the effective radius of the major axis, defined in \protect\cite{atlas_xv}. The dashed black lines are lines of constant velocity dispersion at (left to right) 50, 100, 200, 300, 400, 500 km s$^{-1}$. The red straight line is the zone of exclusion, defined in \protect\cite{atlas_xv}. The labels `inner' and `outer' refer, respectively, to the corotation of the gas with the stellar disk prevailing the inner or outer regions of the velocity field. Misaligned galaxies are those with $30^\circ \leq$ $\Delta$ PA $\leq 150^\circ$. Red and cyan symbols are for galaxies that have the gas velocity field corotating with either the younger or the older stellar disk, respectively, while unknown cases are black.}
\label{fig:MS_stargas}
\end{figure}

In Figure \ref{fig:MS_stargas} we show the results on the comparison between the velocity fields of stars and gas by plotting the 42 CRDs for which velocity maps are available for both (section \ref{sect:stargaspa}) on the mass-size diagram, introduced in \cite{atlas_xx}. We found that in most of the cases (33 on 42), the gas velocity field is aligned with one of the stellar disks, and, in particular, in 15 cases the gas is corotating with the inner disk, and in 18 cases with the outer one. Therefore, there is no preferable corotation with the inner or the outer disk.
 
Of these 33 galaxies with stars and gas aligned, 18 also have age maps that distinctly change from the inner to the outer stellar disk. In 16 cases the gas corotates with the younger disk, while in 2 cases it corotates with the older disk. The comparison between stellar and gas velocity fields with age maps suggests that the distinction into inner and outer disk is just a matter of spatial distribution, and it is not helpful in determining which disk is the primary (pre-existing) and which the secondary (acquired). 
 
For 9 CRDs, namely 21\% of the 42 CRDs with velocity maps for stars and gas, the gas rotation is misaligned with respect to both stellar disks. From the SDSS images, four of these galaxies have neighbouring objects or surrounding blobs, which could be hints of recent gas accretion, and two also have evidences of ongoing star formation; on the other hand, the remaining galaxies appear isolated and undisturbed. In Figure \ref{fig:MS_stargas}, we can see that misaligned galaxies typically have masses less than the break mass M$_{\mbox{\scriptsize{b}}} = 3 \times 10^{10}$~M$_\odot$ (see Figure 23 of C16), with the exception of one galaxy. In the latter case, the misalignment could be related to the central black hole activity \citep{Starkenburg_19, Duckworth_20}; in fact, the H$\alpha$ velocity field has a large stellar velocity dispersion in the center, suggesting the presence of an AGN \citep{Law21}.\\

\subsection{Global stellar population properties}
Except for three galaxies, our sample of CRDs is made of ETGs; to compare our results with the literature, we considered the results of \cite{Li_18} (hereafter, Li+18) on $\sim$900 ETGs from the MaNGA sample. The definition of the global properties, as well as the procedure to calculate gradients, are the same. The only significant difference with Li+18 is that they fitted stellar population properties using a reddening curve instead of multiplicative polynomials. The results by Li+18 are a useful benchmark as they were independently tested in detail by \citet[her figure~5]{Liu2020} who found an excellent agreement.

\begin{figure}
\centering
\includegraphics[width=\columnwidth]{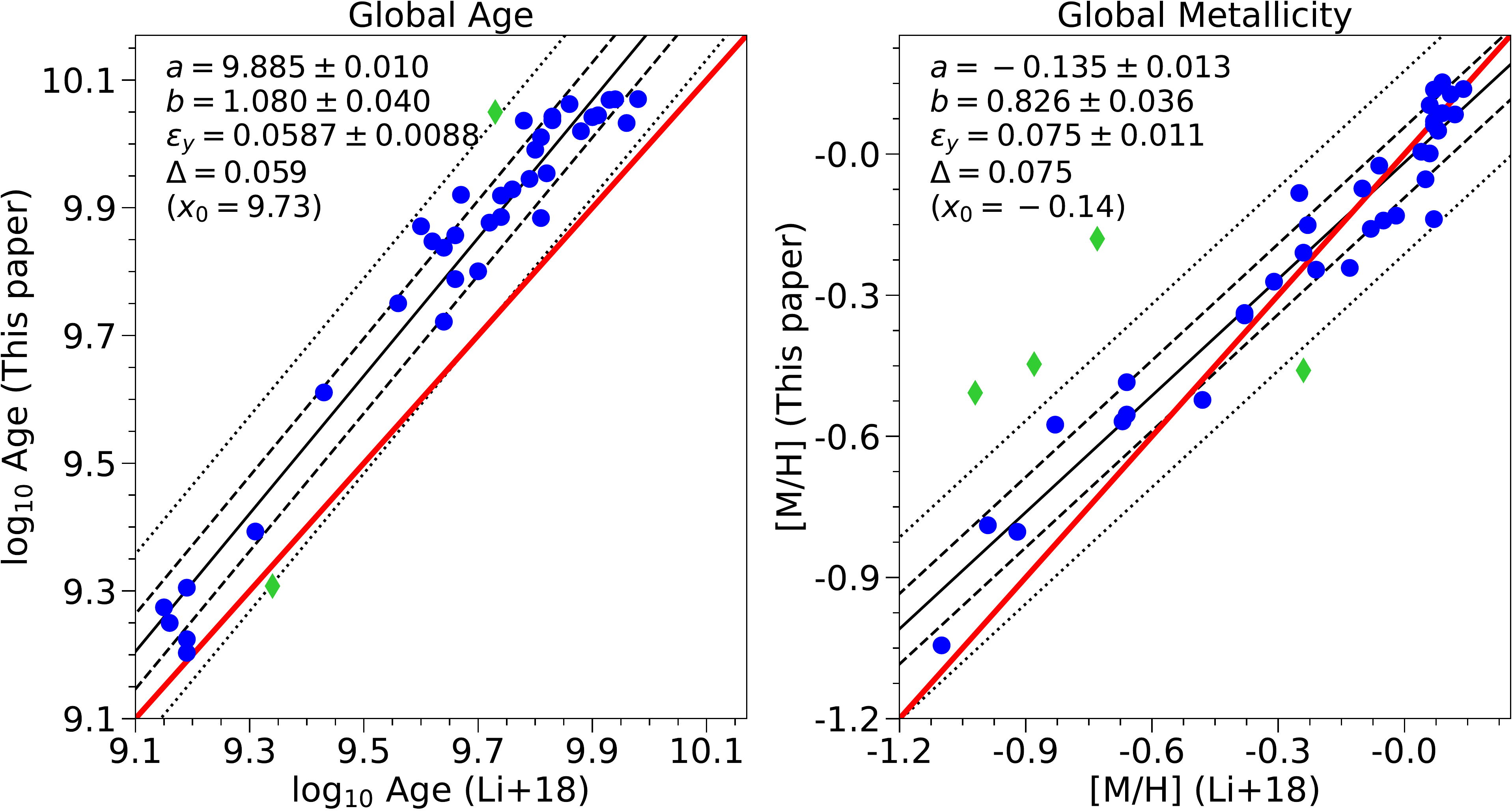}
\caption{Comparison of global age and metallicity for 39 CRDs in common with the sample of Li+18. The blue dots are the values used to fit the linear relation $y = a + b(x - x_0)$, with $x_0$ being the mean value, while green dots are outliers excluded by the routine during the fit. The black dashed and dotted lines are the 1$\sigma$ and 2.6$\sigma$ confidences, respectively. The red straight line is the $y=x$ line. The best-fit coefficients are texted on the plots; $\varepsilon_y$ and $\Delta$ are the intrinsic and observed scatters, respectively.}
\label{fig:crd_vs_li}
\end{figure}

The sample of Li+18 includes 39 of our CRDs; to evaluate how consistent our results on stellar population are with respect to Li+18, we compared the properties of the galaxies in common by fitting the relation $y = a + b(x - x_0)$, where $x_0$ is the median value. As shown in Figure \ref{fig:crd_vs_li}, ages and metallicities show a good correlation; however, the difference in ages is systematic, with our ages being typically $\sim$0.15dex older than those calculated by Li+18. We attribute this difference to the difference in the fitting procedure mentioned above. Given the good correlation between the two samples, we assume the typical errors for the age and metallicity values of all CRDs to be $\Delta/\sqrt{2}$, where $\Delta$ is the observed scatter in the linear fit; then we consider 0.04 dex and 0.05 dex as the typical errors for age and metallicity, respectively. 

On the other hand, gradients do not correlate very well (the correlation coefficients for age and metallicity gradients are $r = 0.42$ and $r=0.39$, respectively, with associated correlation probabilities $p = 0.01$ and $p = 0.02$). We ascribe the lack of correlation of gradients to the fact that they are intrinsically difficult to determine, and the method used by us and Li+18 to estimate them is sensitive to small differences in the fitting procedure of the stellar population properties of single bins, as well as the fit of the profiles.  

\begin{figure}
\includegraphics[width=\columnwidth]{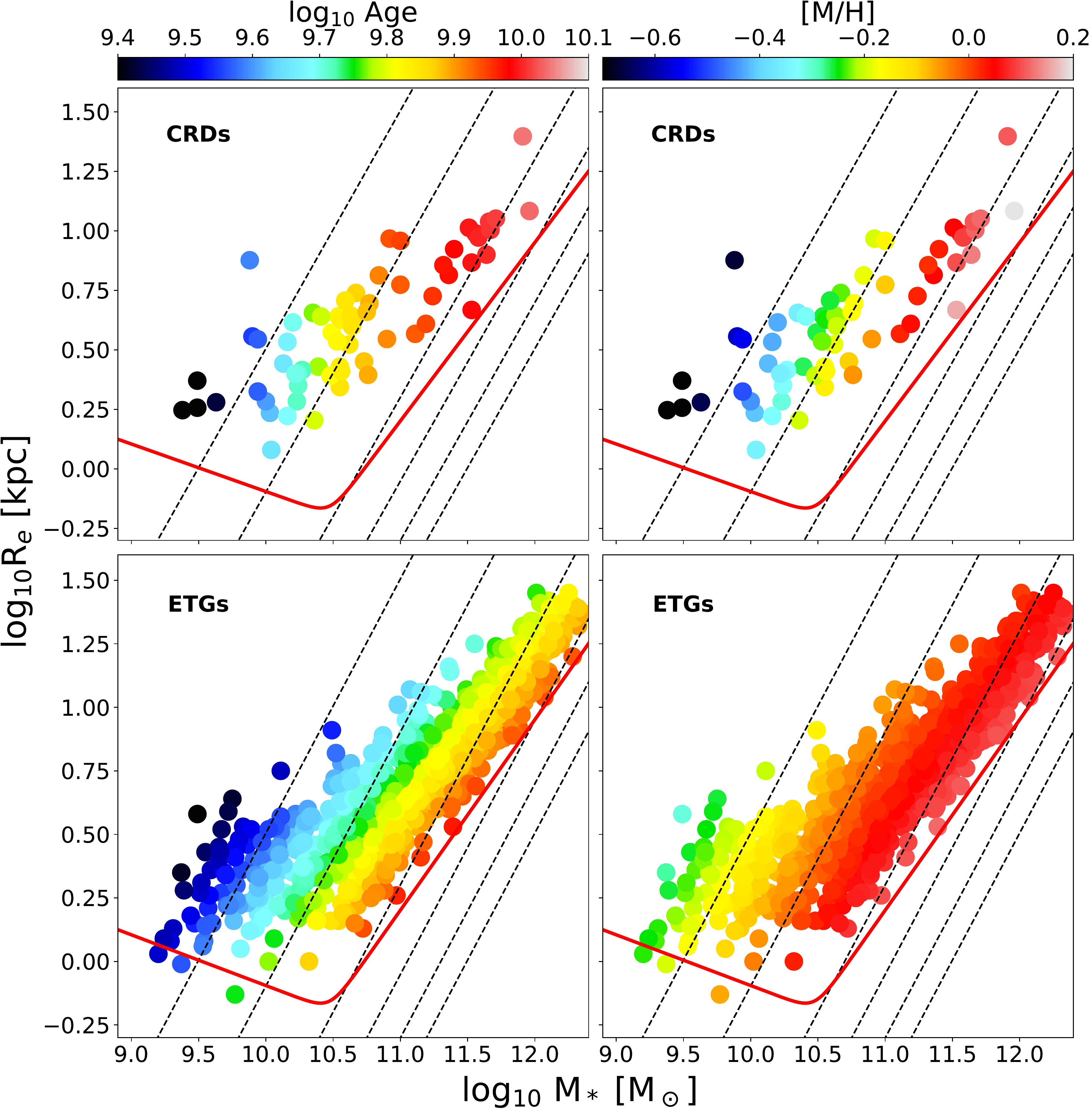}
\caption{LOESS-smoothed log$_{10}$Age (left columns) and metallicity [M/H] (right columns) of CRDs (upper panels) and ETGs from Li+18 (lower panels). Lines are the same of Figure \ref{fig:MS_stargas}. Note: masses of CRDs are calculated from the photometry, using equation (2) of \protect\cite{Cap_letter}; instead, masses in Li+18 are calculated from dynamical models. }
\label{fig:mass-size}
\end{figure}
\begin{figure}
\includegraphics[width=\columnwidth]{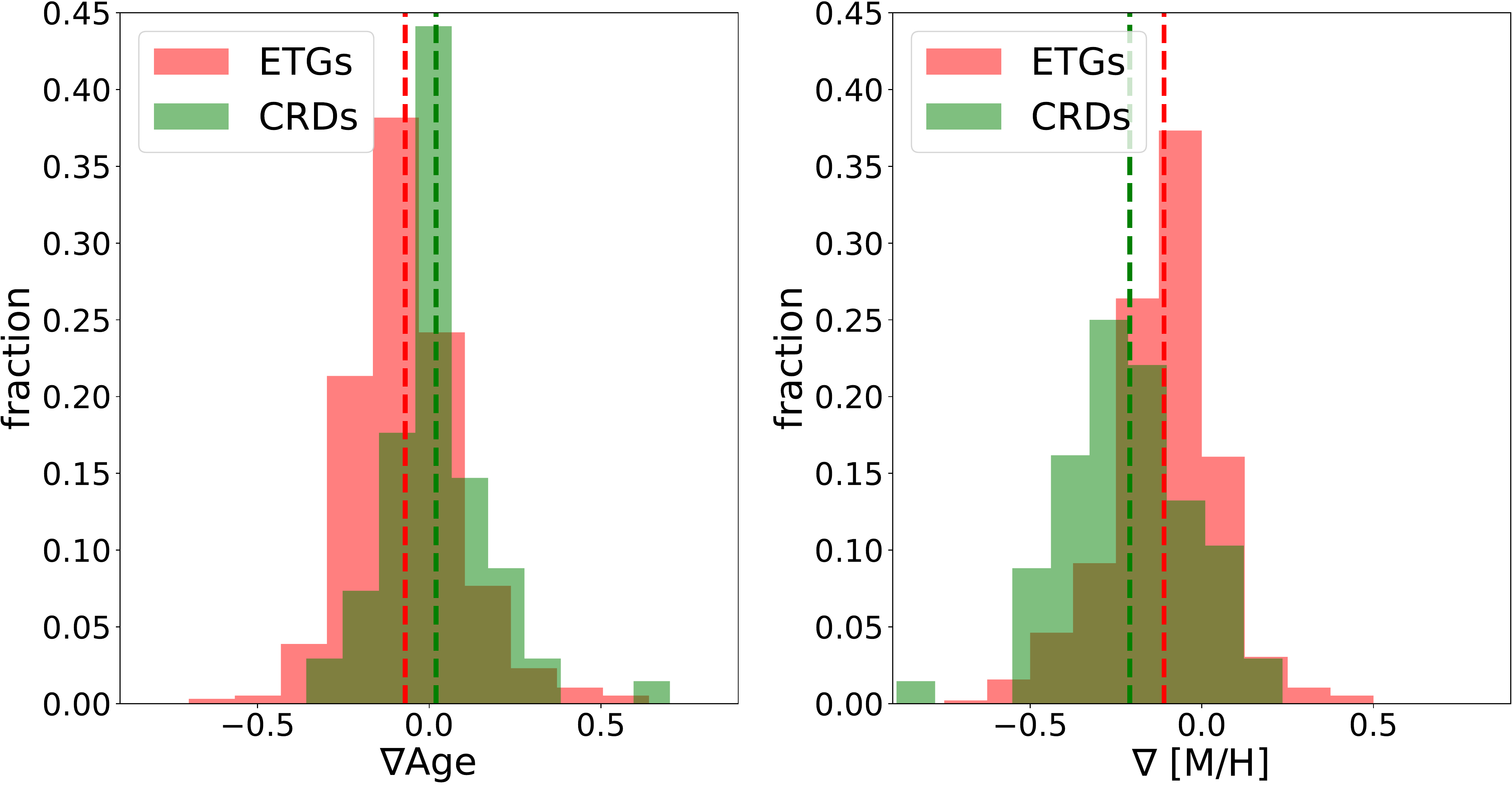}
\caption{Comparison of age and metallicity gradients between ETGs from Li+18 (red shaded) and CRDs (green shaded) from this work. Gradients of CRDs are calculated using equations \eqref{eq:gradage} and \eqref{eq:gradmet}. The red and green dashed lines are the mean values of ETGs and CRDs gradients, respectively.}
\label{fig:frac_grads}
\end{figure}

In Figure \ref{fig:mass-size} we show the stellar population properties of CRDs on the mass-size diagram, and compare them to ETGs from Li+18. It is evident that CRDs follow the same trend in age and metallicity as ETGs. Even though CRDs appear systematically older than ETGs, the age difference is of the same order of that in Figure \ref{fig:crd_vs_li}; therefore we attribute this difference to the difference in the fitting procedures, and conclude that CRDs and ETGs are consistent with having similar ages. From the comparison of the metallicities, instead, it seems that, while having similar metallicities at high masses, CRDs are significantly less metallic at low masses. We verified that this difference is true, and not due to a bad LOESS smoothing (e.g. due to the presence of few outliers with very low metallicities). However, the dispersion of metallicity values for ETGs at low $\sigma_e$ (i.e. low masses) is large (see Figure 5 of Li+18), and such that their trend may be consistent with that of CRDs; additionally, the statistics of low mass CRDs is low, and a larger number may result in a trend similar to that of ETGs. We then conclude that, from this study, CRDs appear to be less metallic than ETGs at the lower stellar masses; however, a more thorough statistical analysis is necessary to confirm or disprove this statement.

In Figure \ref{fig:frac_grads} we compare gradients of CRDs and ETGs. Metallicity gradients of CRDs, albeit slightly steeper on average, are similar to those of ETGs, and also cover the same range of values; similarly, age gradients of CRDs are typically flatter, but cover a comparable range of values to ETGs.

It should be pointed out that the definition we gave of the global stellar population properties, as well as the relative gradients, refer to the spatial region within the effective radius, namely the typical region where the spectrum is dominated by the inner disk. This implies that results on the stellar population properties can be biased towards the major contributor of the inner regions; in fact, as discussed in section \ref{sect:gaspop}, many age maps show an abrupt change of the stellar age from one disk to the other. We verified that results on stellar population properties, especially on age gradients, can be significantly different when extended to larger radii; however, since the external spatial bins generally have low S/N, and fits can be unreliable, we are not confident in presenting such results, and only point out that a more accurate analysis of the stellar population properties of CRDs requires good quality observations out to the most external regions, or the spectroscopic disentangling of the two disks.

\subsection{Unimodal, multimodal and star-forming CRDs}

\begin{figure}
\centering
\includegraphics[width=\columnwidth]{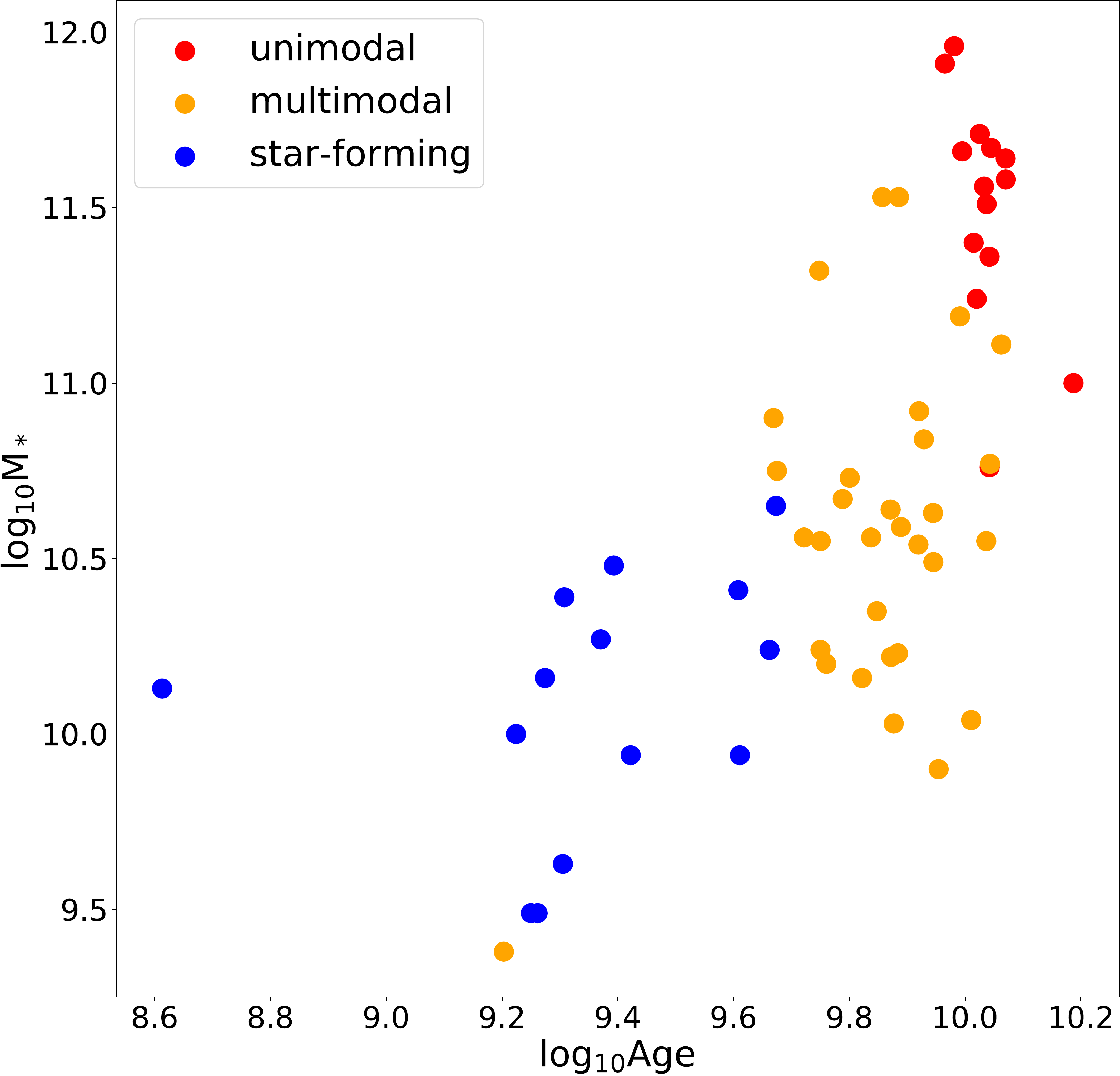}
\caption{CRDs on the mass vs. age diagram. The distinction into unimodal, multimodal and star-forming is based on the weights maps of regularised fits (section \ref{sect:regul}) and was possible for 49 CRDs. Unimodal and multimodal galaxies are those exhibiting the same single or multiple blobs all over the galaxy, while those labelled as star-forming have multimodal maps that change at different spatial regions of the galaxy.}
\label{fig:regul_massage}
\end{figure}

By performing regularised fits, we distinguished galaxies into unimodal, multimodal and star-forming, based on the analysis of the weights map of the stellar population fits. We found 14 CRDs exhibiting unimodality, 31 CRDs exhibiting multimodality, and 14 star-forming CRDs, plus 5 uncertain cases. In Figure \ref{fig:regul_massage} we show them in a mass vs. age diagram.  

From the diagram, we can see a clear distinction between the three classes. All unimodal galaxies are old and massive. Although most of them are confined within the region of slow rotators \citep{Emsellem_2011} on the ($\lambda_{R_e}, \varepsilon$) diagram, the fact that most of these galaxies exhibit both counter-rotation and the two $\sigma_\ast$ peaks at large radii, and that three of them have been spectroscopically found to be counter-rotating, persuade us that they are not misclassified CRDs. On the contrary, star-forming galaxies are the youngest and have low masses ($<$M$_{\rm b}$, for most of them); interestingly, 6 of the 8 `CRDs in formation' (section \ref{sect:selection}) are included in this class (the other 2 are multimodal), which further support that these CRDs have formed just recently. Finally, multimodal galaxies have intermediate ages, while spanning a large range of masses. 

\section{Discussion}\label{sect:discussion}

In light of the results of this work, we discuss now the applicability of the formation scenarios for CRDs mentioned in the Introduction.

The general statistics of CRDs among MaNGA galaxies is relatively low, in agreement with other galaxy surveys or samples with different targeting criteria, like ATLAS$^{\mbox{\scriptsize{3D}}}$ (where CRDs constitute the $4\%$ of the sample). This generally low statistics of CRDs presumably reflects the intrinsic difficulty of building up a massive and extended counter-rotating stellar disk, rather than a low detection rate due to instrumental limits (even though a significant fraction of CRDs is probably missing because of it). Our results, in agreement with previous studies, suggest that the statistics of CRDs decrease from early to late morphological types of the Hubble sequence. This can be interpreted as evidence that the gas accretion scenario is, in general, the most common route to produce a CRD; in fact, to form a counter-rotating disk from an external gas supply, any pre-existing gas in the progenitor galaxy must be swept away before the counter-rotating disk forms. Therefore, it is easier for a CRD to form in a gas-poor galaxy than in a gas-rich one; this would also explain why no late spirals have ever been observed hosting a CRD since they typically have a larger fraction of gas than early spirals and ETGs.

The formation of a counter-rotating stellar disk by accretion of external gas in a retrograde orbit always implies that the secondary disk is also the younger one, and that it corotates with the gaseous disk (e.g., \citealt{Coccato_2012}). The fact that in most of the CRDs with the gaseous and stellar disks aligned and distinct ages for the stellar disks (18 cases) the gas corotates with the younger disk (16 cases) supports this formation scenario against the others. Additionally, 14 of these 16 cases have multimodal or star-forming weights maps (the other 2 cases are unknown), implying the coexistence of multiple stellar populations with different ages and metallicities in the same spatial regions, and all over the galaxy. 

The same scenario of an origin of the counter-rotating disk by retrograde gas accretion may be likely also for those galaxies with misalignment between gas and stars, since this is an evidence of recent gas accretion (note however that the misalignment could be due to an accretion event subsequent and unrelated to the formation of the counter-rotating disk). The fraction of misaligned CRDs ($\sim21\%$) is lower than that of misaligned fast rotators ($\sim36\%$ in ATLAS$^{\mbox{\scriptsize{3D}}}$, \citealt{Davis_2011}), suggesting that CRDs are more likely to have the gas and stars aligned; in fact, if the counter-rotating stellar disk forms via gas accretion, we expect the gaseous disk to be aligned with it. Misalignment as a sign of recent accretion, and linked with the early stages of the formation of the counter-rotating disk, would be the case for three of the 9 misaligned CRDs in our sample with ongoing star formation, as resulting from the weights map, from the SDSS images, and from the disturbed stellar kinematics. Further, if a CRD formed via episodic accretion of gas, misalignment could be due to a recent episode of accretion. For the other cases, misalignment is probably unrelated to the formation of the CRD. To further investigate the relation between misalignment and the origin of CRDs, it would be useful to know the dynamical settling time of the gas, to be compared with the duration of gas accretion; the latter could be difficult to determine, though, since it depends on many circumstances, as the source of external gas \citep{Thakar_1997}, the accretion modalities \citep{Thakar_1996}, and  the angular momentum of the gas \citep{Bryant_19}. Note that cosmological simulations of a massive ETG in a realistic environment show that the evolution of misaligned disks can be complex, with the misalignment persisting several ($\sim$ 2 to 5) Gyr \citep{vandeVoort_15,DB_16}.

For the two cases where the gas corotates with the older disk, the CRD formation via gas accretion is unlikely; the difference in age between the two disks also discards the formation by internal instabilities. One can advocate a scenario in which a disk galaxy in a prograde orbit merges with a gas-poor galaxy with a younger stellar population that rotates in a retrograde orbit; the latter would then settle into a counter-rotating stellar disk, while the gas, originally associated with the primary disk, keeps corotating with it. 

As evident from Figure \ref{fig:regul_massage}, all CRDs exhibiting unimodality are also the oldest and most massive. At such old ages, it is not straightforward to determine whether the two disks actually have the same stellar population, or it is rather the technical difficulty of separating the spectra of two very old populations with similar metallicities that makes them appear as one. Therefore, we cannot rely on this analysis to discuss which scenario is most favourable. However, a detailed study on these CRDs would be of great interest; in fact, if these galaxies truly have a single, old stellar population, the two disks must have formed from the same chemical mixture and in the same event of star formation that gave origin to the galaxy, thus supporting the formation scenario by internal processes, so far disproved by observations.

Finally, even the CRD formation in a major merger event seems viable, since it is consistent with the properties observed for some CRDs. For example, a recent merger could explain the highly disturbed kinematics and morphology shown by a number of CRDs; and even the old massive unimodal CRDs could be the result of a major merger that took place a long time ago, in which the stellar disks of the progenitors had the same ages and metallicities (though this is unlikely to apply to all the 14 unimodal CRDs).

To further investigate the formation scenarios of CRDs, a spectral decomposition of the two stellar disks is needed. This analysis would provide additional information on the kinematics, as the velocity dispersions of the two disks, and on the stellar population properties, as the metallicity, the age, and the metallicity gradient separately for the two disks.

If many (or all) of the proposed scenarios are statistically relevant, the major role in the formation of a CRD is played by its environment. For example, merger and interaction may be a more likely origin for CRDs in relatively dense environments, where gas stripping and galaxy motions make gas accretion difficult. In low-density environments and in the field,  instead, galaxies can acquire undisturbed a large amount of gas from, e.g., cosmological filaments; for a CRD in such an environment, then, formation via gas accretion seems more probable. In conclusion, a study of CRDs in relation with their environment is of great importance to determine the relevance and the statistics of the many proposed formation scenarios.

Overall, this work supports the idea that galaxies with counter-rotating stellar disks form primarily via a retrograde accretion of gas, although other formation channels cannot be excluded in a number of cases. 

\section{Summary and Conclusions}\label{sect:conclusions}

We used data from the DR16 of the MaNGA survey to construct a sample of galaxies with counter-rotating stellar disks. After excluding problematic cases (section \ref{sect:selection}), we visually inspected the kinematic maps of about 4000 galaxies, and selected those that exhibit at least one of the two kinematic features characterising counter-rotators (counter-rotation or two peaks in $\sigma_\ast$); we designated these galaxies with the acronym `CRD'. We studied the stellar and gas kinematics, and the stellar population properties of these CRDs. We summarise our results as follows.  

\begin{itemize}

\item The sample consists of 64 CRDs, 61 of which are ETGs (38 E and 23 S0), 1 is a spiral and 2 have uncertain morphology. CRDs constitute, at 95\% confidence,  $<3$\% of ETGs (in particular, $<5$\% of E and $<3$\% of S0), and $<1$\% of spirals, in agreement with previous estimates.\\

\item From two-components kinematic fits, we obtained spectroscopic confirmation that 17 CRDs comprise two counter-rotating stellar components.\\

\item We calculated the difference in the PAs of the gas and the stellar velocity fields of 42 CRDs, and found that in 33 cases the gas is aligned with one of the two stellar disks, without preference for the disk prevailing in the inner or in the outer region. By comparing with the age maps, we were able to qualitatively determine that in 16 cases the gas corotates with the younger disk, and in 2 cases with the older one. In 9 cases, the gaseous disk is misaligned, and in 8 of these cases, the stellar mass is lower than the break mass M$_{\mbox{\scriptsize{b}}}$.\\

\item The analysis of the stellar population properties indicates that CRDs have similar trend in ages and metallicities as ETGs, but CRDs appear less metallic at low masses. CRDs also have similar age and metallicity gradients, while being, on average, typically flatter the former and steeper the latter.\\

\item We distinguished CRDs based on the number of different stellar populations present in the weights maps of the regularised fit of various spatial bins. We found that CRDs exhibiting unimodality (14) are the oldest and most massive. CRDs with a multimodal and variable weights map, that we labelled as star-forming (14), are the youngest and least massive. Finally, CRDs with the same multimodal weights maps all over the galaxy (31) have intermediate ages, and span a wide range of masses.

\end{itemize}

Based on these results on the first, relatively large sample of CRDs, we conclude that the unique kinematic class of galaxies with counter-rotating stellar disks form primarily via external counter-rotating gas accretion. However, Nature is rarely captured by a simple scenario, and also in the case of CRDs other mechanisms are as well possible in specific cases. Older and more massive unimodal CRDs may have formed via internally driven mechanisms; their formation via major merger, albeit unlikely, can not be a priori ruled out. 

To further investigate the CRDs in our sample, observations with higher resolution and S/N are needed. Additionally, the spectral decomposition of the two disks for each galaxy would provide cogent evidences to constrain the proposed formation scenarios. Finally, a study of the environment of CRDs would help understand how it correlates with the observed CRD properties, and then its possible role in the different formation scenarios.

\section*{Data availability}

The data cubes of the DR16 are publicly available, as well as the kinematic maps produced by the DAP, and can be found at \url{https://www.sdss.org/dr16/manga/}. The softwares used in this paper are all public. The kinematic and stellar population maps of the whole sample of CRDs are provided in an electronic appendix. Values tabulated in Table \ref{tab:crd} are available also in electronic format. Data used for Figure \ref{fig:lamelgraham} are available as supplementary electronic material.

\section*{Acknowledgement}

We want to thank the anonymous referee for the constructive comments and suggestions that helped improving this paper.

\appendix
\section{Maps of Counter-Rotating Disks}\label{app:crd}
In this appendix, we show the kinematic and stellar population maps of the CRDs in our sample. Table 1 of the main text includes the MaNGA IDs of the CRDs, and their main properties used here. 

In Figure \ref{fig:crd_all} we show the SDSS image, the kinematic maps and the stellar population maps of CRDs. Each galaxy includes six images, distributed in six rows. The first rows show the SDSS images, with the MaNGA ID of the galaxy overwritten; the second and third rows show the stellar velocity ($V_\ast$) and velocity dispersion ($\sigma_\ast$) maps, in km s$^{-1}$, produced as described in section 3.2 of the main paper; the fourth rows show the velocity fields, in km s$^{-1}$, of the ionised gas ($V_{H\alpha}$), as extracted by the DAP; the fifth and sixth rows show the r-band luminosity weighted age (log$_{10}$Age) and metallicity ([M/H]) maps, produced as described in section 4.1. In all maps, ticks are in arcsec.

\begin{figure*}
\begin{subfigure}{\textwidth}
\centering
\includegraphics[scale=.38]{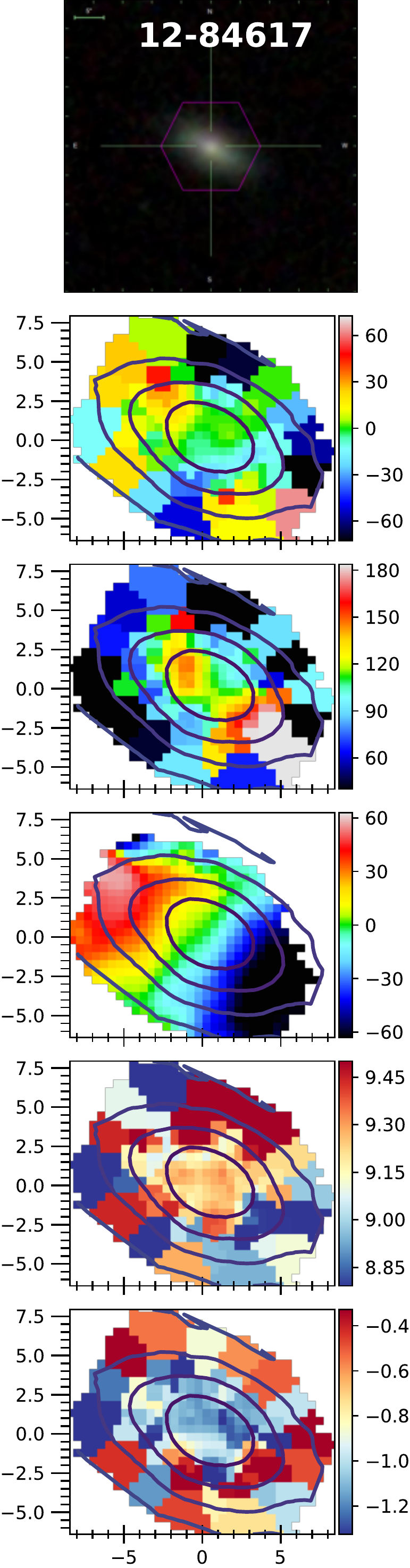}
\includegraphics[scale=.38]{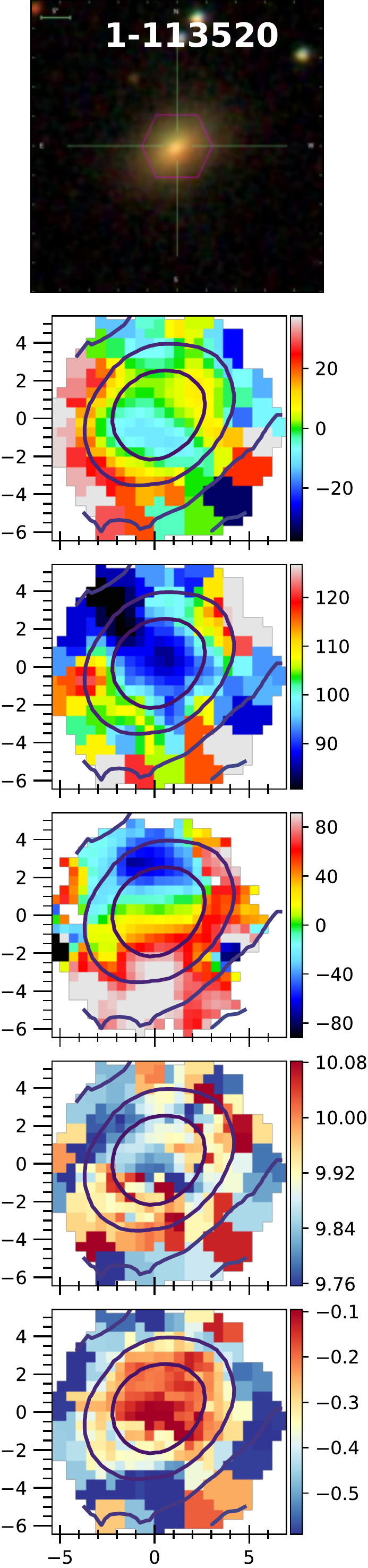}
\includegraphics[scale=.38]{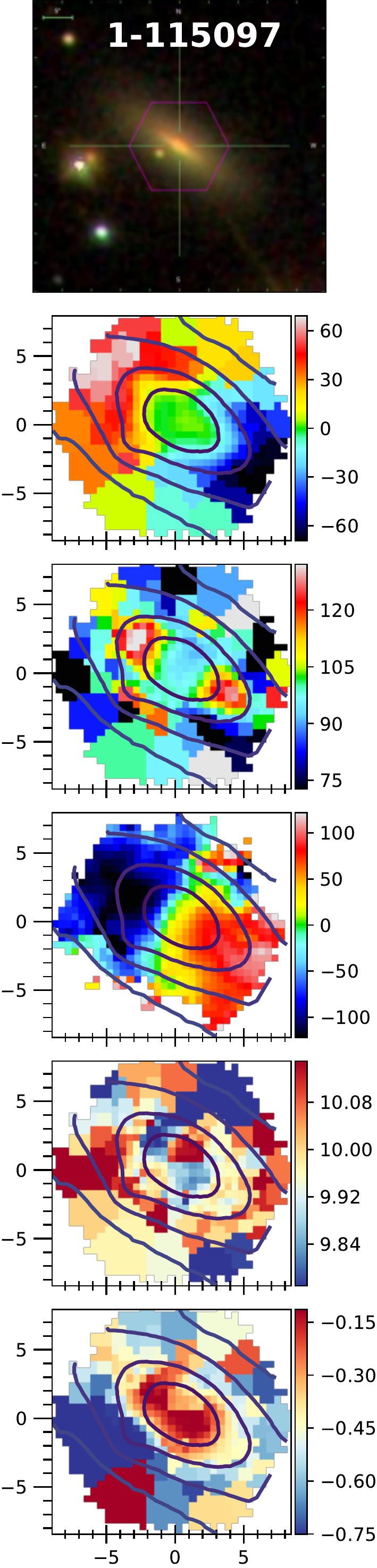}
\includegraphics[scale=.38]{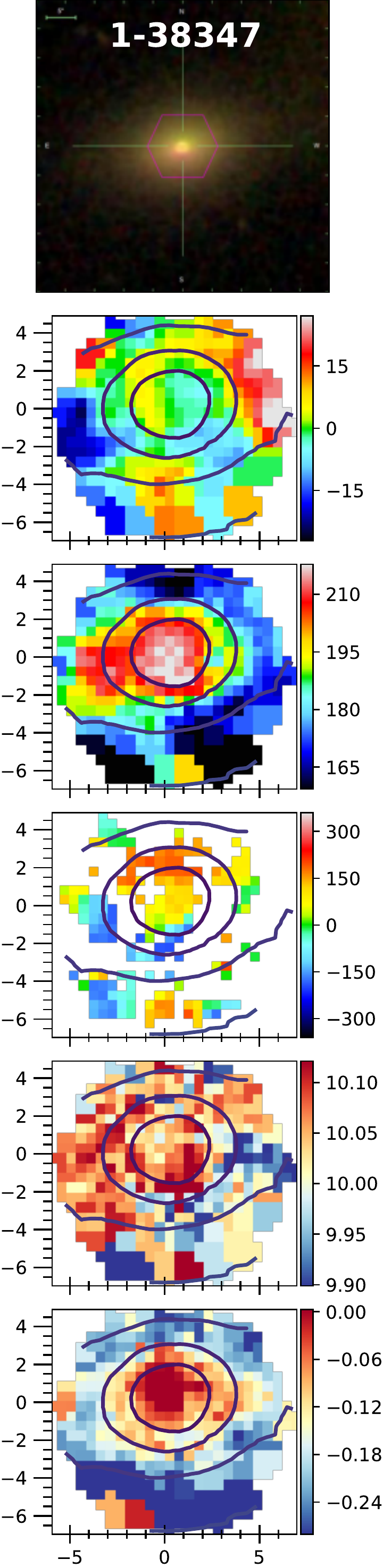}
\includegraphics[scale=.38]{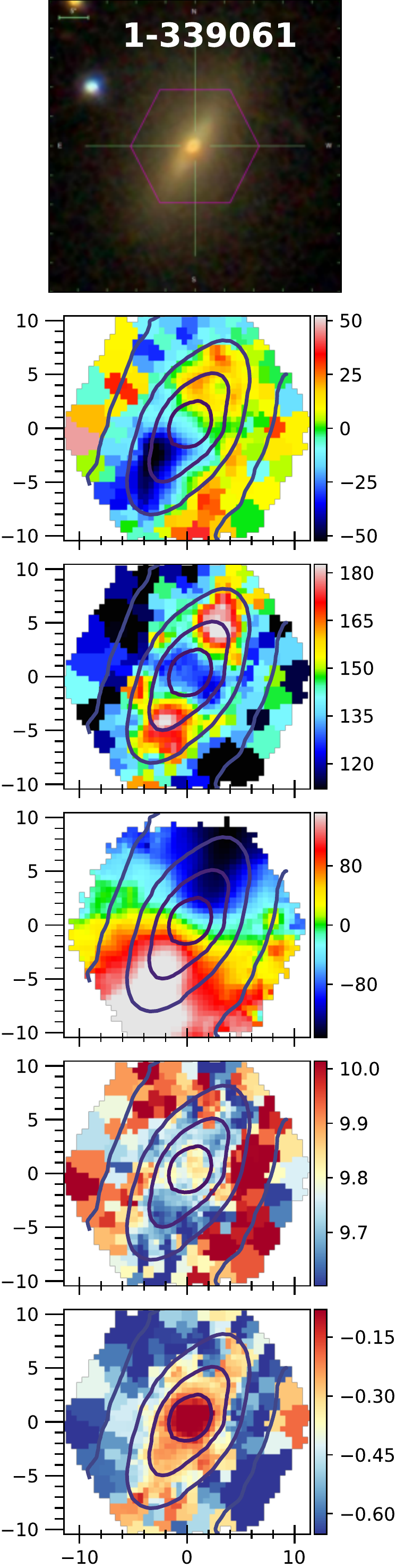}
\includegraphics[scale=.38]{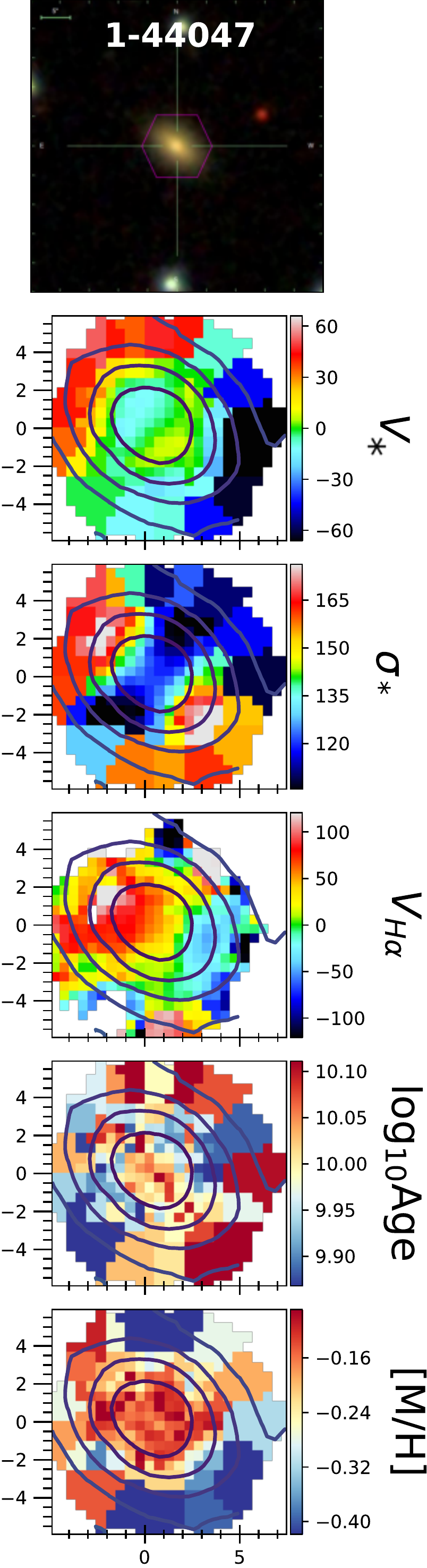}
\includegraphics[scale=.38]{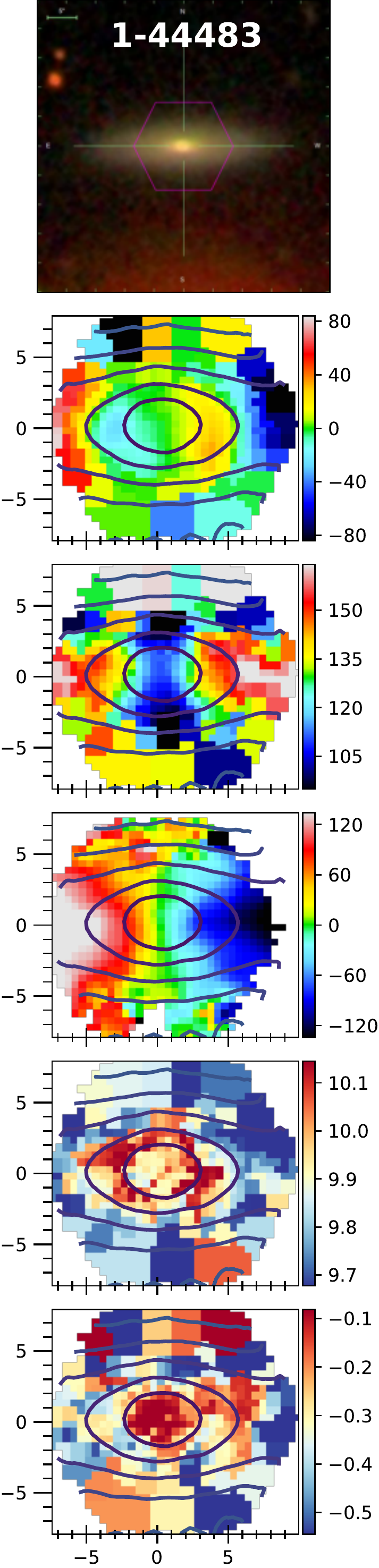}
\includegraphics[scale=.38]{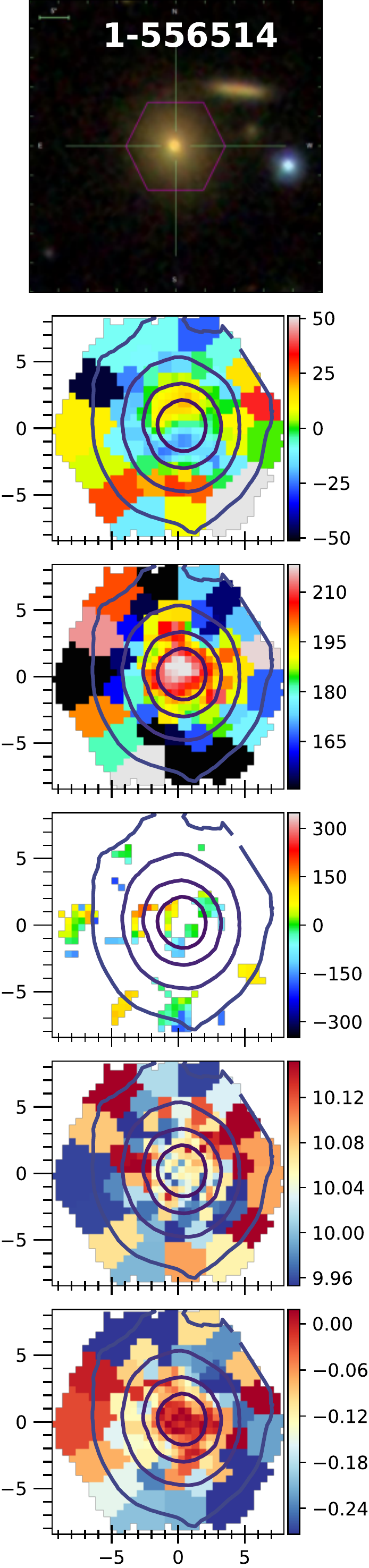}
\includegraphics[scale=.38]{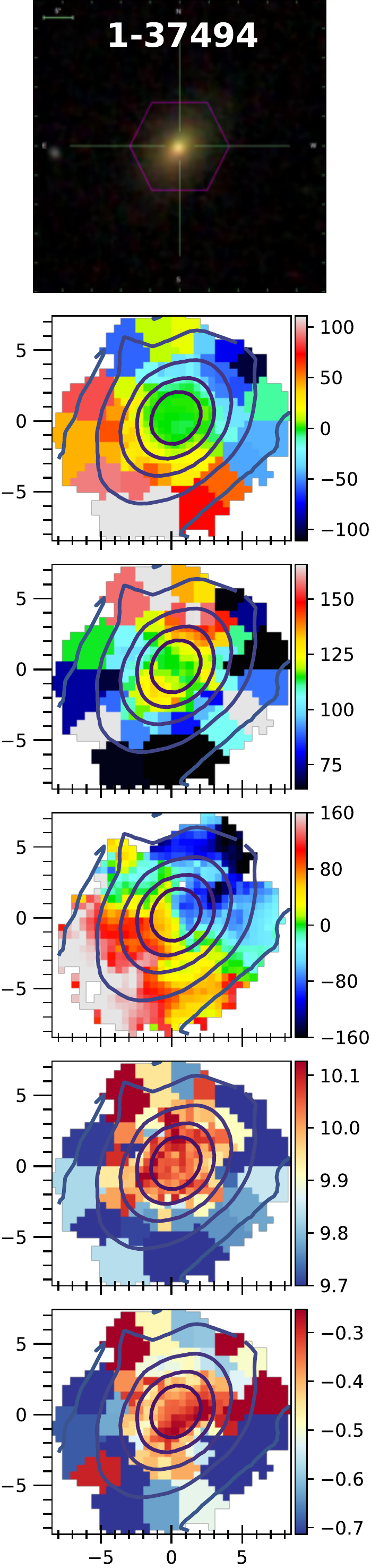}
\includegraphics[scale=.38]{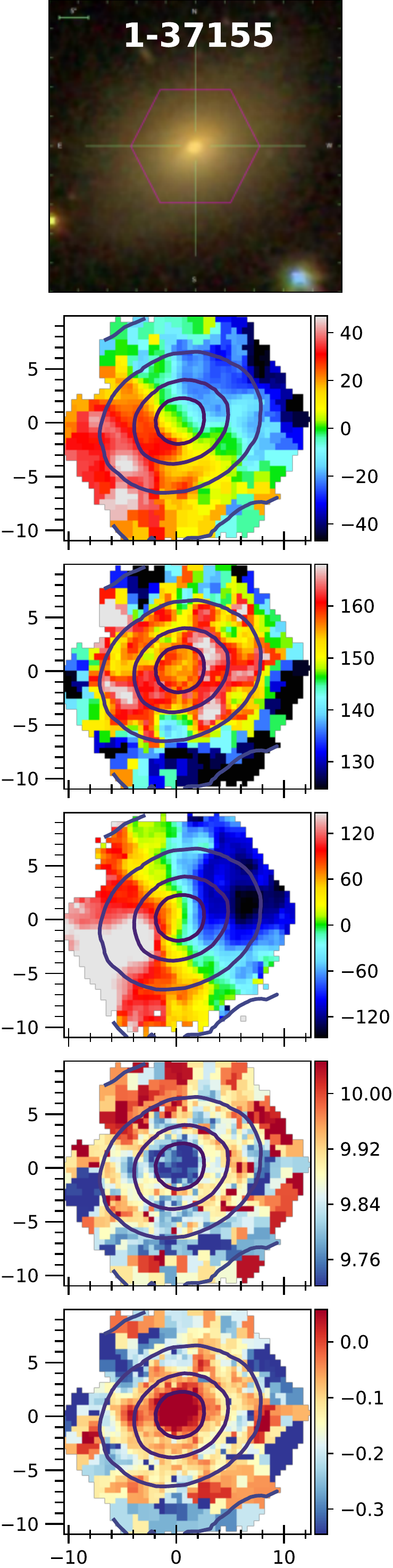}
\includegraphics[scale=.38]{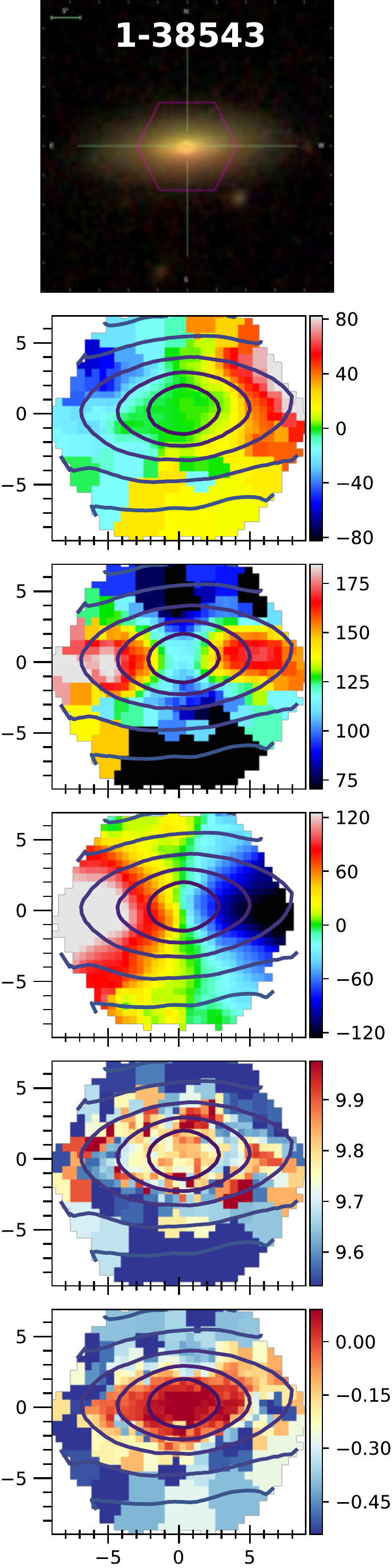}
\includegraphics[scale=.38]{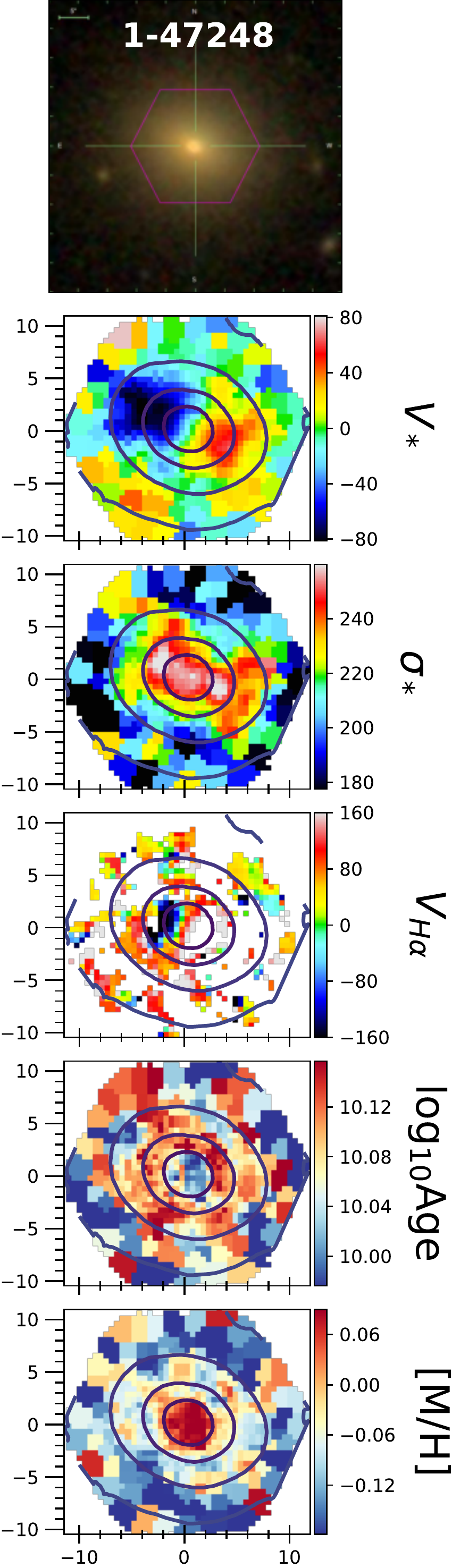}
\end{subfigure}
\caption{SDSS and maps of all the sample of 64 CRDs. Rows: (1) SDSS image with MaNGA ID overwritten; (2) and (3) Stellar velocity and velocity dispersion maps ([km s$^{-1}$]); (4) H$\alpha$ velocity map ([km s$^{-1}$]); (5) age map in log$_{10}$(Age/yr); (6) metallicity map in dex.}
\label{fig:crd_all}
\end{figure*}
\begin{figure*}
\ContinuedFloat
\begin{subfigure}{\textwidth}
\centering
\includegraphics[scale=0.38]{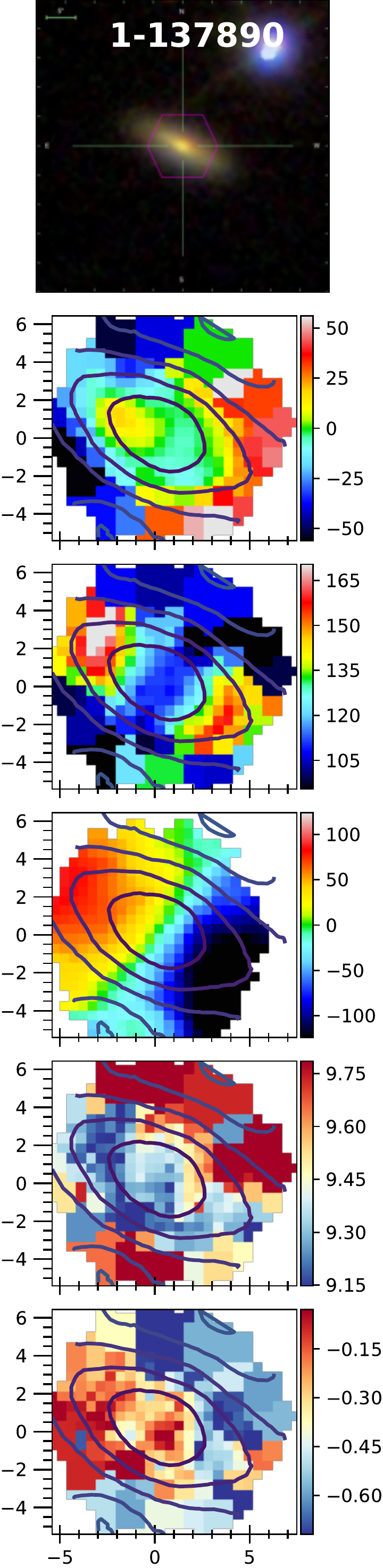}
\includegraphics[scale=0.38]{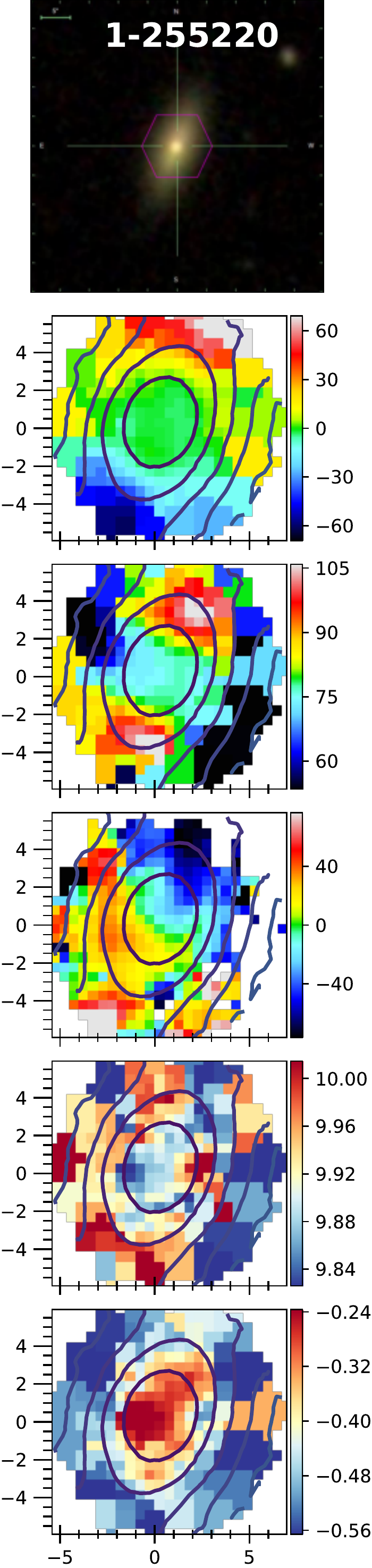}
\includegraphics[scale=0.38]{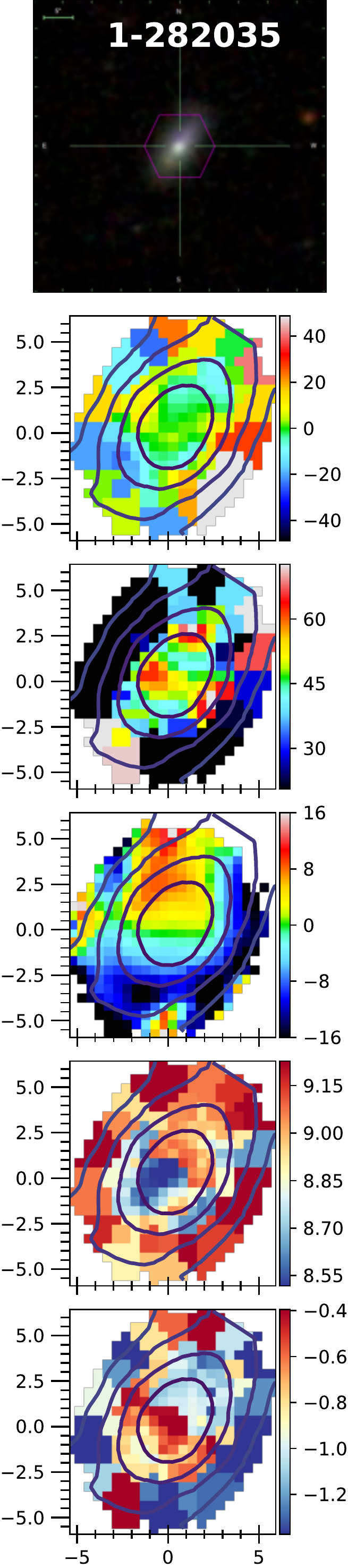}
\includegraphics[scale=0.38]{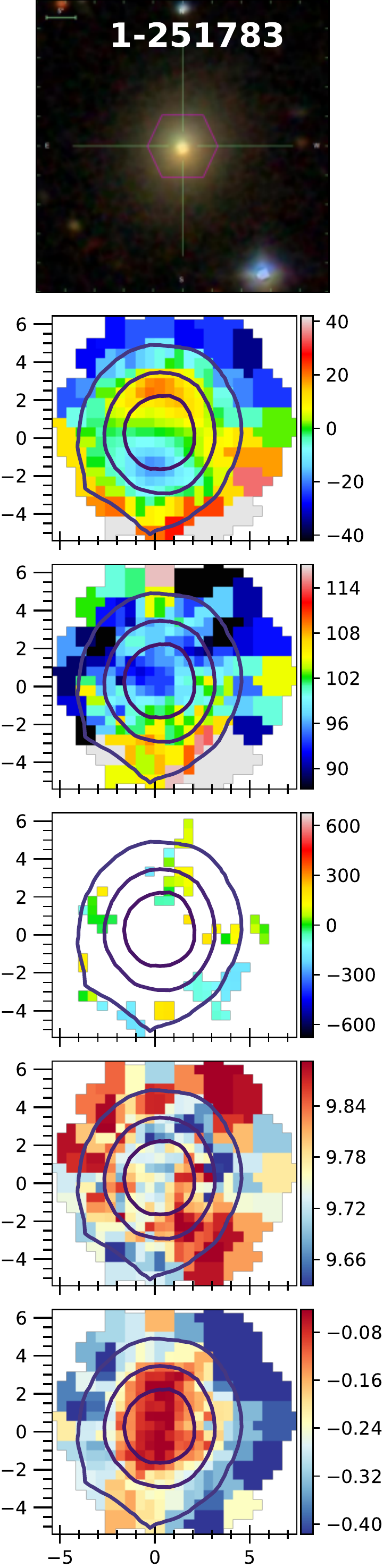}
\includegraphics[scale=0.38]{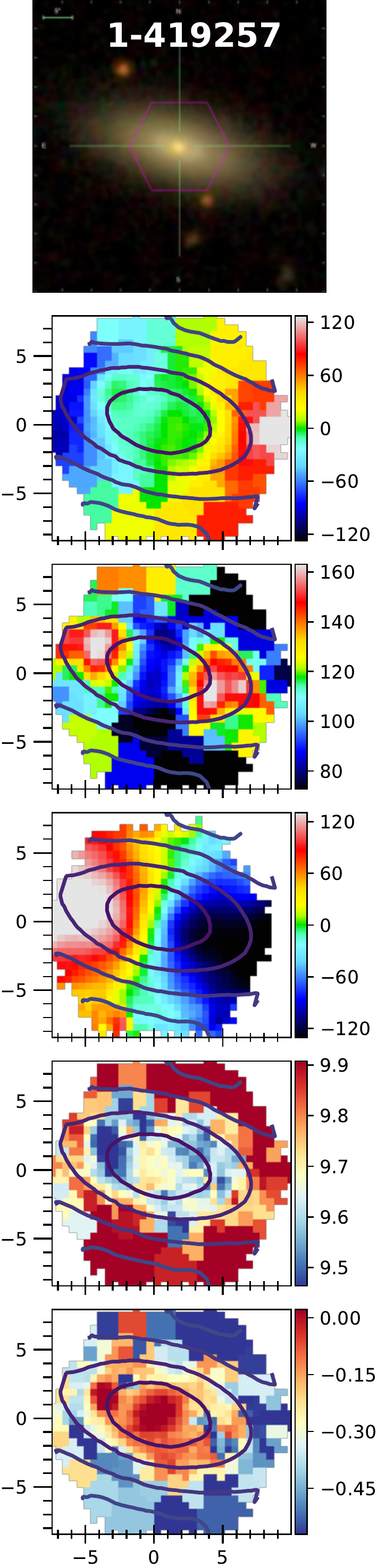}
\includegraphics[scale=0.38]{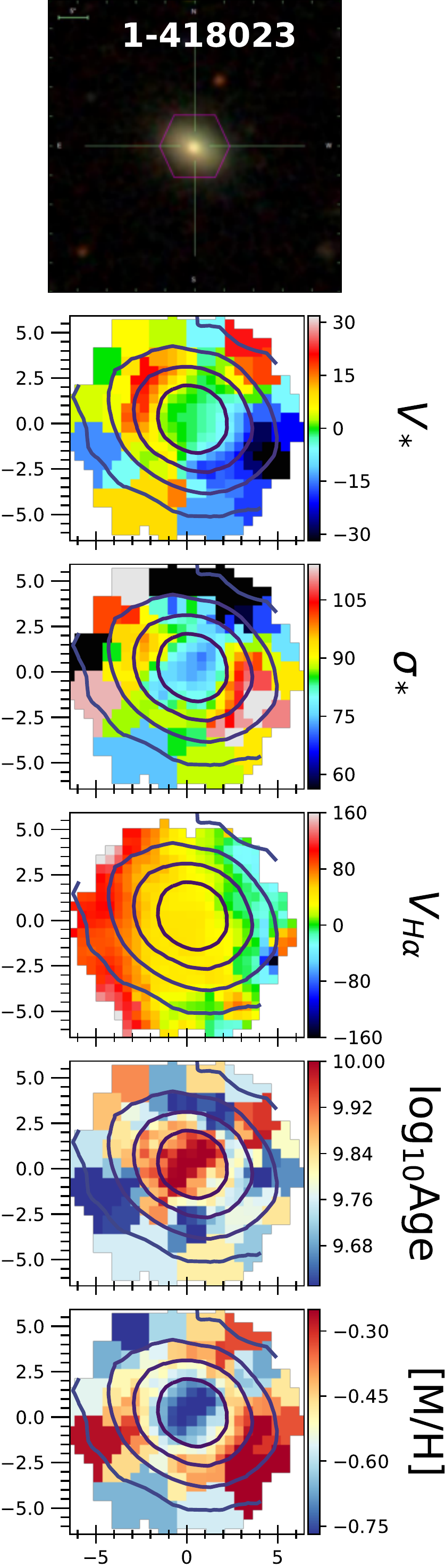}
\includegraphics[scale=0.38]{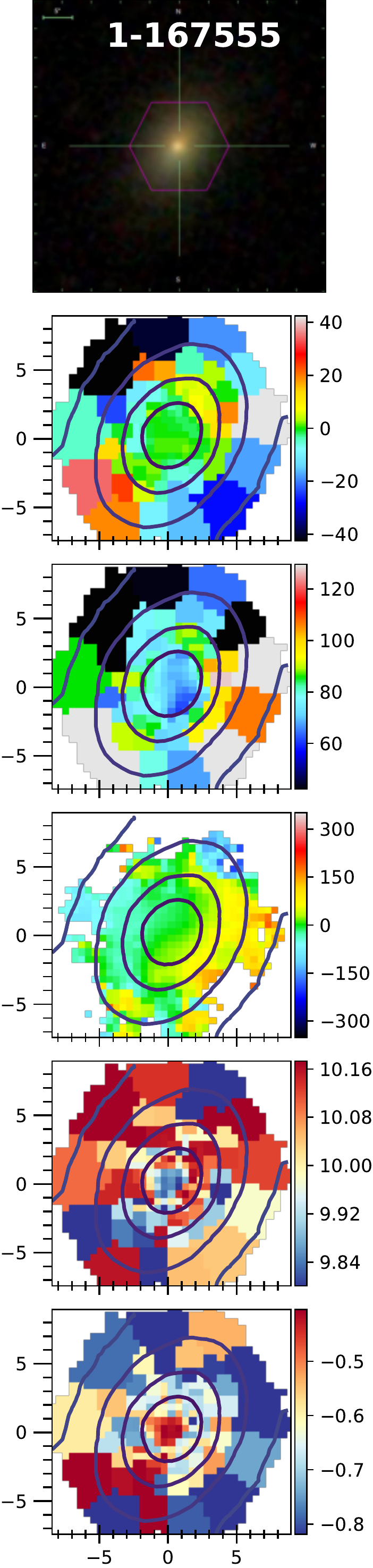}
\includegraphics[scale=0.38]{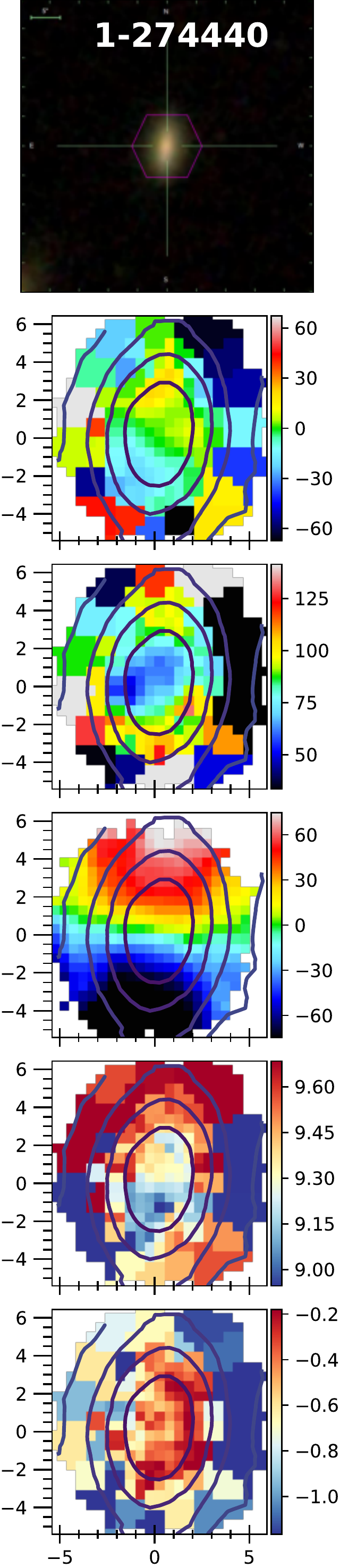}
\includegraphics[scale=0.38]{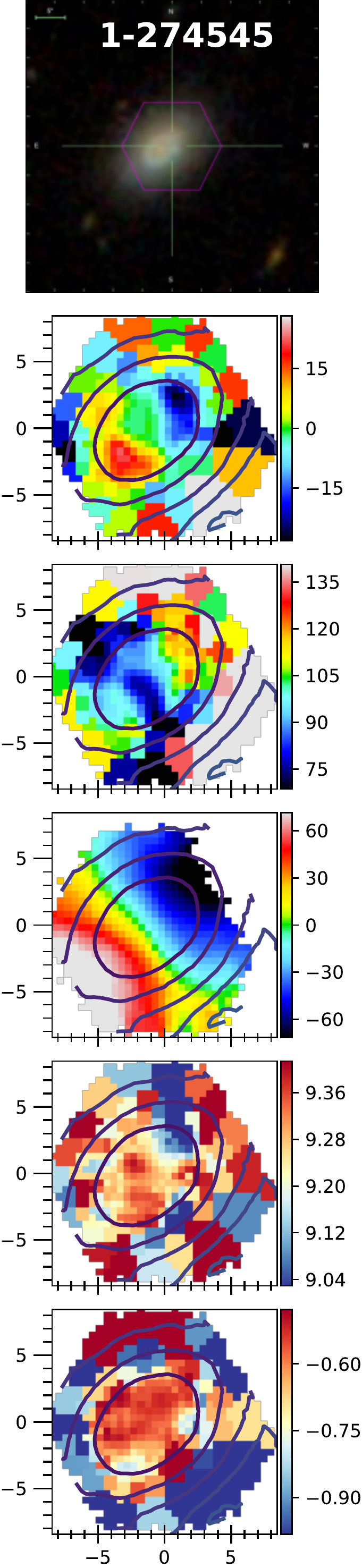}
\includegraphics[scale=0.38]{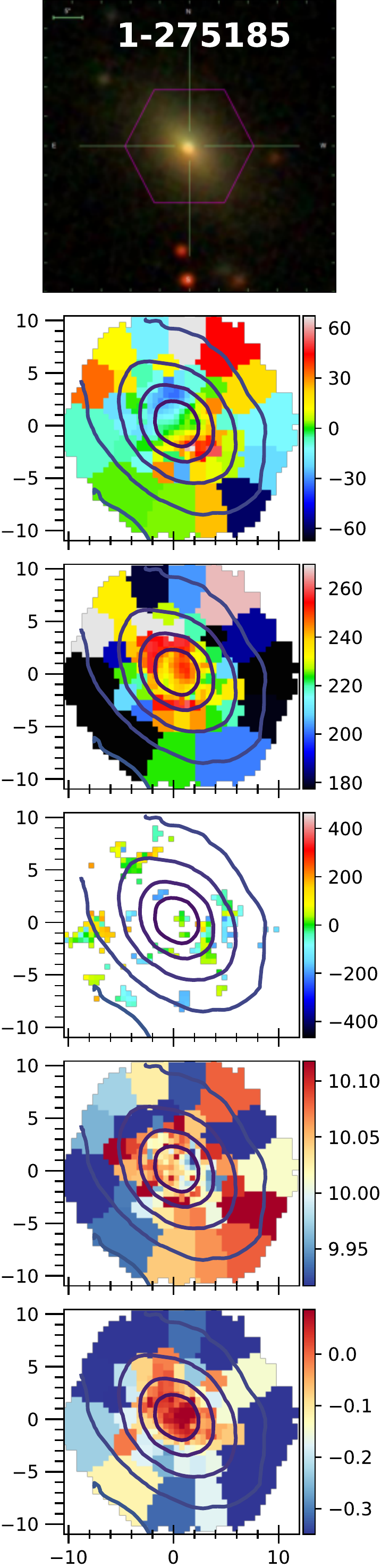}
\includegraphics[scale=0.38]{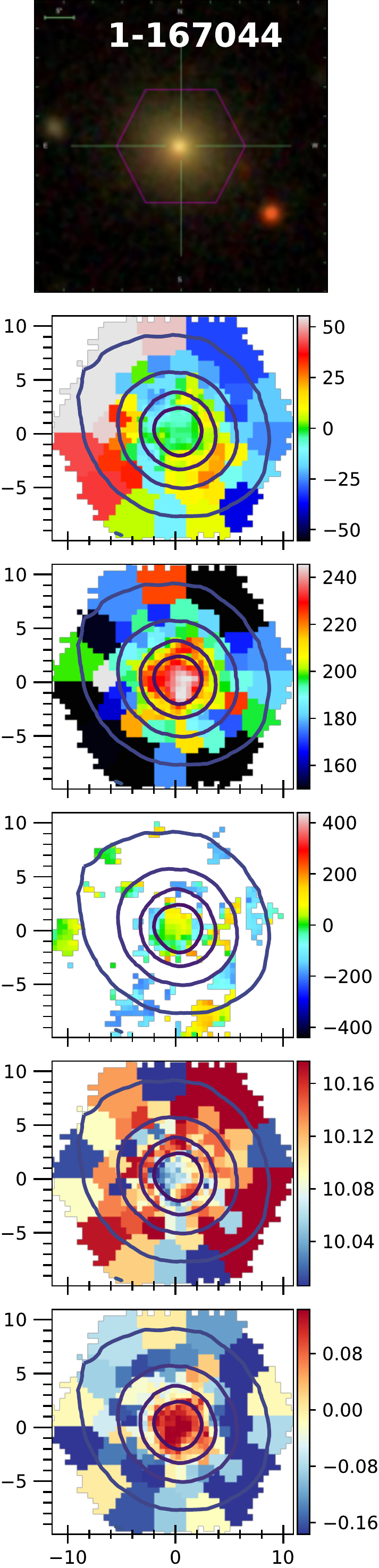}
\includegraphics[scale=0.38]{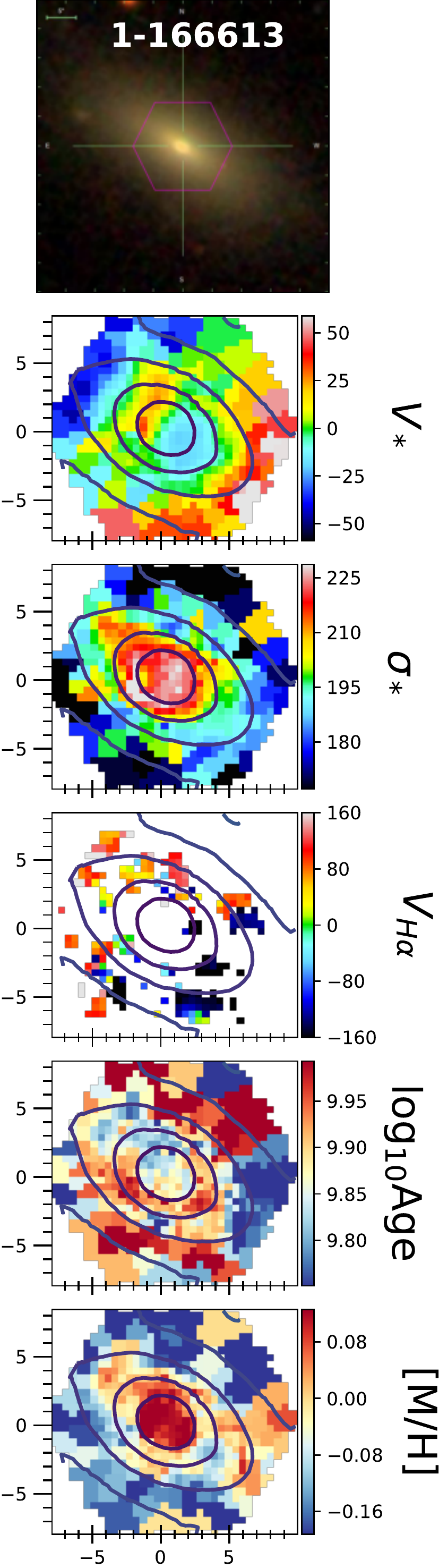}
\end{subfigure}
\caption{\textit{(continue)}}
\end{figure*}
\begin{figure*}
\ContinuedFloat
\begin{subfigure}{\textwidth}
\centering
\includegraphics[width=0.17\textwidth , height=.48\textheight]{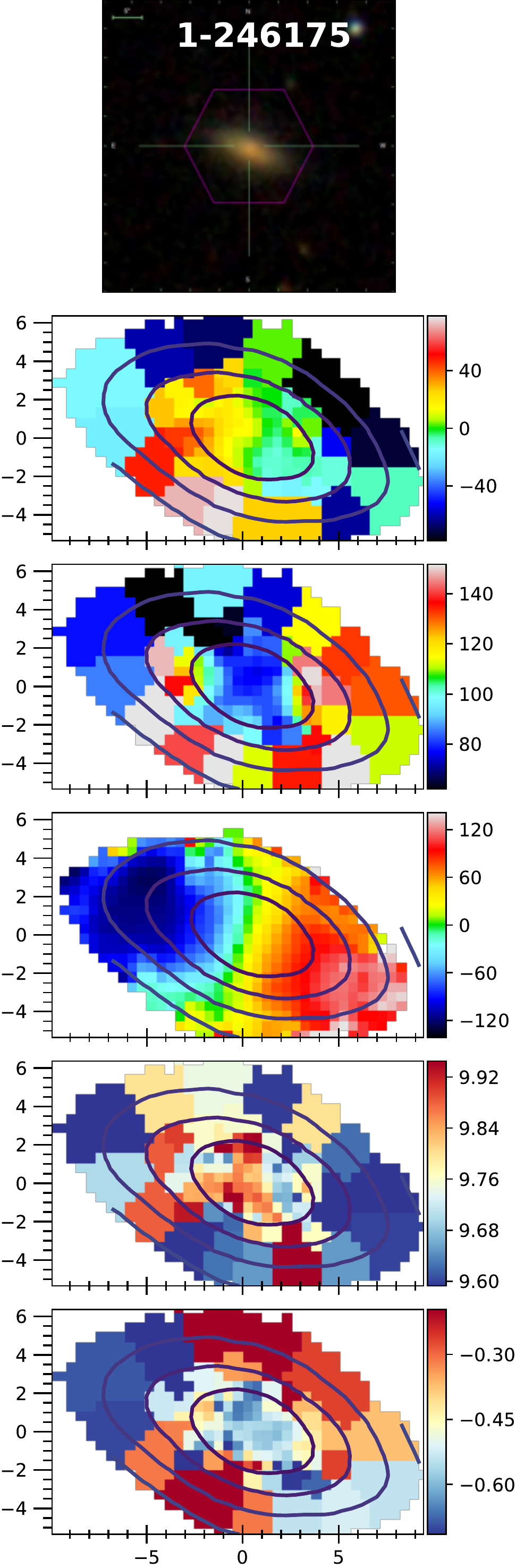}
\includegraphics[scale=0.38]{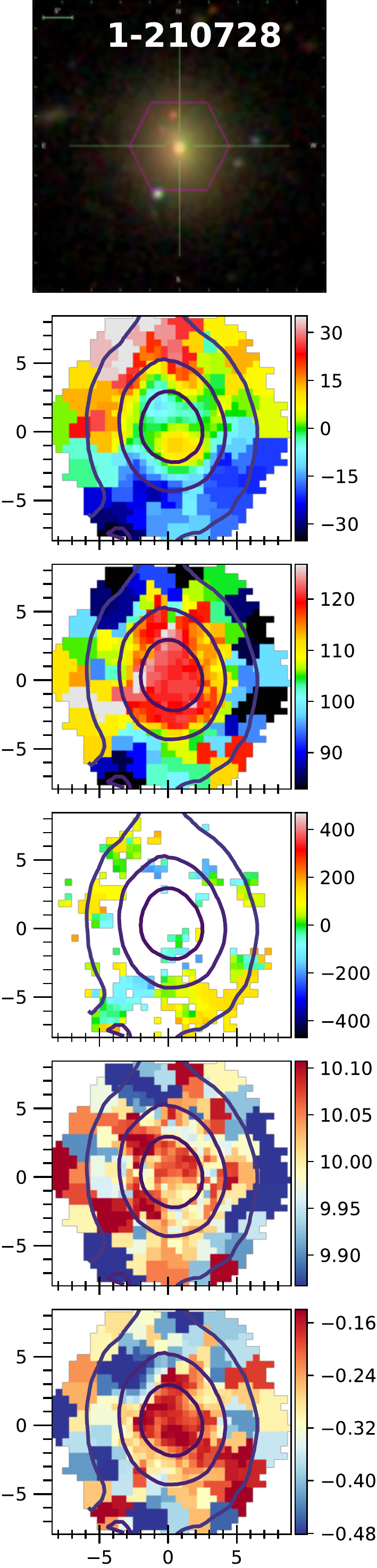}
\includegraphics[scale=0.38]{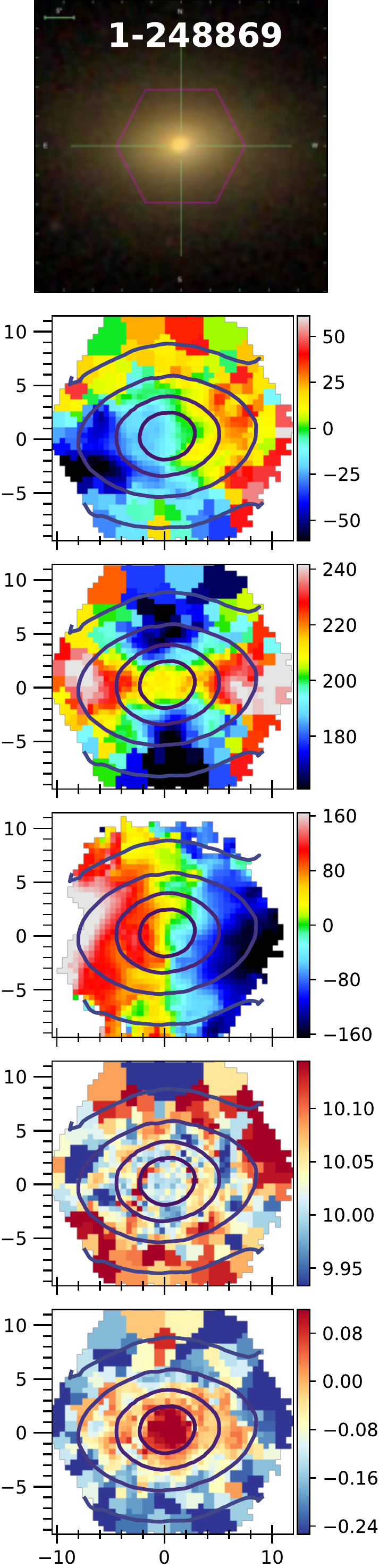}
\includegraphics[scale=0.38]{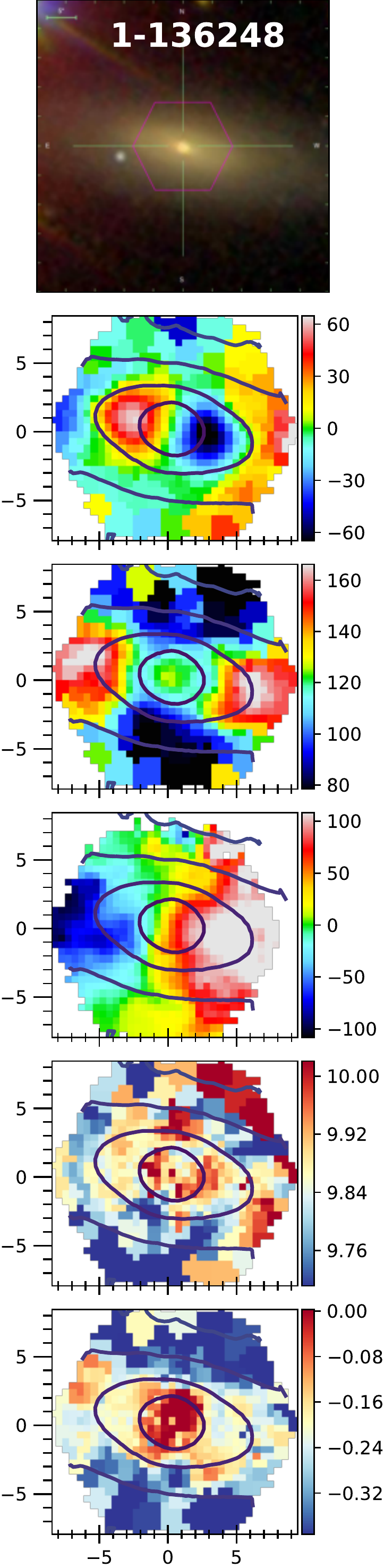}
\includegraphics[scale=0.38]{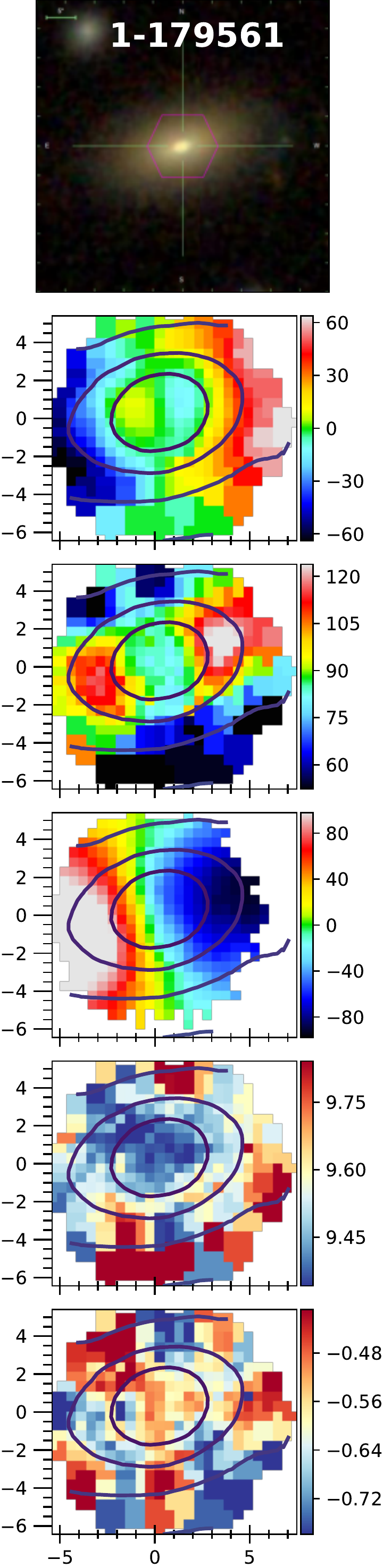}
\includegraphics[scale=0.38]{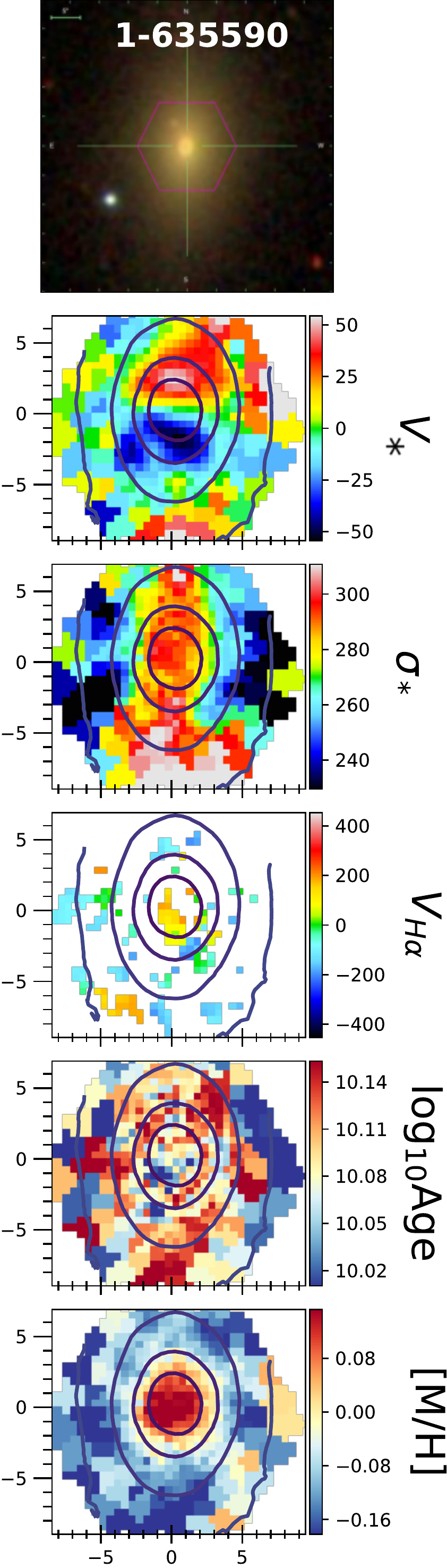}
\includegraphics[scale=0.38]{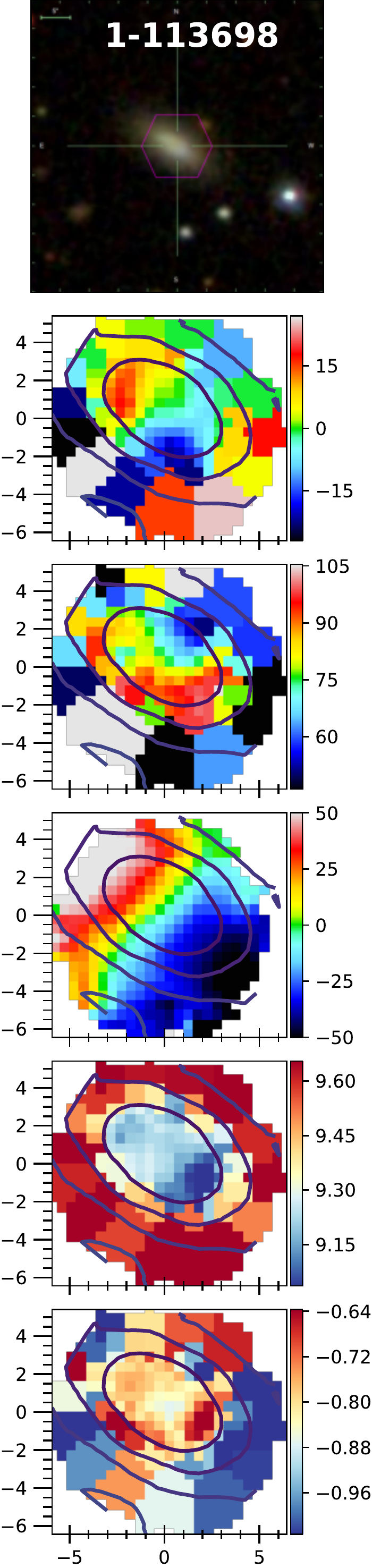}
\includegraphics[scale=0.38]{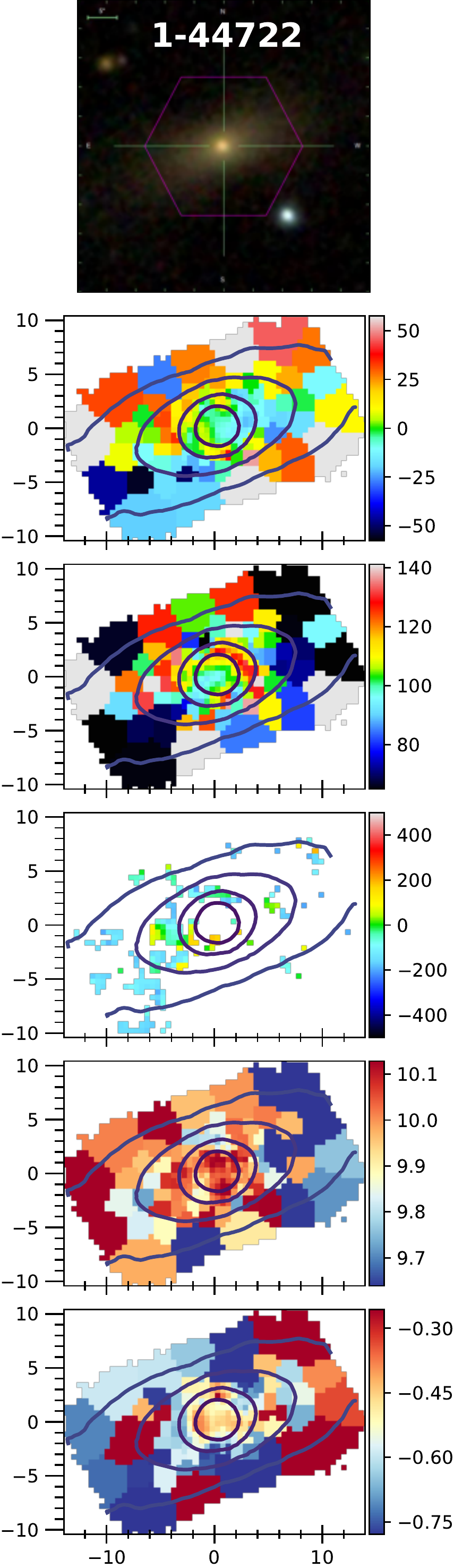}
\includegraphics[scale=0.38]{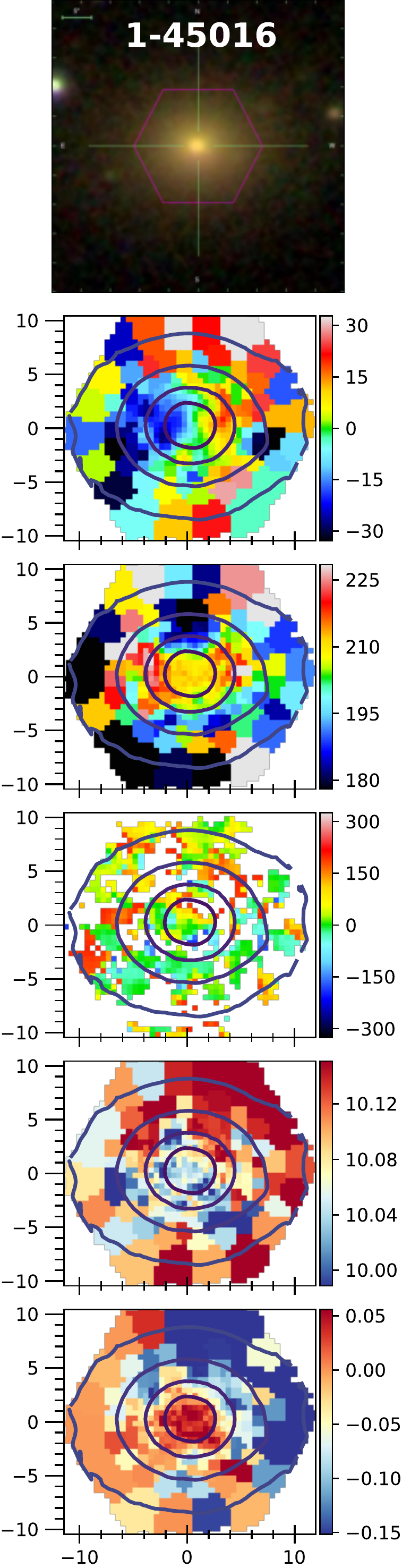}
\includegraphics[scale=0.38]{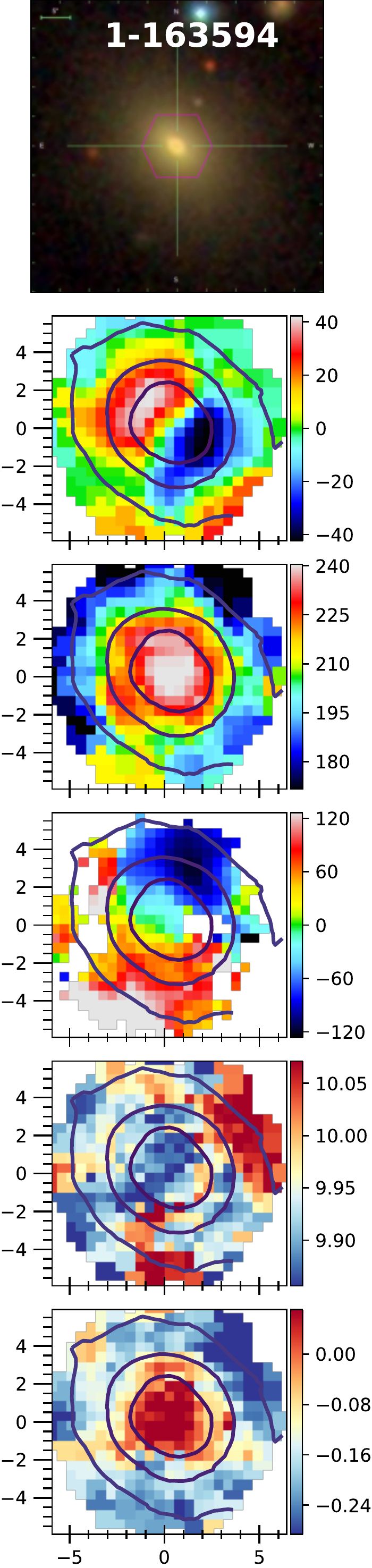}
\includegraphics[scale=0.38]{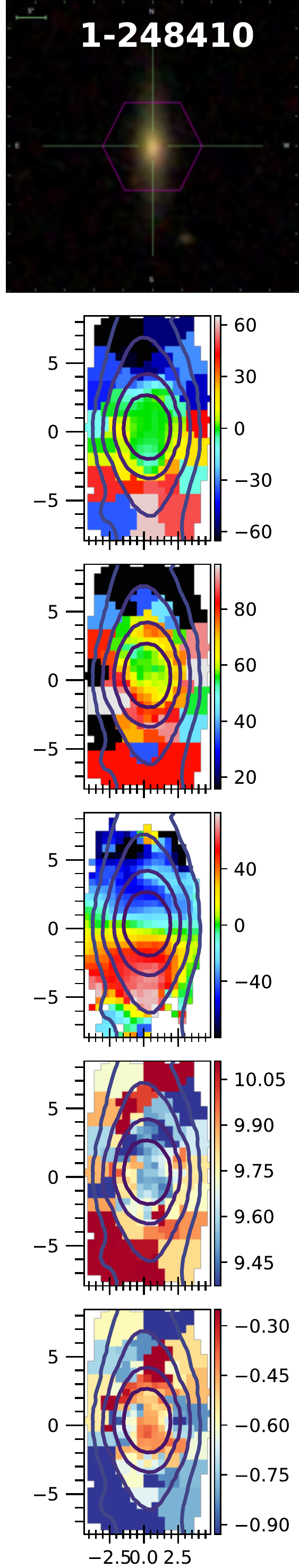}
\includegraphics[scale=0.38]{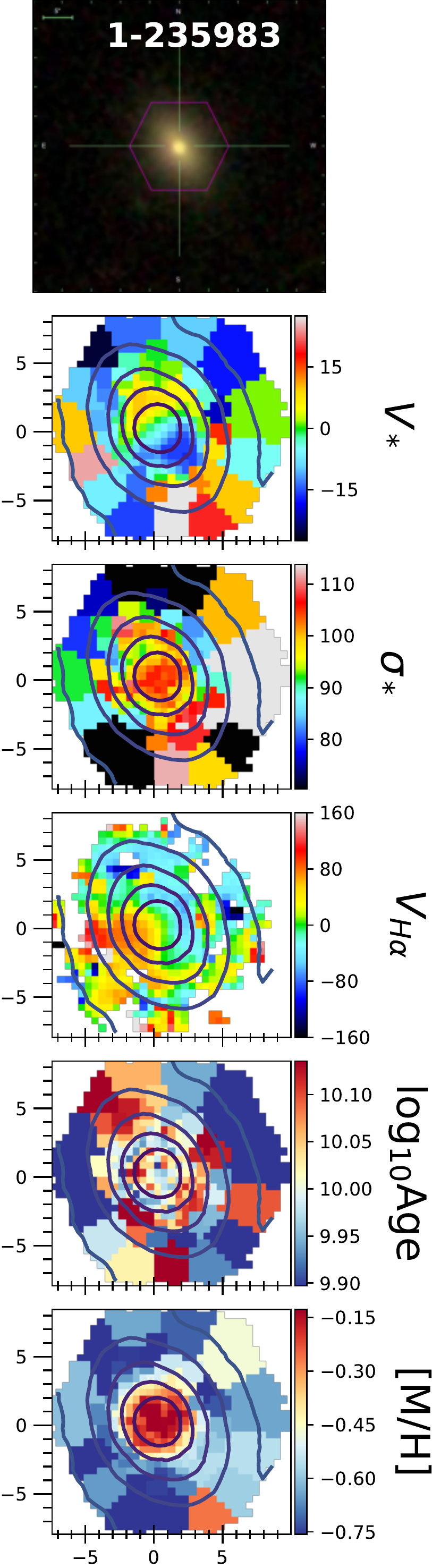}
\end{subfigure}
\caption{\textit{(continue)}}
\end{figure*}
\begin{figure*}
\ContinuedFloat
\begin{subfigure}{\textwidth}
\centering
\includegraphics[scale=0.38]{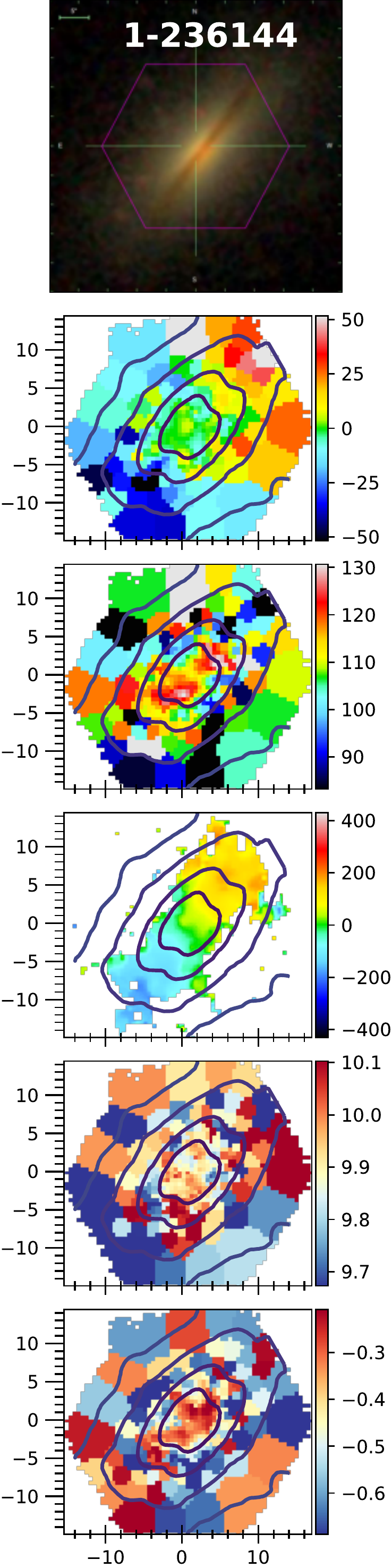}
\includegraphics[scale=0.38]{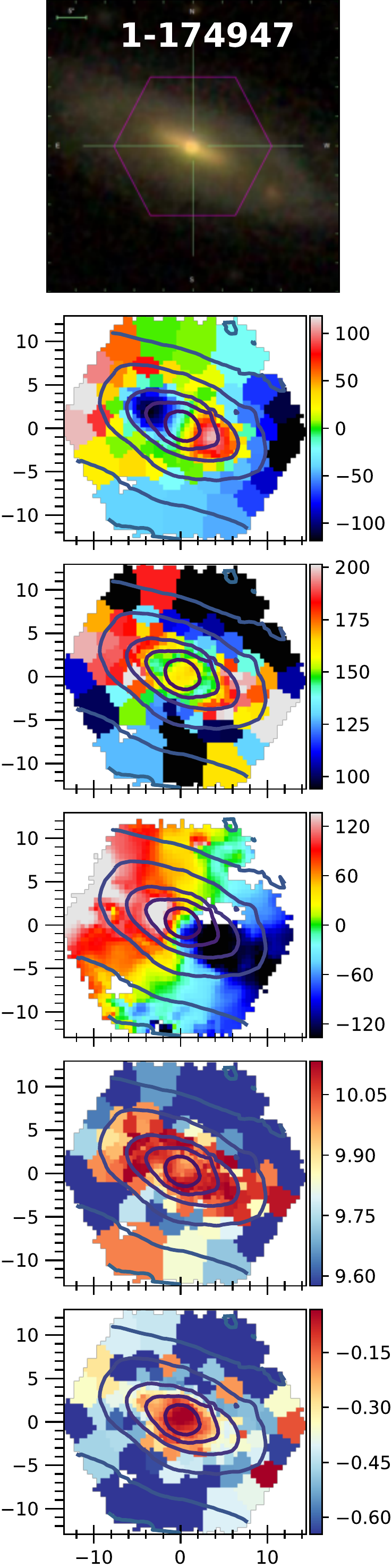}
\includegraphics[scale=0.38]{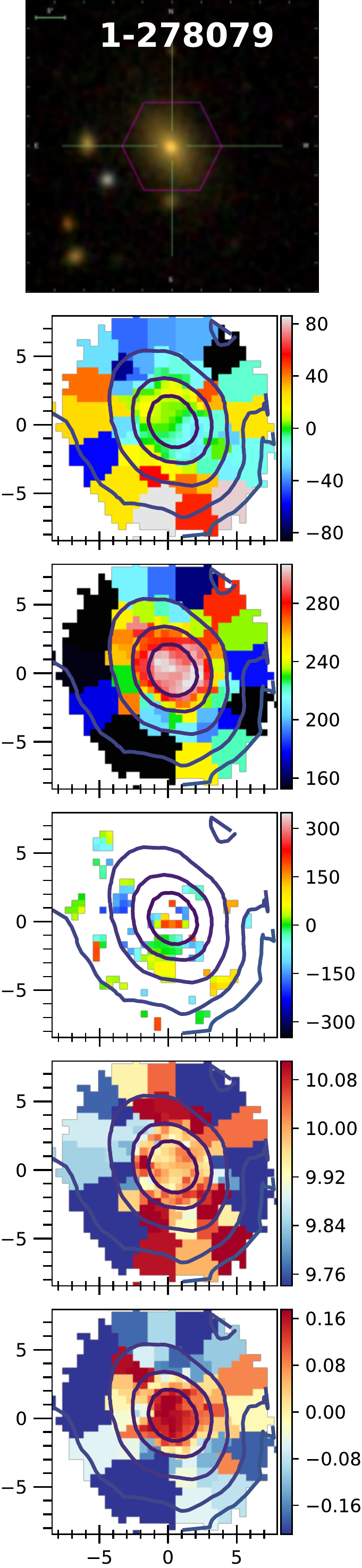}
\includegraphics[scale=0.38]{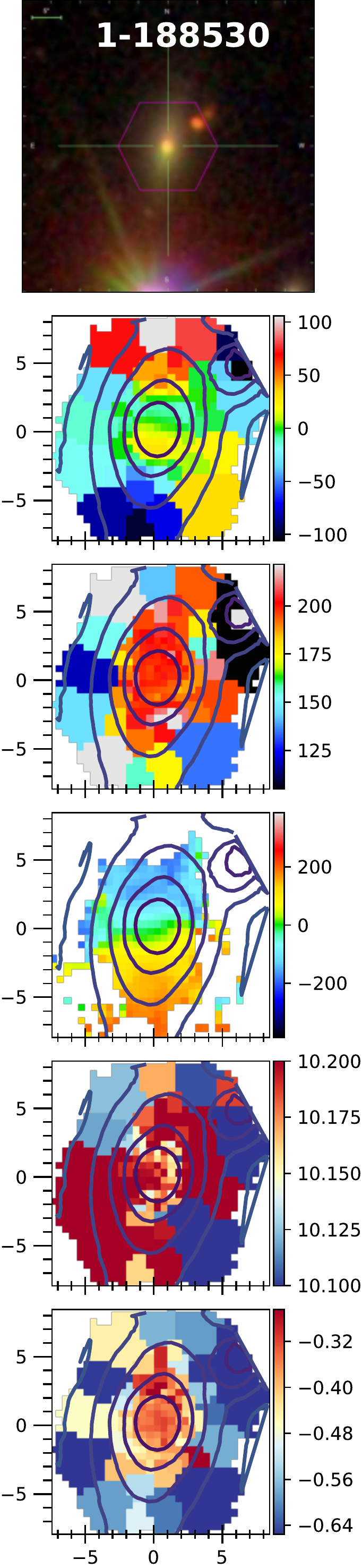}
\includegraphics[scale=0.38]{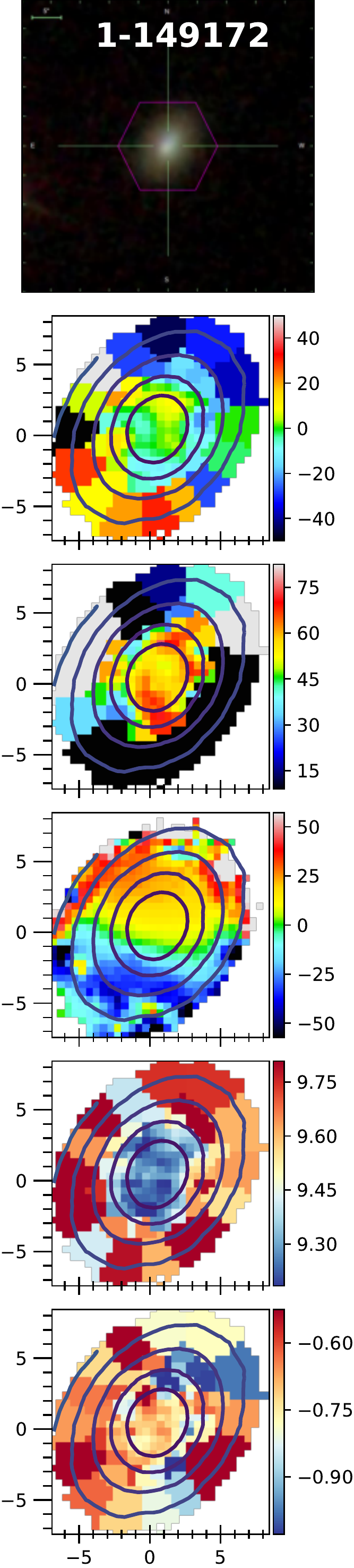}
\includegraphics[scale=0.38]{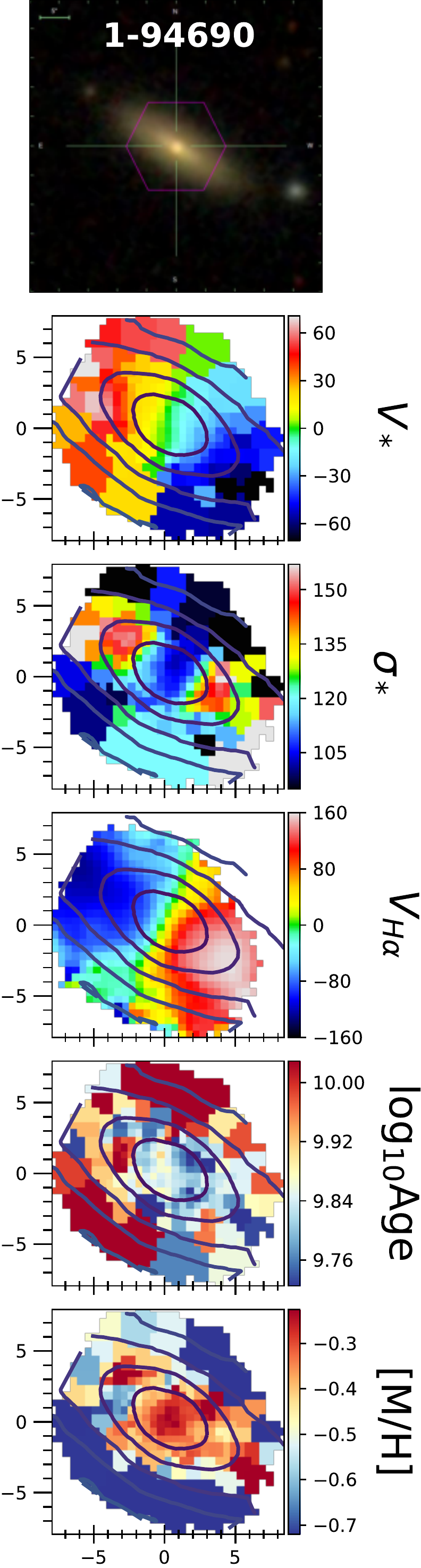}
\includegraphics[scale=0.38]{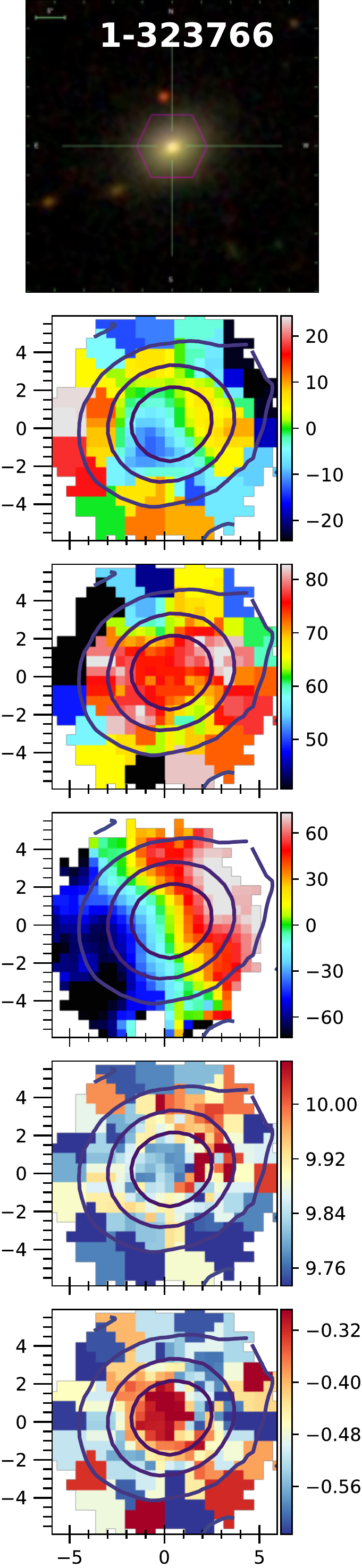}
\includegraphics[scale=0.38]{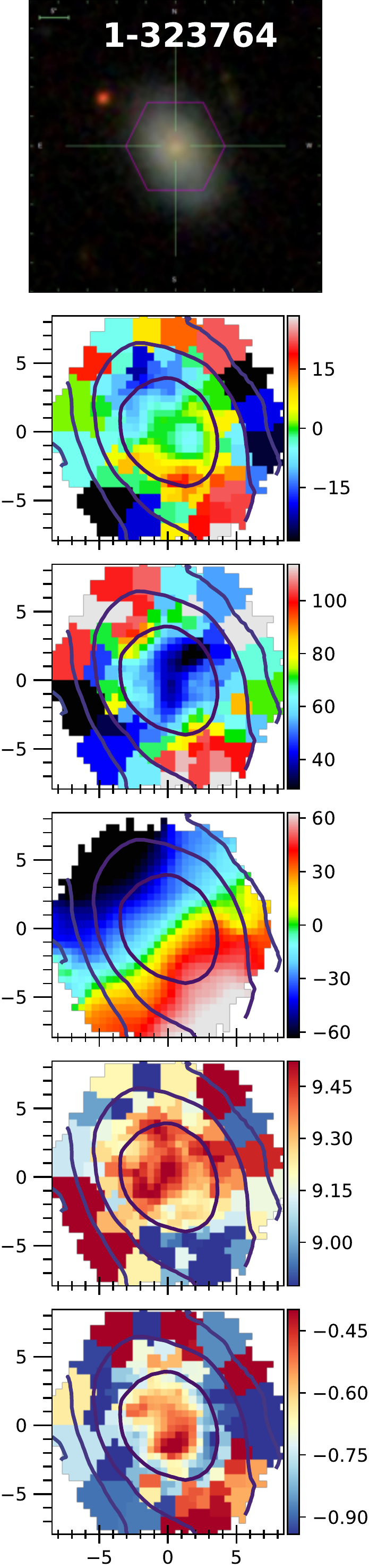}
\includegraphics[scale=0.38]{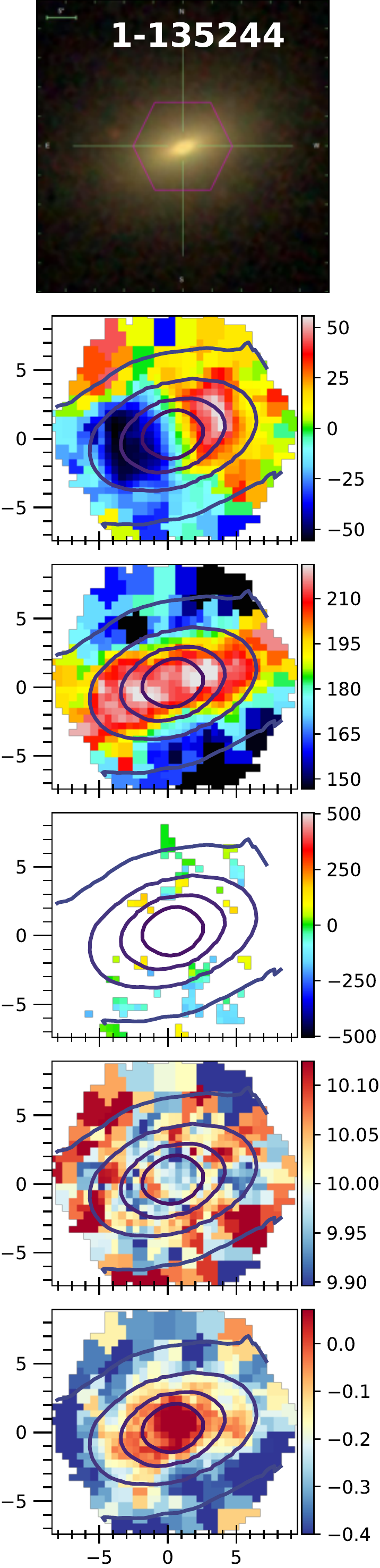}
\includegraphics[scale=0.38]{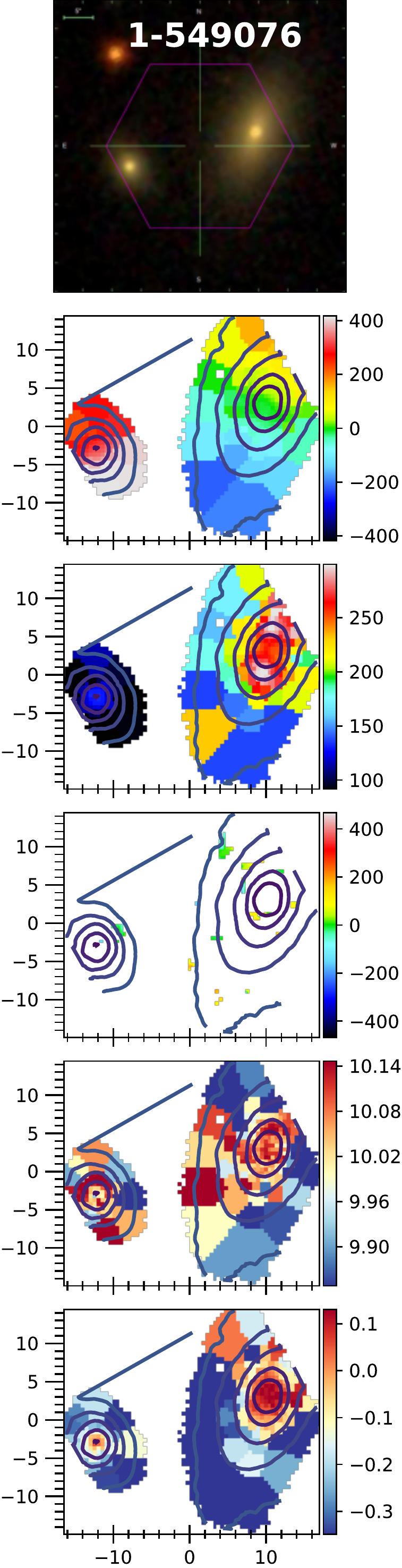}
\includegraphics[scale=0.38]{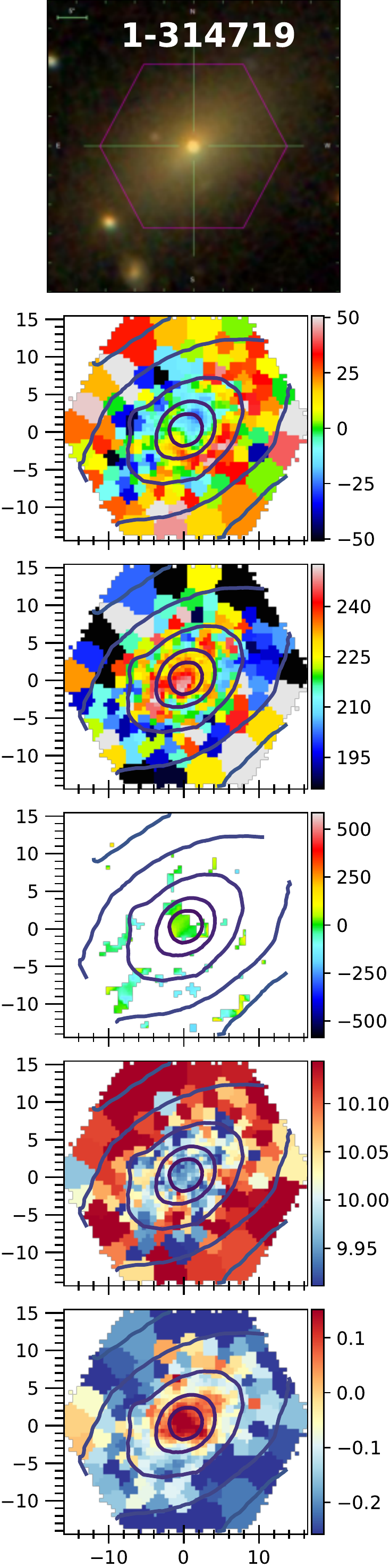}
\includegraphics[scale=0.38]{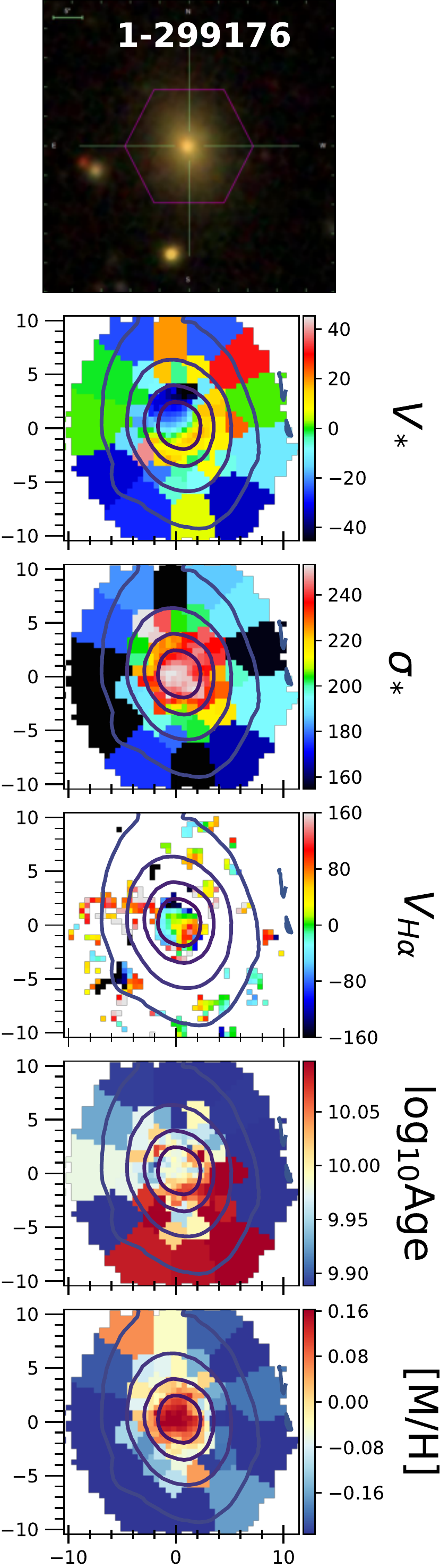}
\end{subfigure}
\caption{\textit{(continue)}}
\end{figure*}
\begin{figure*}
\ContinuedFloat
\begin{subfigure}{\textwidth}
\centering
\includegraphics[scale=0.37]{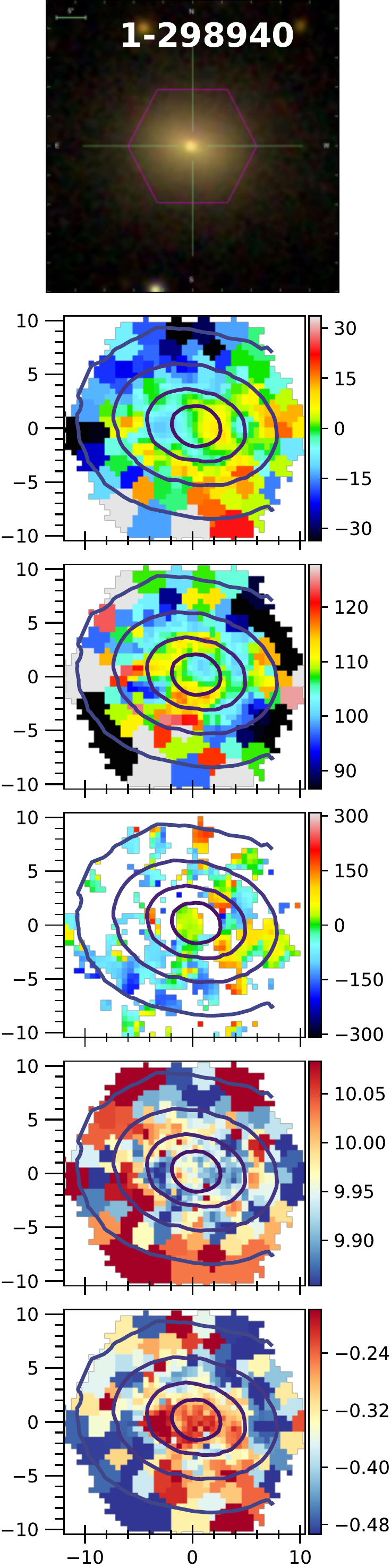}
\includegraphics[scale=0.37]{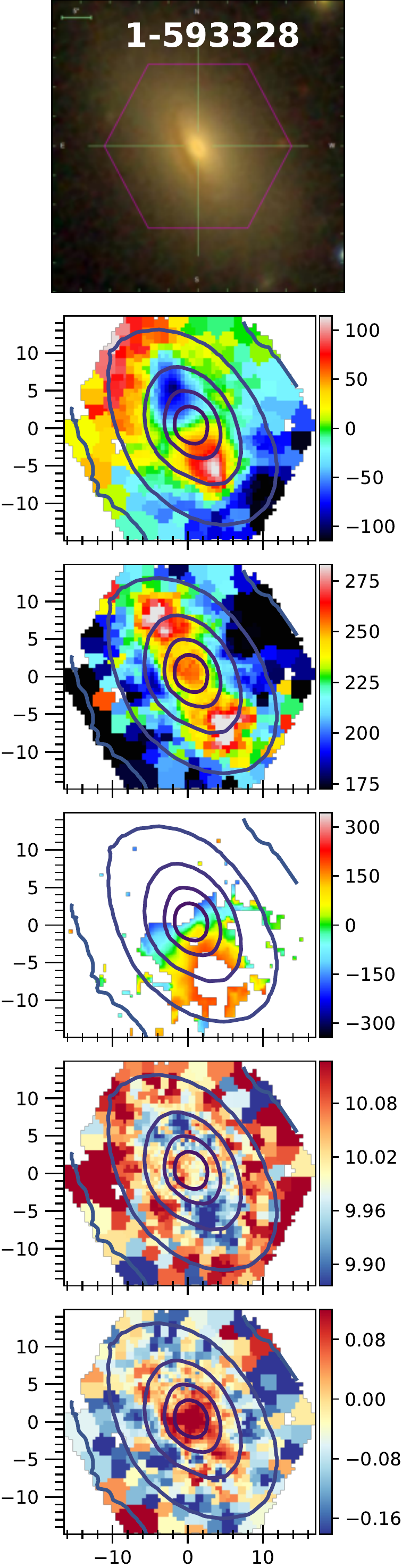}
\includegraphics[scale=0.37]{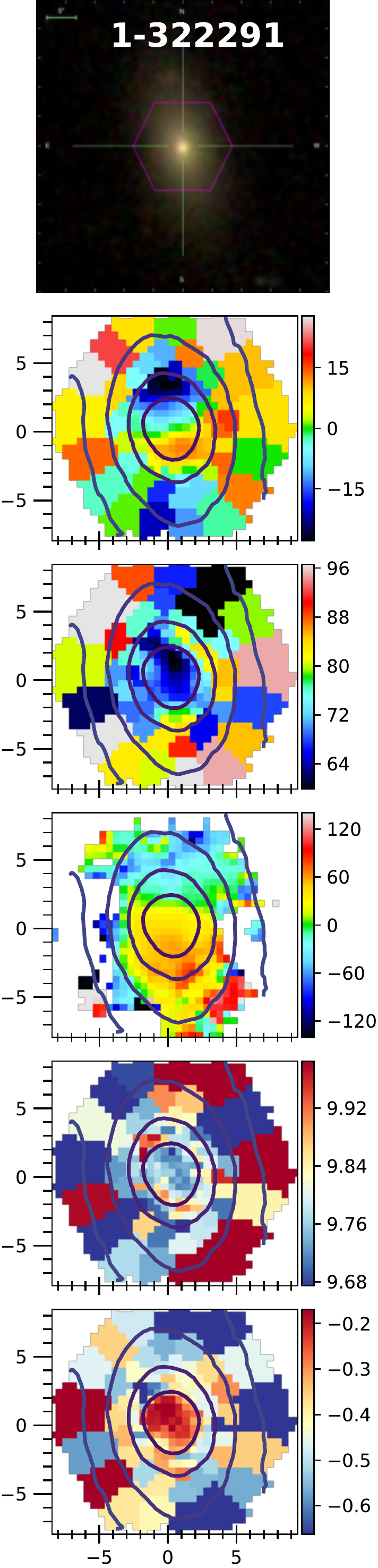}
\includegraphics[scale=0.37]{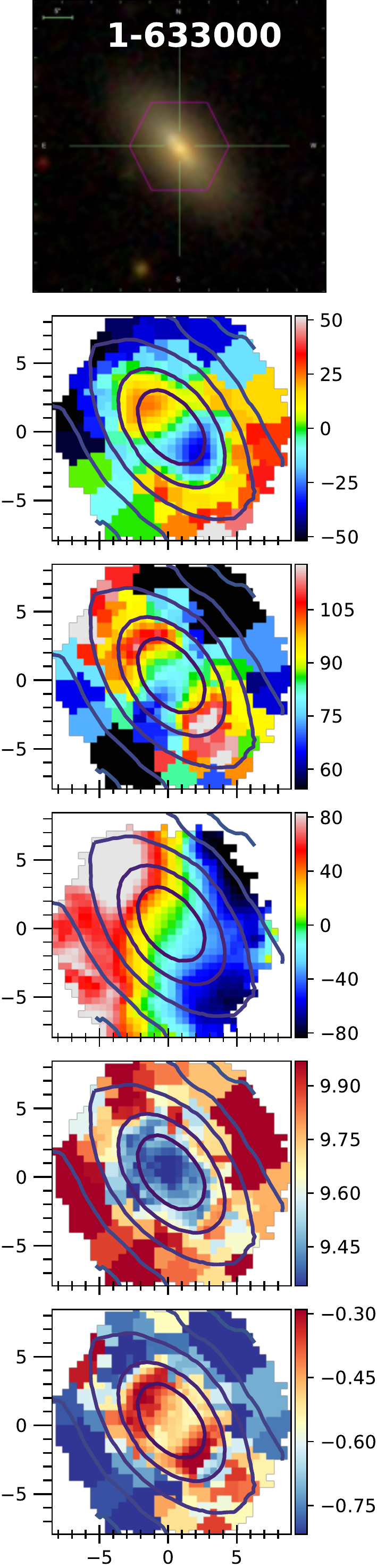}
\includegraphics[scale=0.37]{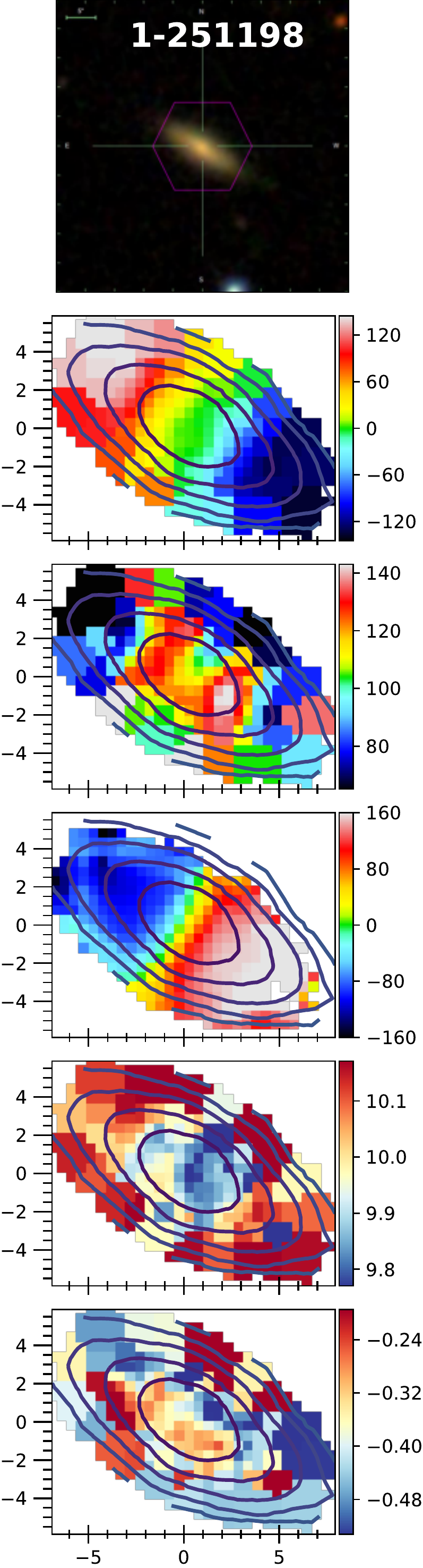}
\includegraphics[scale=0.37]{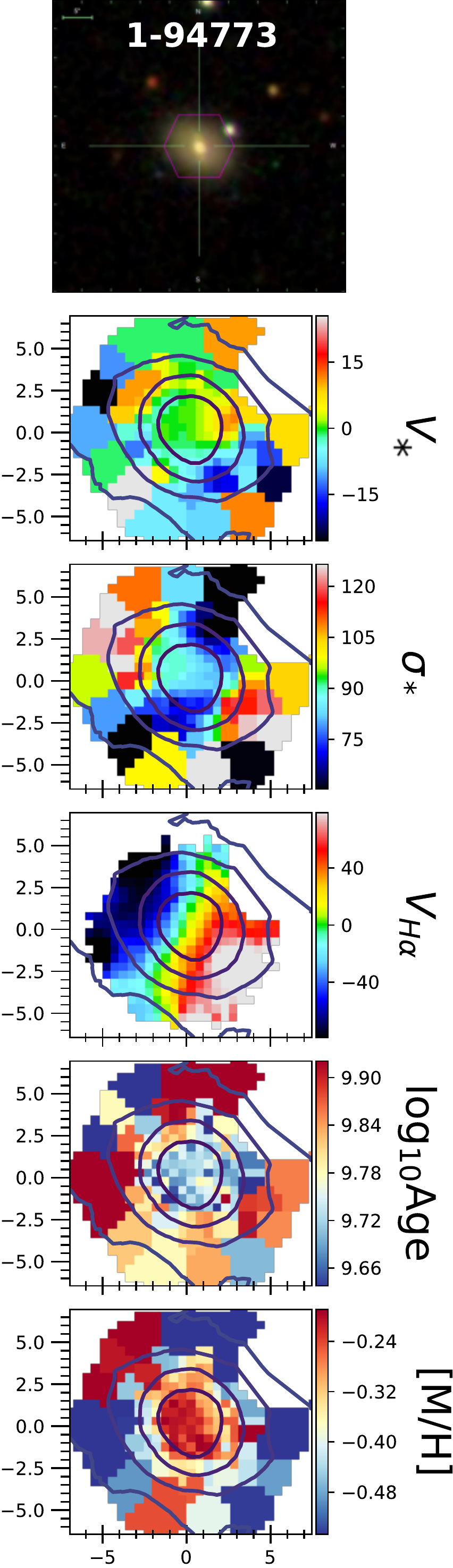}
\includegraphics[scale=0.37]{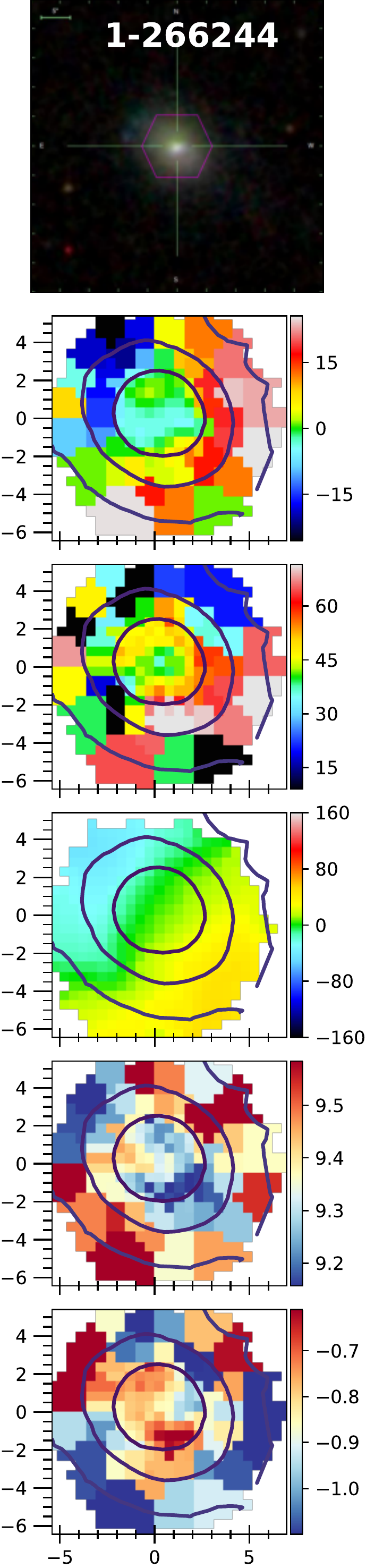}
\includegraphics[scale=0.37]{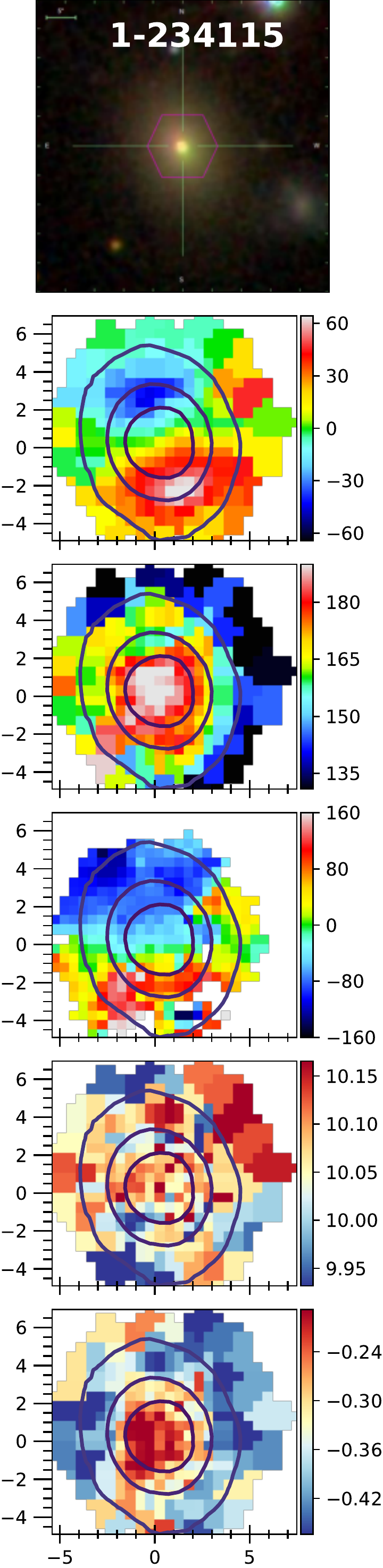}
\includegraphics[scale=0.37]{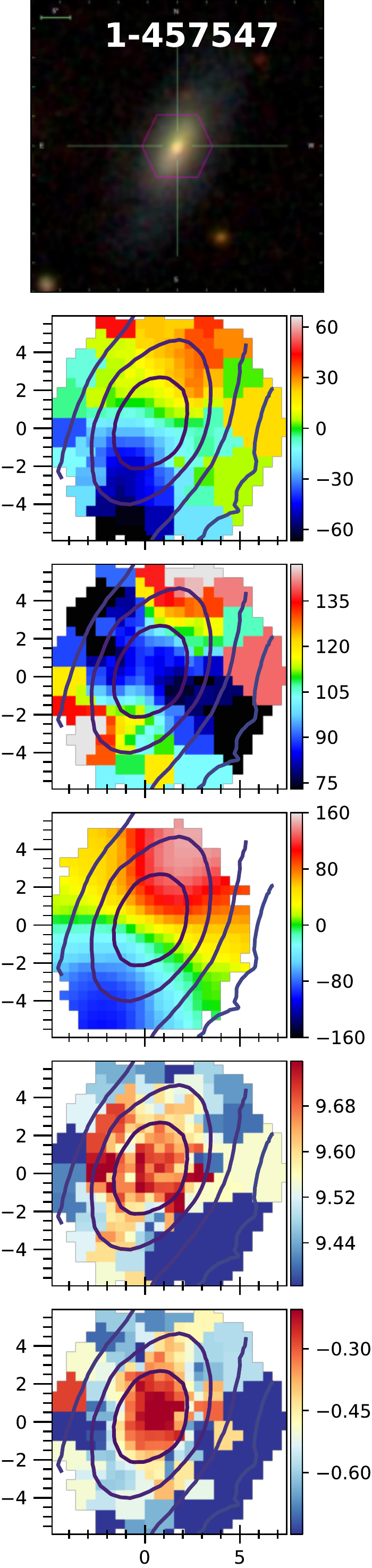}
\includegraphics[scale=0.37]{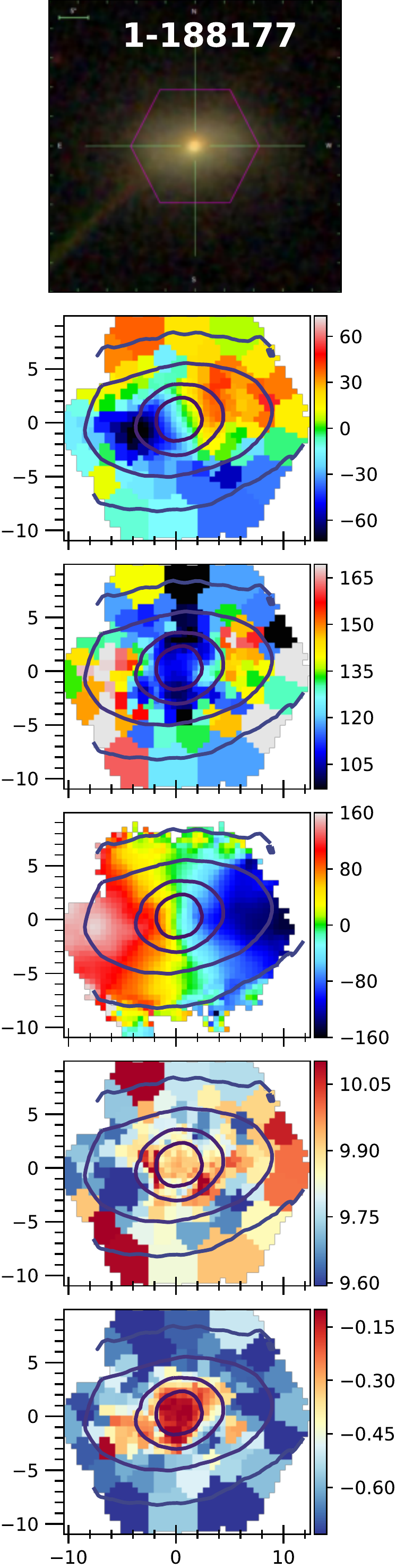}
\includegraphics[scale=0.37]{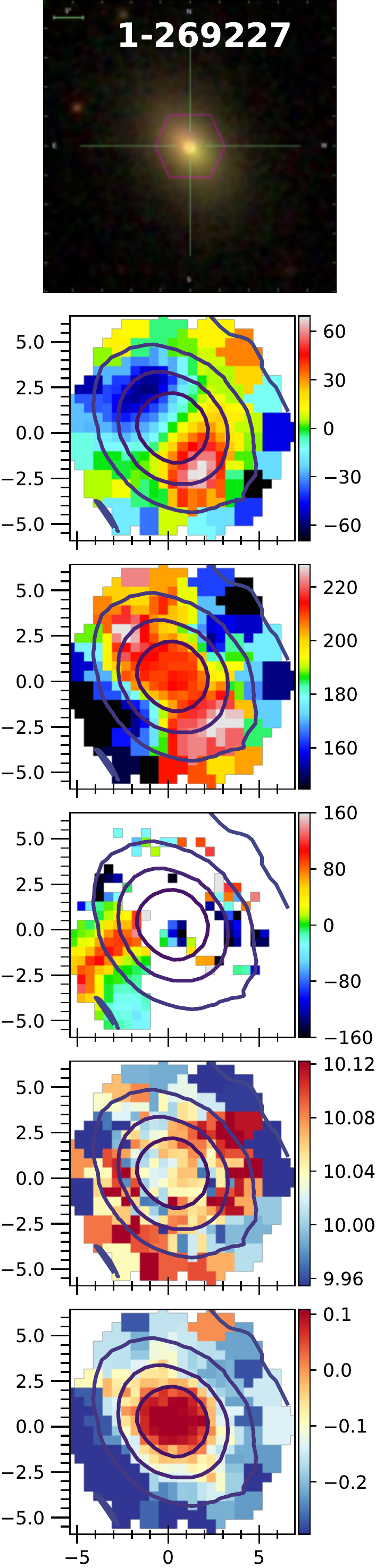}
\includegraphics[scale=0.37]{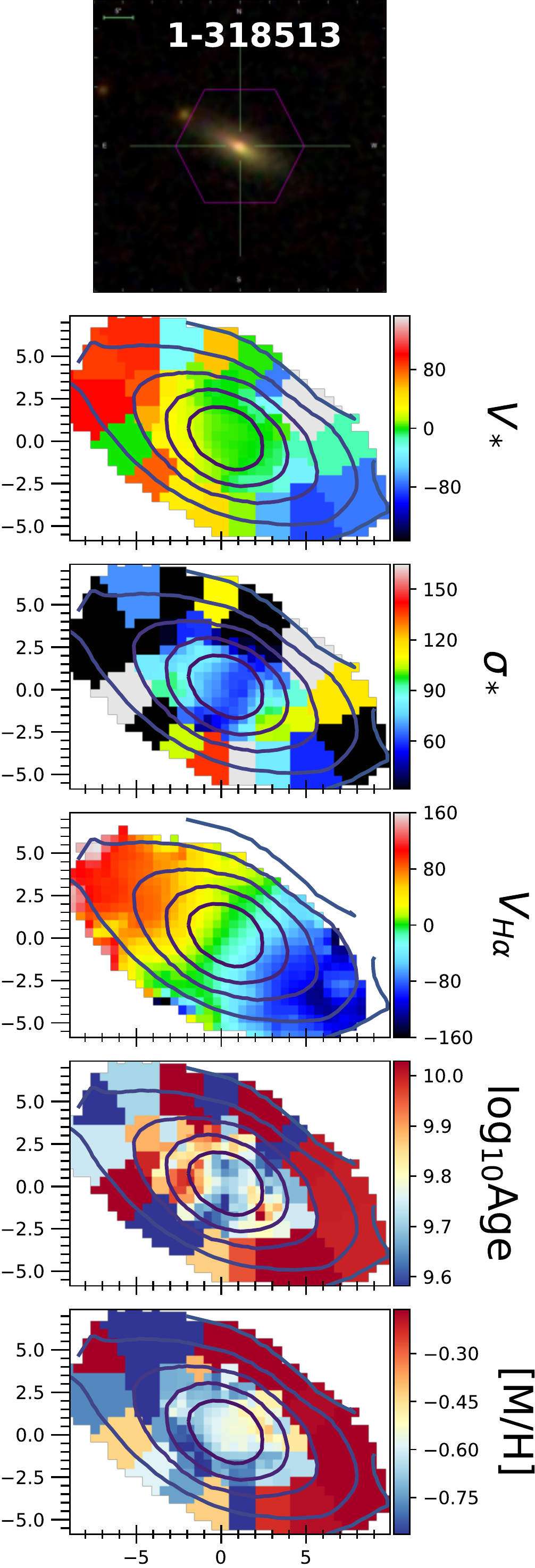}
\end{subfigure}
\caption{\textit{(continue)}}
\end{figure*}

\begin{figure*}
\ContinuedFloat
\begin{subfigure}{\textwidth}
\centering
\includegraphics[scale=0.38]{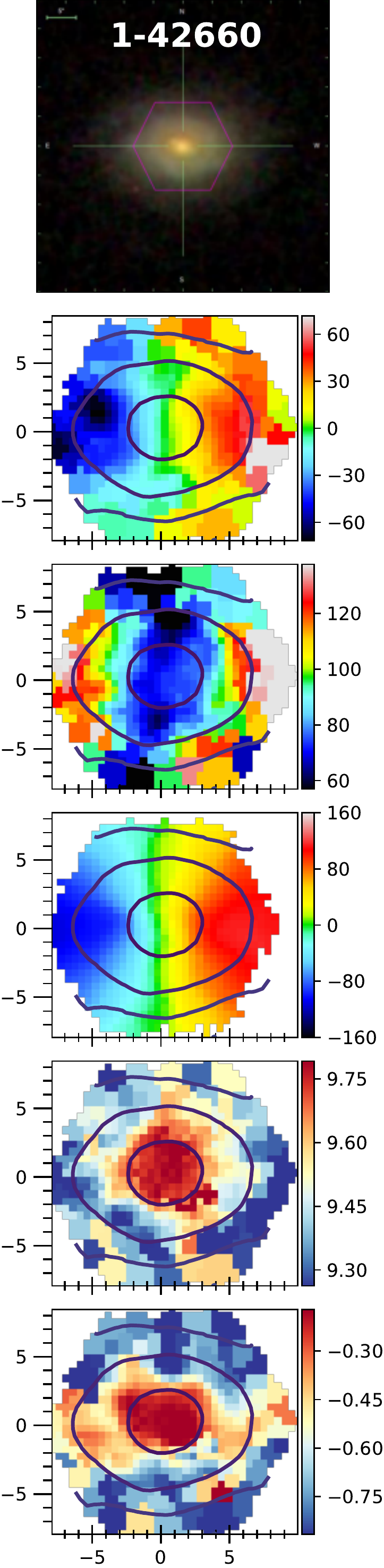}
\includegraphics[scale=0.38]{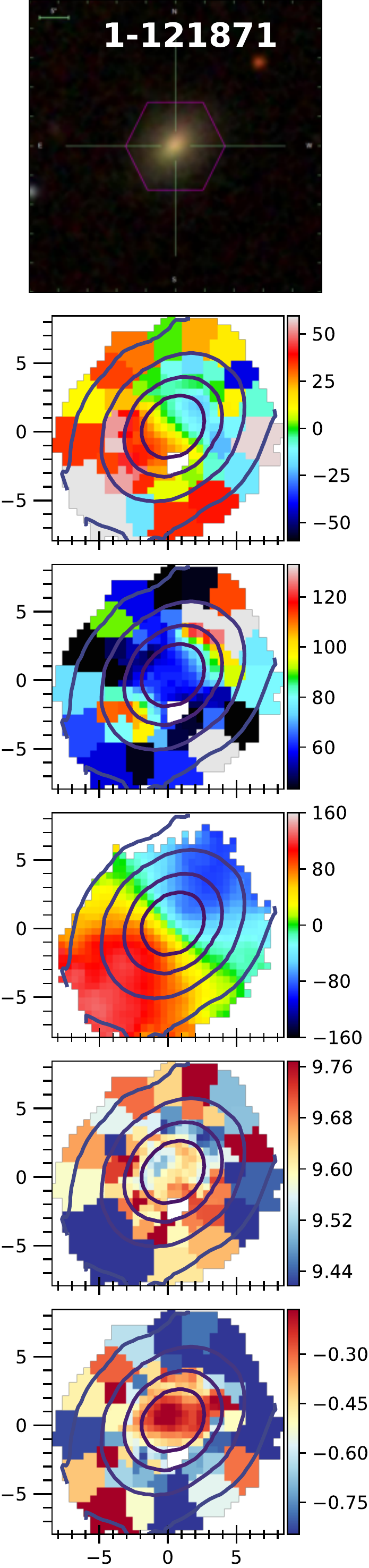}
\includegraphics[scale=0.38]{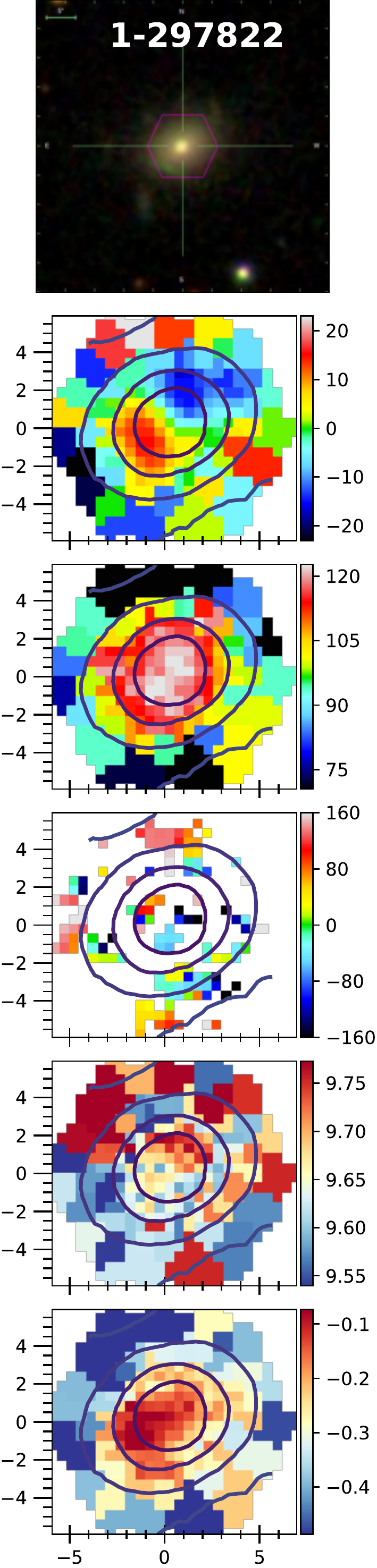}
\includegraphics[scale=0.38]{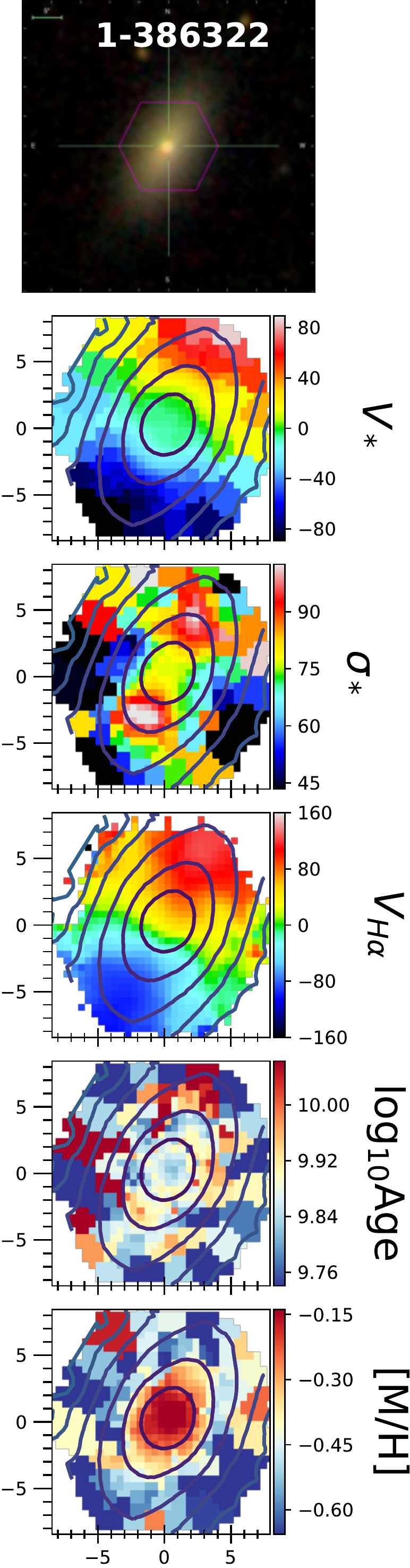}
\end{subfigure}
\caption{\textit{(continue)}}
\end{figure*}

\section{$\chi^2$ maps for the spectral fits}\label{app:chi2}
\begin{figure}
\centering
\includegraphics[width=.49\columnwidth]{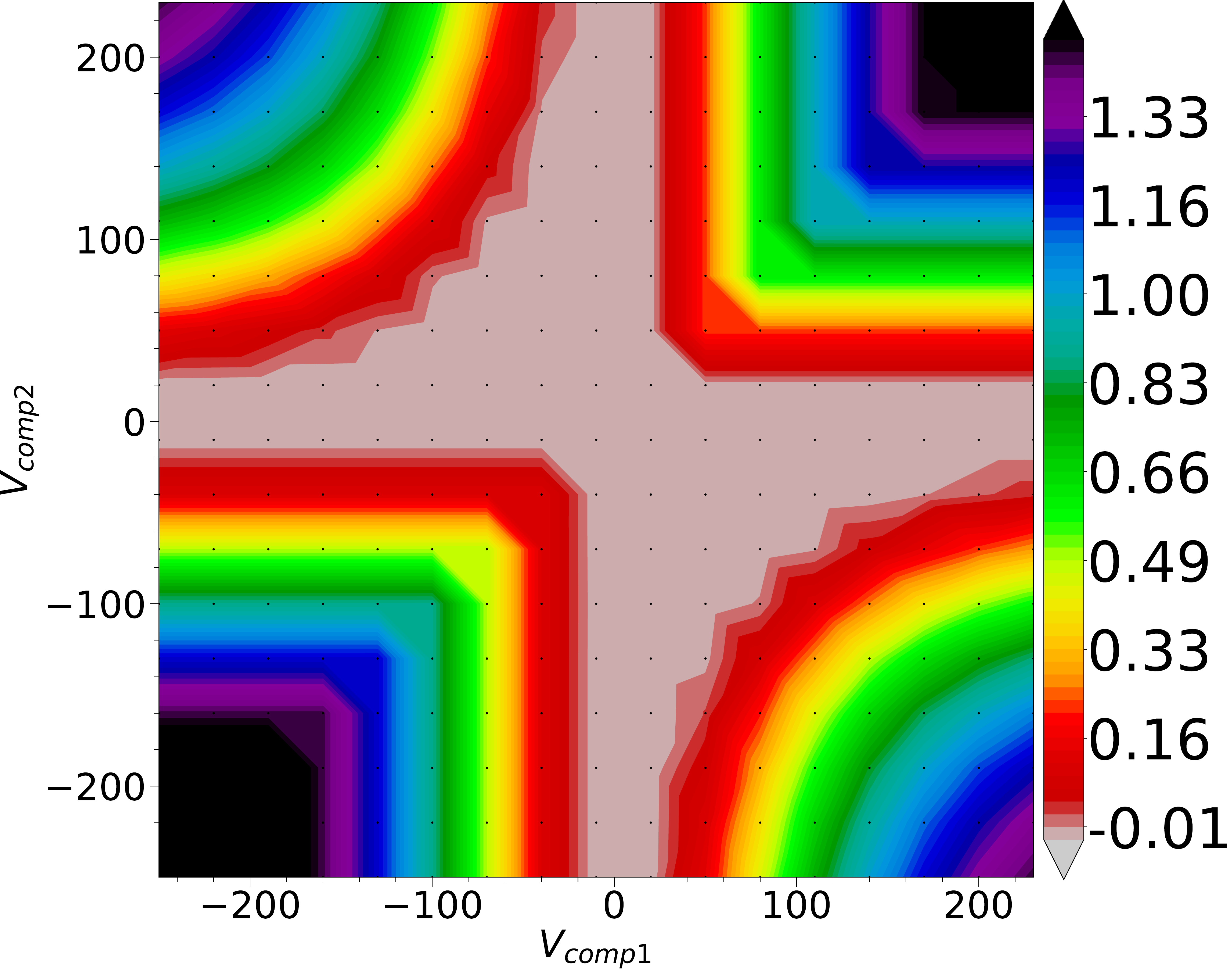}
\includegraphics[width=.49\columnwidth]{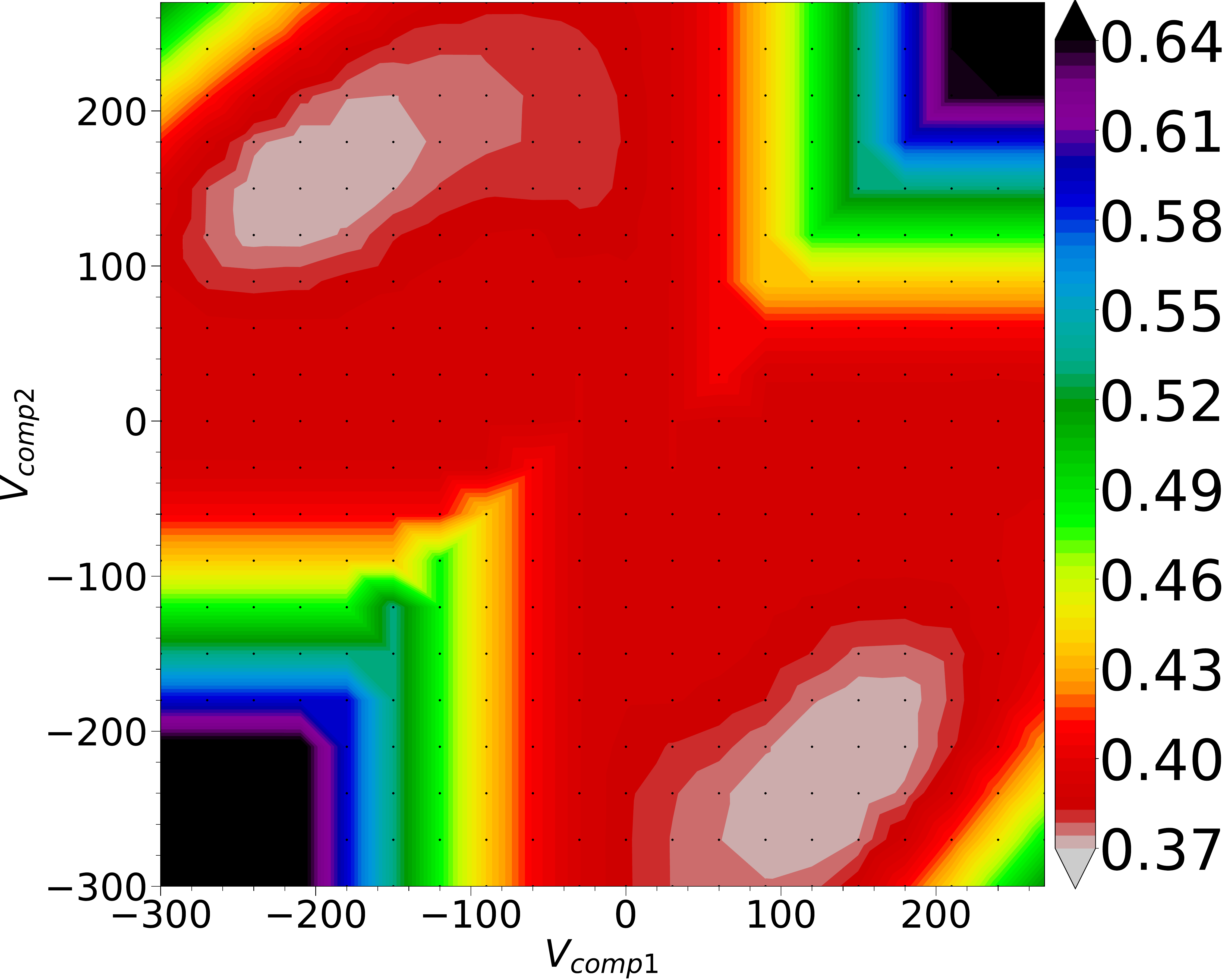}
\caption{Two examples of $\chi^2$ maps. The x and y axes are the fitted velocities $V_{\mbox{\scriptsize{comp1}}}$ and $V_{\mbox{\scriptsize{comp2}}}$, respectively, ranging between $\pm 300$ km s$^{-1}$. Black dots correspond to the velocity couples where the local $\chi^2$ minima are evaluated. The colorbar is the $\log_{10}$($\chi^2$/DOF). \textit{Left:} $\chi^2$ map of a synthetic galaxy composed of a single spectrum with velocity $V_{\mbox{\scriptsize{synth}}}=+150$ km s$^{-1}$. When fitting two components to a galaxy with only one kinematic component, or with two spectroscopically indistinguishable components, the minimum values of $\chi^2$ are those for which one fitted component has the velocity of the single-component solution ($\approx V_{\mbox{\scriptsize{synth}}}$); the other component, instead, can have any other velocities, for it has no weight in the fit. The global minimum will then result in a cross-like structure centered in $V_{\mbox{\scriptsize{comp1}}} = V_{\mbox{\scriptsize{comp2}}} \approx V_{\mbox{\scriptsize{synth}}}$. Here, the cross is centered at $\approx 0$ km s$^{-1}$ for the plotted velocities have been subtracted to $V_{\mbox{\scriptsize{synth}}}$. \textit{Right:} $\chi^2$ map of a synthetic galaxy composed of two spectra with velocities V$_{\mbox{\scriptsize{1}}} = -200$ km s$^{-1}$ and V$_{\mbox{\scriptsize{2}}} = +150$ km s$^{-1}$. When the two kinematic components are spectroscopically distinguishable, the best fit solutions will be those with $V_{\mbox{\scriptsize{comp1}}} \approx V_{\mbox{\scriptsize{1}}}$ and $V_{\mbox{\scriptsize{comp2}}} \approx V_{\mbox{\scriptsize{2}}}$, and with $V_{\mbox{\scriptsize{comp1}}} \approx V_2$ and $V_{\mbox{\scriptsize{comp2}}} \approx V_{\mbox{\scriptsize{1}}}$, thus resulting in two specular regions of minimum $\chi^2$ (butterfly-shape).}
\label{fig:chi2synth}
\end{figure}

\begin{figure}
\centering
\begin{subfigure}{0.45\columnwidth}
\includegraphics[width=\columnwidth]{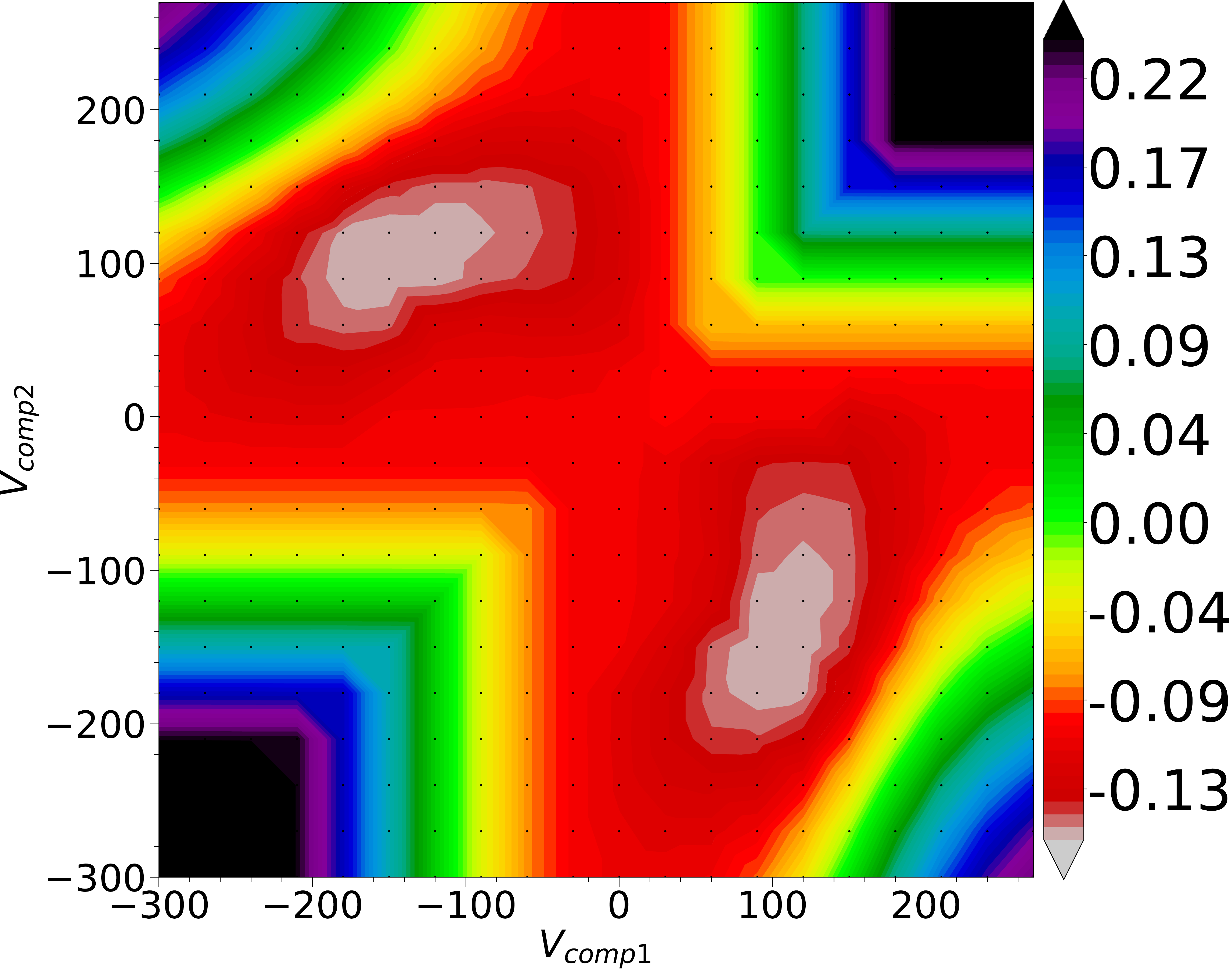}
\caption{}
\end{subfigure}
\begin{subfigure}{0.45\columnwidth}
\includegraphics[width=\columnwidth]{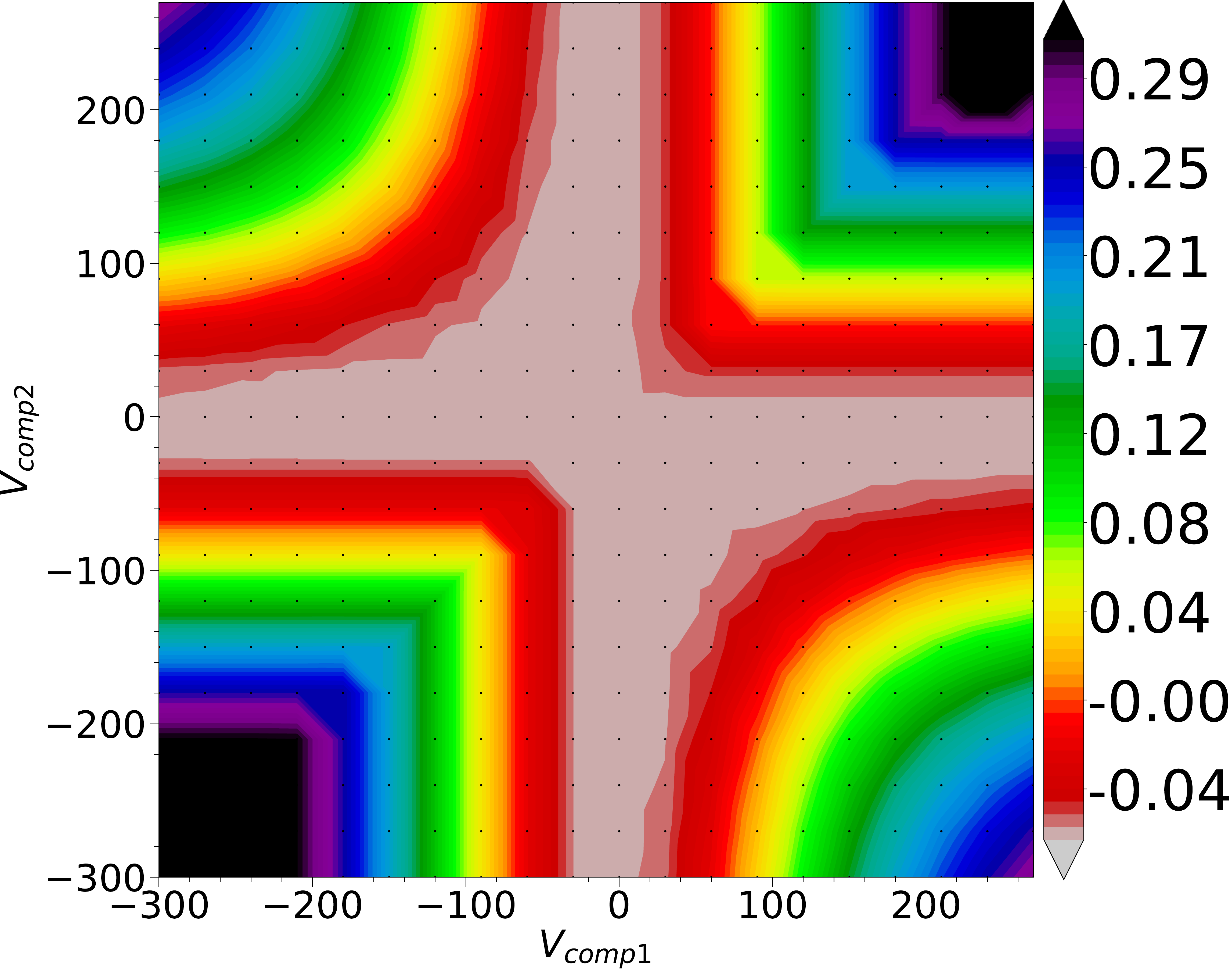}
\caption{}
\end{subfigure}\\
\begin{subfigure}{0.45\columnwidth}
\includegraphics[width=\columnwidth]{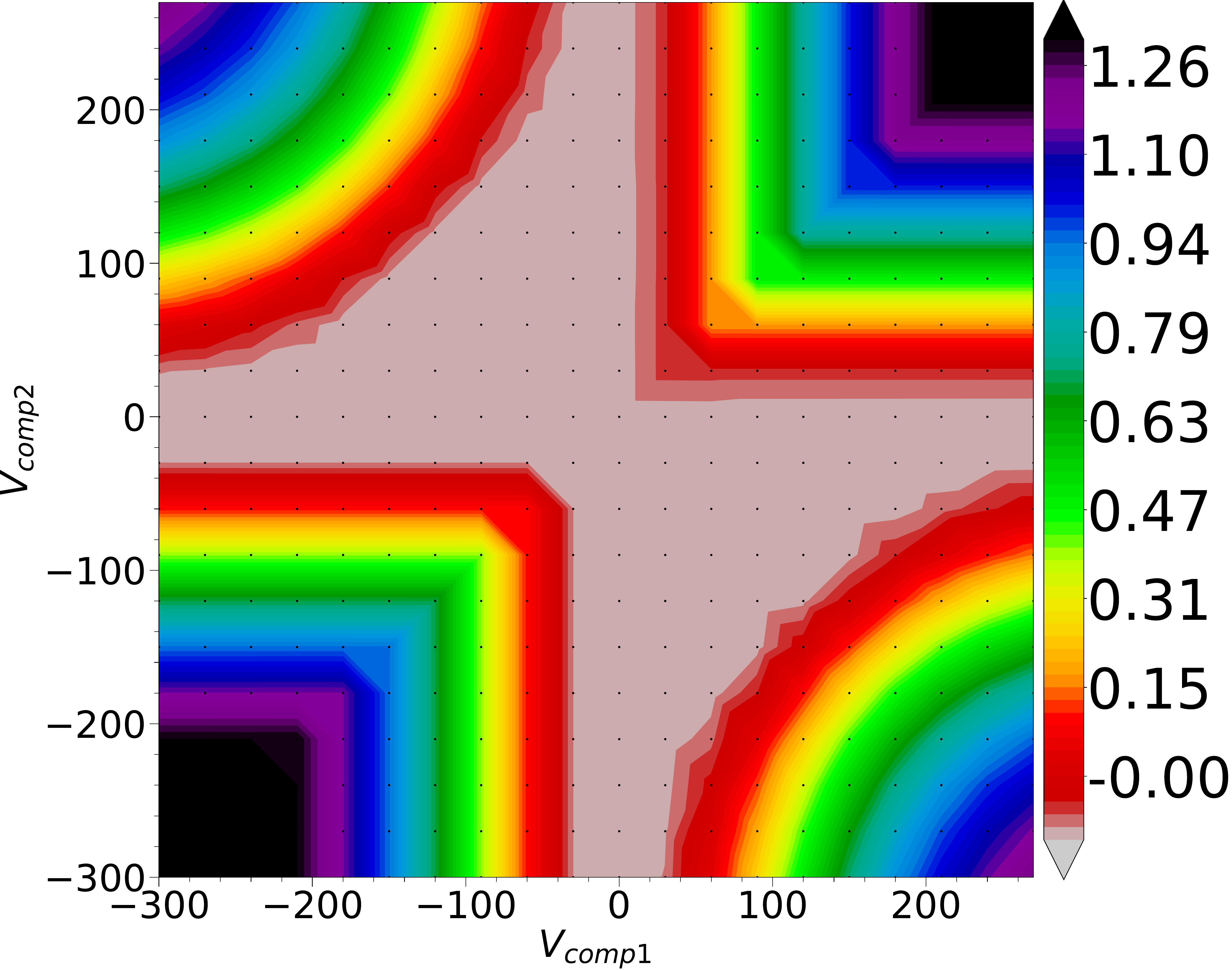}
\caption{}
\end{subfigure}
\begin{subfigure}{0.45\columnwidth}
\includegraphics[width=\columnwidth]{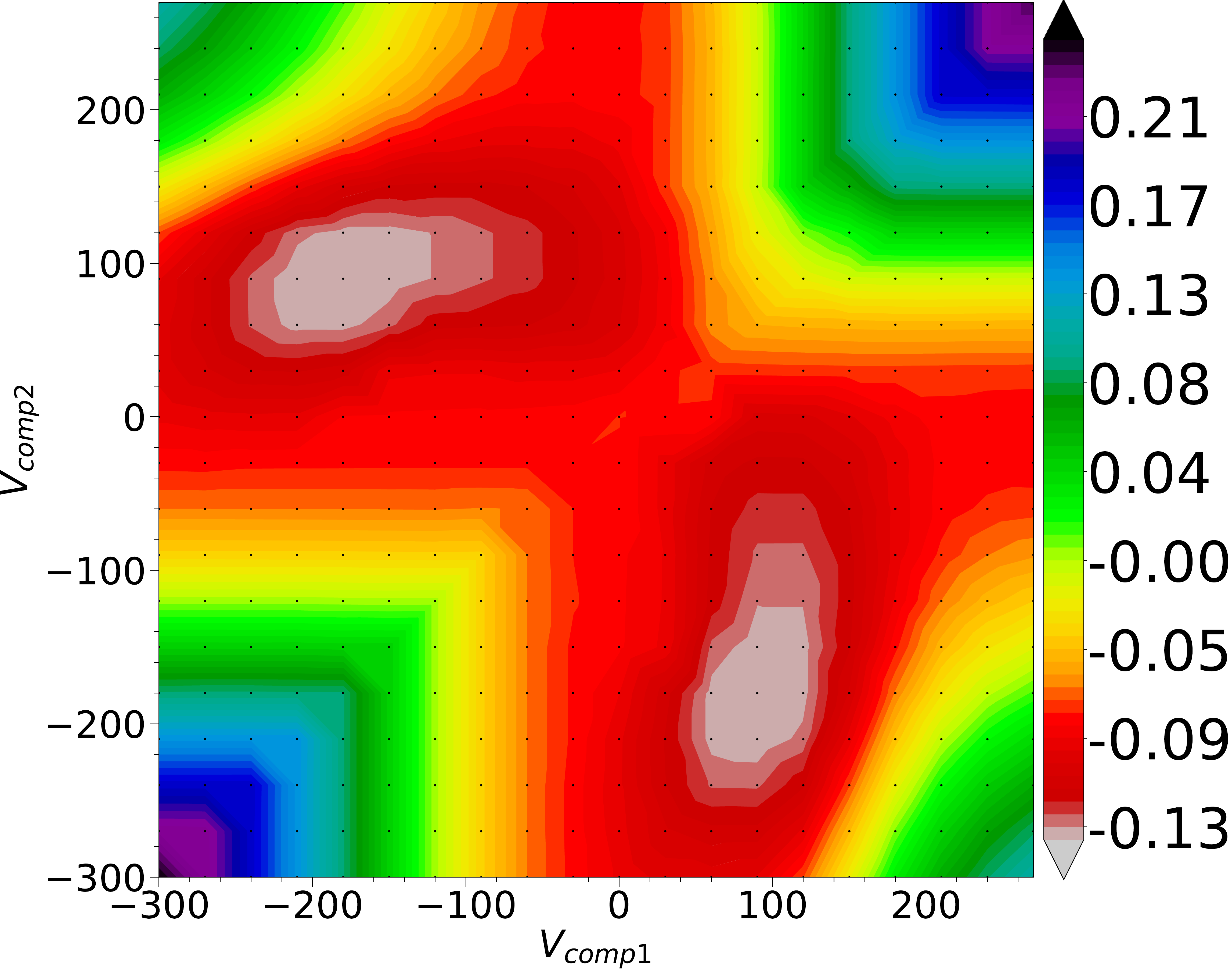}
\caption{}
\end{subfigure}
\caption{$\chi^2$ maps for the galaxy with MaNGA ID: 1-38543. (a) $\chi^2$ map extracted around one of the two $\sigma_\ast$ peaks, as described is section 3.3 of the main paper; (b) $\chi^2$ map extracted in a region along the minor axis; (c) $\chi^2$ map produced using the sum of the spectra of all spatial bins of the galaxy as input spectrum; (d) $\chi^2$ map produced using all spectra of MILES-HC as templates.}
\label{fig:chi2tests}
\end{figure}

In the following, we first describe how the $\chi^2$ maps are produced, and then how to read them. In the next section, we discuss our method, and present some tests we made to improve it.

When fitting the spectrum of a CRD, if the two stellar disks are not spectroscopically distinguishable, the two-component fit will return two best fit solutions as follows: for one component, the velocity will be the same as the single-component fit; the other component, instead, can be assigned with any velocity (in the imposed velocity range), for it has no weight in the fit. This will result in a `cross-like' structure aligned with the coordinate axes of minimum $\chi^2$ in the maps, which is the same result we expect from a galaxy composed of a single disk. An example of this cross degeneracy is shown in the left panel of Figure \ref{fig:chi2synth}, which shows the $\chi^2$ map of a synthetic galaxy spectrum composed of a single kinematic component. On the other hand, when the two disks are spectroscopically distinguishable, the global minimum will be found in two specular regions in the $\chi^2$ map, which presents a characteristic `butterfly-shape' with two distinct regions of minimum $\chi^2$. As an example, the right panel of Figure \ref{fig:chi2synth} shows the $\chi^2$ map of a synthetic galaxy composed of two kinematic components. From this picture, we can clearly distinguish two specular regions of minimum $\chi^2$: in these regions, both components are weighted with a significant fraction of the total weights, i.e. they both contribute importantly to the fit.

The first step to produce optimal $\chi^2$ maps is to properly choose the spectrum to be fitted. We constructed a `collapsed spectrum' by summing all spectra the region with one of the two $\sigma_\ast$ peaks or, equivalently, the inversion of the velocity field. We then performed a single component fit on the collapsed spectrum, with the same parameters of the first single component fit (section 3.2 of the main paper). From this fit, we get the $\sigma_{\mbox{\scriptsize{std}}}$, that we use both to mask the > 3$\sigma_{\mbox{\scriptsize{std}}}$ outliers and as the input noise for the two-component fit, and the best-fit spectrum, calculated as the weighted sum of all the template spectra, that will be used as template to fit both components in the two-components fit. The $\chi^2$ maps are finally produced by fitting two components to the collapsed spectrum as follows. For each component, we consider a range of starting velocities between $\pm$ 300 km s$^{-1}$ with respect to the single component solution for the velocity (resulting in a map centered in $V_{comp1} = V_{comp2} = 0$ ), separated by a fixed velocity step $V_{step}$ = 30 km s$^{-1}$, and, for each couple of starting velocities, we compute the $\chi^2$/DOF and plot it on the map. In each fit, we impose the velocities to lie within $V_{i,start} - V_{step} /2 \leq V_i \leq V_{i,start} + V_{step} /2$ (where i stands for either comp1 or comp2), to force the best fit solution to lie in the interval around the starting values, while keeping the velocity dispersions unconstrained. A $\chi^2$ map then show the resulting minimum $\chi^2$ values, calculated at different couples of fitted velocities. Note that 'comp1' and 'comp2' are just nominal attributions, i.e. they do not specifically designate one of the two kinematic components, and the two fitted components are interchangeable for \texttt{pPXF}.\\

\subsection{Tests on two-components fits}\label{sect:chi2test}
In this section we discuss some tests we made to improve the method of $\chi^2$ maps for the recovery of the two kinematic components. In the following, we discuss these tests using a single representative galaxy, whose $\chi^2$ maps, extracted as described in the previous section, is shown in Figure \ref{fig:chi2tests} (a). However, these tests have been made on many galaxies exhibiting the cross-degeneracy, the two minima, or uncertain maps, and the results are roughly the same for all galaxies tested, with no significant differences.

The choice of considering, for building the $\chi^2$ maps, only the regions where the CRD features appear is justified by the fact that those are the regions where we expect the separation between the two components to be spectroscopically more evident. Extracting $\chi^2$ maps at different locations shows that the distinction of the two $\chi^2$ minima, while evident in the collapsed spectrum of the region of the $\sigma_\ast$ peak, gets less clear as one moves away from the peak, and soon disappears. An example of a $\chi^2$ map extracted along the kinematic minor axis is shown in Figure \ref{fig:chi2tests} (b). We also performed the two-component fits by considering the sum of all spatial bins of the galaxy as the input spectrum; though, such spectrum results biased towards the major contributor to the total flux of the two disks, thus resulting in a cross-degeneracy, notwithstanding the higher S/N, as shown in Figure \ref{fig:chi2tests} (c). By testing these possibilities, we confirmed that the best input spectrum to be fitted for the recovery of the two kinematic components is that of those bins around the regions of the peaks.

The choice of fitting two kinematic components with the same template spectrum could introduce a bias if the two stellar disks had different intrinsic spectra (e.g. because of significantly different ages and metallicities). A simple alternative we tested was to calculate two different best-fit spectra from the single component fits, with the whole MILES-HC library at two distinct spatial regions of the galaxy, where the inner and the outer disk prevail. Then two best-fit templates (i.e. the weighted sums of MILES-HC relative to the two single-component fits) are used to fit the two components. We found that, in general, this approach gives the same qualitative results of our former approach; however, fits of the latter are generally worse, primarly because the collapsed spectrum of the outer region (which is typically outside $R_e$) is often noisy, and the resulting best-fit spectrum is often not well representative of the collapsed spectrum considered to fit the two components.

The best approach to fit the two components would be to use the whole template library, despite the longer computational time required. We tested such approach to see how significant the improvement would be on the recovery of the two components, and we found that, although the fits are generally better, the $\chi^2$ maps are qualitatively the same. An example is shown in Figure \ref{fig:chi2tests} (d). Thus, to save computational time, we did not use the whole library for the two-components fit.
 
The recovery of the two components also depends on the fraction of the total flux with which each disk contributes. With \texttt{pPXF} it is possible to enforce linear constraints on the template weights during the fit; in other words, it is possible to constrain the fluxes of the two components to contribute a certain fraction of the total flux. This should in principle avoid the cross-degeneracy in the $\chi^2$ maps, because none of the two components is allowed to be weighted as zero, and both components are forced to contribute to the fit. We tested the method of the linear constraints on many galaxies, but, although the cross-shape disappears, the degeneracy remains. As a general result, galaxies with two minima in the $\chi^2$ maps turn out to show a clearer distinction when increasing the (imposed) contribution to the flux of the fitted component, while galaxies with the cross lose the cross-shape, but no distinction of the two minima was obtained.

%%%%%%%%%%%%%%%%%%%%%%%%%%%%%%%%%%%%%%%%%%%%%%%%%%
\bibliographystyle{mnras}
\bibliography{biblio}

%\begin{thebibliography}{99}
%\end{thebibliography}

% Don't change these lines
\bsp	% typesetting comment
\label{lastpage}
\end{document}